\documentclass[5p, authoryear, twocolumn]{elsarticle}   	% use "amsart" instead of "article" for AMSLaTeX format, 5p or review
\usepackage{geometry}                		% See geometry.pdf to learn the layout options. There are lots.
\usepackage{graphicx}				% Use pdf, png, jpg, or eps� with pdflatex; use eps in DVI mode
\usepackage{amssymb}

\usepackage{subfigure}

\usepackage{wrapfig}
\usepackage{lscape}
\usepackage{rotating}
\usepackage{epstopdf}
\usepackage{afterpage}
\usepackage{flafter}

 \usepackage{mathtools}
 \usepackage{bm}
 
\usepackage{csvsimple}
\usepackage{longtable}
\usepackage{booktabs}
\usepackage{array}

\usepackage{pdflscape}
 
 \usepackage{subfig}
 
 \usepackage{chapterbib}
 
\renewcommand{\vec}[1]{\bm{#1}}
 
 \usepackage[switch]{lineno}
\newcommand*\patchAmsMathEnvironmentForLineno[1]{%
  \expandafter\let\csname old#1\expandafter\endcsname\csname #1\endcsname
  \expandafter\let\csname oldend#1\expandafter\endcsname\csname end#1\endcsname
  \renewenvironment{#1}%
     {\linenomath\csname old#1\endcsname}%
     {\csname oldend#1\endcsname\endlinenomath}}% 
\newcommand*\patchBothAmsMathEnvironmentsForLineno[1]{%
  \patchAmsMathEnvironmentForLineno{#1}%
  \patchAmsMathEnvironmentForLineno{#1*}}%
\AtBeginDocument{%
\patchBothAmsMathEnvironmentsForLineno{equation}%
\patchBothAmsMathEnvironmentsForLineno{align}%
\patchBothAmsMathEnvironmentsForLineno{flalign}%
\patchBothAmsMathEnvironmentsForLineno{alignat}%
\patchBothAmsMathEnvironmentsForLineno{gather}%
\patchBothAmsMathEnvironmentsForLineno{multline}%
}
 
\journal{JGR: Planets}

\begin{document}\sloppy

%\linenumbers
\clearpage
\clearpage

\begin{frontmatter}

\author[harv]{Simon J. Lock \corref{cor}}
\author[dav]{Sarah T. Stewart}
\cortext[cor]{Corresponding author: slock@fas.harvard.edu}
\address[harv]{Department of Earth and Planetary Sciences, Harvard University, 20 Oxford Street, Cambridge, MA 02138, U.S.A.}
\address[dav]{Department of Earth and Planetary Sciences, U. California Davis, One Shields Avenue, Davis, CA 95616, U.S.A.}

\title{The structure of terrestrial bodies: Impact heating, corotation limits and synestias}

%xxxxxxxxxxxxxxxxxxxxxxxxxxxxxxxxxxxxxxxxxxxxxxxxxxxxxxxxxxxxxxxxxxxxxxxxxxxxxxxxxxxxxxxxxxxxxxxxxxxxxxxxxxxxxxxxxxxxxxxxxxxxxxxxxxxxxxxx
%xxxxxxxxxxxxxxxxxxxxxxxxxxxxxxxxxxxxxxxxxxxxxxxxxxxxxxxxxxxxxxxxxxxxxxxxxxxxxxxxxxxxxxxxxxxxxxxxxxxxxxxxxxxxxxxxxxxxxxxxxxxxxxxxxxxxxxxx
%xxxxxxxxxxxxxxxxxxxxxxxxxxxxxxxxxxxxxxxxxxxxxxxxxxxxxxxxxxxxxxxxxxxxxxxxxxxxxxxxxxxxxxxxxxxxxxxxxxxxxxxxxxxxxxxxxxxxxxxxxxxxxxxxxxxxxxxx
\begin{abstract}
During accretion, terrestrial bodies attain a wide range of thermal and rotational states, which are accompanied by significant changes in physical structure (size, shape, pressure and temperature profile, etc.). However, variations in structure have been neglected in most studies of rocky planet formation and evolution. Here, we present a new code, HERCULES, that solves for the equilibrium structure of planets as a series of overlapping constant-density spheroids. Using HERCULES and a smoothed particle hydrodynamics code, we show that Earth-like bodies display a dramatic range of morphologies. For any rotating planetary body, there is a thermal limit beyond which the rotational velocity at the equator intersects the Keplerian orbital velocity. Beyond this corotation limit (CoRoL), a hot planetary body forms a structure, which we name a synestia, with a corotating inner region connected to a disk-like outer region. By analyzing calculations of giant impacts and models of planet formation, we show that typical rocky planets are substantially vaporized multiple times during accretion. For the expected angular momentum of growing planets, a large fraction of post-impact bodies will exceed the CoRoL and form synestias. The common occurrence of hot, rotating states during accretion has major implications for planet formation and the properties of the final planets. In particular, the structure of post-impact bodies influences the physical processes that control accretion, core formation and internal evolution. Synestias also lead to new mechanisms for satellite formation. Finally, the wide variety of possible structures for terrestrial bodies also expands the mass-radius range for rocky exoplanets.

\end{abstract}

\end{frontmatter}

%xxxxxxxxxxxxxxxxxxxxxxxxxxxxxxxxxxxxxxxxxxxxxxxxxxxxxxxxxxxxxxxxxxxxxxxxxxxxxxxxxxxxxxxxxxxxxxxxxxxxxxxxxxxxxxxxxxxxxxxxxxxxxxxxxxxxxxxx
%xxxxxxxxxxxxxxxxxxxxxxxxxxxxxxxxxxxxxxxxxxxxxxxxxxxxxxxxxxxxxxxxxxxxxxxxxxxxxxxxxxxxxxxxxxxxxxxxxxxxxxxxxxxxxxxxxxxxxxxxxxxxxxxxxxxxxxxx
%xxxxxxxxxxxxxxxxxxxxxxxxxxxxxxxxxxxxxxxxxxxxxxxxxxxxxxxxxxxxxxxxxxxxxxxxxxxxxxxxxxxxxxxxxxxxxxxxxxxxxxxxxxxxxxxxxxxxxxxxxxxxxxxxxxxxxxxx
\section{Introduction}

%{\bf think about where to add citations to Kun's paper and Matija's paper throughout}
The physical structure of a planet is essential information that is needed to investigate all major physical processes during planet formation and evolution. 
The planet's mass, size and shape govern the efficiency of accretion onto the body, and the internal pressure and temperature profiles control the conditions for differentiation and core formation. 
The vigor of convection and the resulting extent of mixing depends upon the physical structure of the body as well as the driving forces. 
The energy deposited by giant impacts, a key stage of terrestrial planet formation in our solar system and exoplanetary systems \citep{Chambers2010}, 
can radically change a body's physical structure.
Significant portions of the silicate mantles of the impacting bodies melt or vaporize \citep{Melosh1990}. In addition, planets can acquire significant angular momentum (AM) via one or more giant impacts \citep{Agnor1999, Kokubo2007, Kokubo2010}. 

Most planet formation studies have implicitly assumed that the lifetime of the transiently hot states generated by giant impacts \citep[100's-1000's~years,][]{Zahnle2007} is short enough to neglect a detailed treatment of partially vaporized rocky bodies.
For some aspects of terrestrial planet formation, this assumption may be justified, and most investigations of post-impact processes begin after the silicate fraction of a body is fully condensed and the planet has a distinct magma ocean and volatile-dominated atmosphere \citep[e.g.,][]{Solomatov2000,LeBrun2013,Hirschmann2012}.
However, the physical structure of a body in the time period between an energetic impact and a magma ocean can influence a number of different processes, including core formation, the reaccretion of impact debris and the formation of satellites.
Furthermore, recent studies have suggested that the Earth's lower mantle records chemical signatures that predate the last giant impact \citep{Rizo2016a,Mukhopadhyay2012,Tucker2012,Parai2012,Peto2013}.
Ascertaining how such reservoirs survived at least the Moon-forming impact event and persisted to the present day requires an understanding of the range of pressure-temperature conditions generated by giant impacts.

The physical structures of bodies after giant impacts have not been well studied with the exception of the Earth after the proposed Moon-forming event \citep{Hartmann1975,Cameron1976}. 
Most studies have assumed that this event set the present AM of the Earth-Moon system, as proposed by \citet{Cameron1976}.
Detailed studies of giant impact outcomes find that this constraint strongly limits the mass ratio and impact parameters of a potential Moon-forming event \citep{Canup2004,Canup2008a}. 
Hence, the canonical model for the Moon-forming giant impact is that of an approximately Mars-mass body obliquely colliding with the proto-Earth near the mutual escape velocity \citep{Canup2001, Canup2004,Canup2008a}. 
This specific giant impact is often used to represent all giant impacts. 
However, impacts similar to the canonical Moon-forming impact represent only a small fraction of the impacts that occur during accretion \citep{Stewart2012}.
Studies of terrestrial planet formation with more realistic collision outcomes have found that giant impacts with much higher specific energies \citep[e.g.,][]{Quintana2016} and AM \citep{Kokubo2010} than the canonical impact are common. 
The resulting post-impact structures have a correspondingly wide range of thermal and rotational states that can be substantially different from that produced by the canonical Moon-forming impact.
Furthermore, the canonical impact model for Moon formation is currently under scrutiny because of the difficulty of explaining the isotopic similarity between the Earth and Moon \citep[e.g.,][]{Melosh2014,Burkhardt2014} and the discovery that AM could have been transferred away from the Earth-Moon system since the last-giant impact  \citep{Cuk2012,Wisdom2015,Cuk2016,Tian2017}. 
If the AM of the Earth-Moon system immediately after the event was different than the present day, then a wider range of giant impact scenarios are possible \citep{Cuk2012,Canup2012,Lock2016LPSC}.
Thus, a quantitative study of the possible physical structures of hot, rotating rocky bodies is needed.  

Exploring the range of physical structures after giant impacts is particularly important for understanding the mechanisms for terrestrial satellite formation, including the origin of our Moon and its chemical relationship with Earth \citep[e.g.,][]{Canup2015,Lock2016LPSC,Wang2016}.
Previous studies of lunar origin have focused on the conditions after the canonical Moon-forming impact, which we summarize here.
After a canonical impact event, the Earth's spin period would have been $\sim$5~hrs, which would have produced minor rotational flattening for a fully condensed body.
There would have been a large angular velocity discontinuity between the corotating planet and a sub-Keplerian disk. The dynamics of this shear boundary have not been well studied, except to consider possible mixing between the planet and disk \citep[e.g.,][]{Pahlevan2007,Desch2013,Melosh2014}.
The material that is injected into orbit in canonical style impacts originates primarily from the antipode hemisphere of the impactor and is less shocked than the impacted hemisphere \citep{Canup2001,Canup2004,Nakajima2014}.
As a result, the disk material would have been less vaporized than the planet's predominantly silicate vapor atmosphere, creating a thermodynamic discontinuity between the Earth and the disk \citep[although the disk vapor may have been continuous with the planet's atmosphere; see][]{Pahlevan2007,Desch2013}.
%Based on the structure produced by the canonical impact and analogies with astrophysical disks, 
Thus, studies of the system after Earth's last giant impact assume that a dynamically and thermodynamically distinct disk orbits a corotating, nearly spherical central body \citep[e.g.,][]{Thompson1988,Canup2001,Canup2004,Machida2004,Cuk2012,Canup2012,Nakajima2014,Charnoz2015}. 
Given the possible range of thermal and rotational states produced by giant impacts, it is necessary to explore satellite formation from a range of different post-impact structures.

The need to understand the physical structures of rocky planets extends beyond accretionary processes.
Internal structure models are the primary tool used to infer the possible compositions of exoplanets from mass and radius observations \citep[e.g.,][]{Valencia2006,Seager2007,Swift2012}.
Most of the discovered exoplanets fall in a mass and radius range between the rocky and gaseous planets in our solar system \citep{Morton2016}. Hence, most exoplanets are unlike any of the planets in our solar system. In addition, because of the biases in the astronomical techniques used to find them, most of the known exoplanets are close to their stars and have high equilibrium surface temperatures. 
However, current internal structure models do not consider a wide range of thermal or rotational states, especially for rocky planets. Most exoplanet structure calculations model non-rotating, relatively cold bodies similar to the planets in the present-day solar system \citep[e.g.,][]{Zharkov1978,Valencia2006, Swift2012, Hubbard2013, Zeng2013, Zeng2016,Unterborn2016}.
In order to understand the possible range of exoplanet compositions, rocky planet models must be extended to include hot and rotating structures.

Here, we examine the structure of terrestrial planets over a wide range of thermal and rotational states.
We used smoothed particle hydrodynamics (SPH) simulations and developed a new code, HERCULES, for calculating the equilibrium structure of bodies based on modeling the body as a series of overlapping spheroids (\S\ref{sec:methods}).
We find that there is a wide range of sizes and shapes for hot, rotating bodies (\S\ref{sec:structure}).
In particular, we show that there is a limit beyond which a body cannot have a single angular velocity. Beyond this limit, a body can exhibit a range of morphologies with disk-like outer regions.
The corotation limit (CoRoL) is a function that depends upon the composition, thermal state, AM and mass of a body (\S\ref{sec:HSSL}).
We show that typical terrestrial planets experience multiple substantially vaporized states during accretion as a result of giant impacts. For the expected range of AM of growing planets, some post-impact states exceed the CoRoL (\S\ref{sec:impacts}).
The range of possible physical and thermal structures for terrestrial planets has significant implications for understanding physical processes during accretion, for example differentiation and satellite formation, as well as interpreting the composition of exoplanets (\S\ref{sec:discussion}). 
The supporting information includes an extended description of the methods used in this work and data tables.

%xxxxxxxxxxxxxxxxxxxxxxxxxxxxxxxxxxxxxxxxxxxxxxxxxxxxxxxxxxxxxxxxxxxxxxxxxxxxxxxxxxxxxxxxxxxxxxxxxxxxxxxxxxxxxxxxxxxxxxxxxxxxxxxxxxxxxxxx
%xxxxxxxxxxxxxxxxxxxxxxxxxxxxxxxxxxxxxxxxxxxxxxxxxxxxxxxxxxxxxxxxxxxxxxxxxxxxxxxxxxxxxxxxxxxxxxxxxxxxxxxxxxxxxxxxxxxxxxxxxxxxxxxxxxxxxxxx
%xxxxxxxxxxxxxxxxxxxxxxxxxxxxxxxxxxxxxxxxxxxxxxxxxxxxxxxxxxxxxxxxxxxxxxxxxxxxxxxxxxxxxxxxxxxxxxxxxxxxxxxxxxxxxxxxxxxxxxxxxxxxxxxxxxxxxxxx
\section{Methods}
\label{sec:methods}

We modeled the equilibrium structure of isolated bodies and the physical states obtained after giant impacts.
Isolated bodies were modeled using both SPH simulations and the HERCULES code, a new method for modeling rapidly rotating bodies based on overlapping, concentric, constant-density spheroids.
Impact simulations were performed using SPH in a manner similar to prior studies of giant impacts \citep{Cuk2012}.
In this section, we describe both methods and compare the results for isolated, corotating bodies. 

%xxxxxxxxxxxxxxxxxxxxxxxxxxxxxxxxxxxxxxxxxxxxxxxxxxxxxxxxxxxxxxxx
%xxxxxxxxxxxxxxxxxxxxxxxxxxxxxxxxxxxxxxxxxxxxxxxxxxxxxxxxxxxxxxxx
\subsection{Smoothed particle hydrodynamics}
\label{sec:method_SPH}

The GADGET-2 SPH code \citep{Springel2005}, modified to include tabulated equations of state (EOS) \citep{Marcus2009,Marcusthesis}, was used to calculate both the structure of isolated bodies and to simulate giant impacts.
We restricted this work to Earth-like planets, and all planets were modeled as differentiated bodies with 2/3 silicate and 1/3 iron by mass.  
As in prior work, the metal core and silicate were modeled as pure iron and pure forsterite, respectively. We used the M-ANEOS model \citep{Melosh2007} to generate the tabulated EOS, which includes two condensed phases and vapor \citep[refer to Supplementary Materials of][]{Canup2012}. 

Non-rotating, isolated bodies were initialized with a given mass and an approximate isentropic thermal profile. The bodies were then equilibrated for 24~hrs with the entropy of the layers imposed at each time step and any particle velocity damped. To produce cold isolated bodies with a mantle entropy of 4~kJ~K$^{-1}$~kg$^{-1}$, a constant angular velocity was imposed to particles in a non-rotating body of the same thermal structure. The structure was then equilibrated with the entropy of the layers imposed at each time step but without any damping of velocities. Rotating bodies with hotter thermal states were produced by starting from a colder planet of the same AM and increasing the specific entropy of the mantle at a rate of 0.25~kJ~K$^{-1}$~kg$^{-1}$~day$^{-1}$ until the desired specific entropy was reached. Isolated bodies had $10^5$ particles, which is comparable to much of the recently published literature on giant impacts using the SPH technique.

Post-impact states were generated by simulations of giant impacts in the same manner as described in \cite{Cuk2012}.  
Briefly, bodies were initialized in hydrostatic equilibrium by forming each body in isolation with isentropic silicate thermal profiles (\S\ref{sec:profiles}) with silicate specific entropy, $S_{\rm silicate}$~$=$~3.2 or 4~kJ~K$^{-1}$~kg$^{-1}$, corresponding to 1~bar potential temperatures of $\sim$1900 and $\sim$3300~K, respectively. 
Each impacting body was composed of $10^{5}$ to $5\times10^{5}$ particles. Impact simulations were calculated for 24 to 48 hrs of simulation time, until the change in bound mass was negligible and the post-impact structure reached a quasi-equilibrium shape. 
The parameters for each simulation are given in Table~\ref{sup:tab:impacts}. 
Part of this suite of simulations (116 impacts) was calculated for \cite{Cuk2012} and is primarily composed of small impactors (of mass $M_{\rm p}$~$\leq$~$M_{\rm Mars}$) onto targets with pre-impact rotation.
We extended the suite of giant impacts (by 46) to include other proposed Moon-forming scenarios \citep{Canup2001,Canup2004,Canup2012} and to sample the full range of giant impacts found in the later stages of planet formation \citep{Raymond2009,Quintana2016}. Our collection of impact outcomes span a range much broader than those proposed for the terminal impact event on Earth. The final bound masses range from 0.45 to 1.1 Earth masses ($M_{\rm Earth}$), but are typically between $0.8$ and $1.1 M_{\rm Earth}$.

%xxxxxxxxxxxxxxxxxxxxxxxxxxxxxxxxxxxxxxxxxxxxxxxxxxxxxxxxxxxxxxxx
%xxxxxxxxxxxxxxxxxxxxxxxxxxxxxxxxxxxxxxxxxxxxxxxxxxxxxxxxxxxxxxxx
\subsection{The HERCULES code}
\label{sec:method_HERCULES}

We developed a new code for calculating the equilibrium structure of rotating planetary bodies based on describing a body as a series of overlapping, concentric, constant-density spheroids.
A model based on the same principle, the concentric Maclaurin Spheroid (CMS) model, was originally developed by \cite{Hubbard2012, Hubbard2013} for planets with slow rotation rates.
The CMS model is a self-consistent field method that iteratively solves for the equilibrium shape of the surfaces of each of the constant-density spheroids.
However, the iterative equation used by \cite{Hubbard2013} is based on an incomplete solution of the Poisson equation that diverges for high degrees of rotational flattening \citep{Kong2013,Hubbard2014}.
An iterative approach for finding the shape of a single Maclaurin spheroid based on the full solution to the Poisson's equation was formulated by \cite{Kong2013} but was not extended to consider planetary structures with multiple layers and variable densities. 
Although the issue of non-convergence is of little concern for slowly rotating planets such as Jupiter, for which the CMS model was designed, it is a significant issue for the rapidly rotating bodies that we consider in this work. 

\begin{figure}
\centering
\includegraphics[scale=0.8333333]{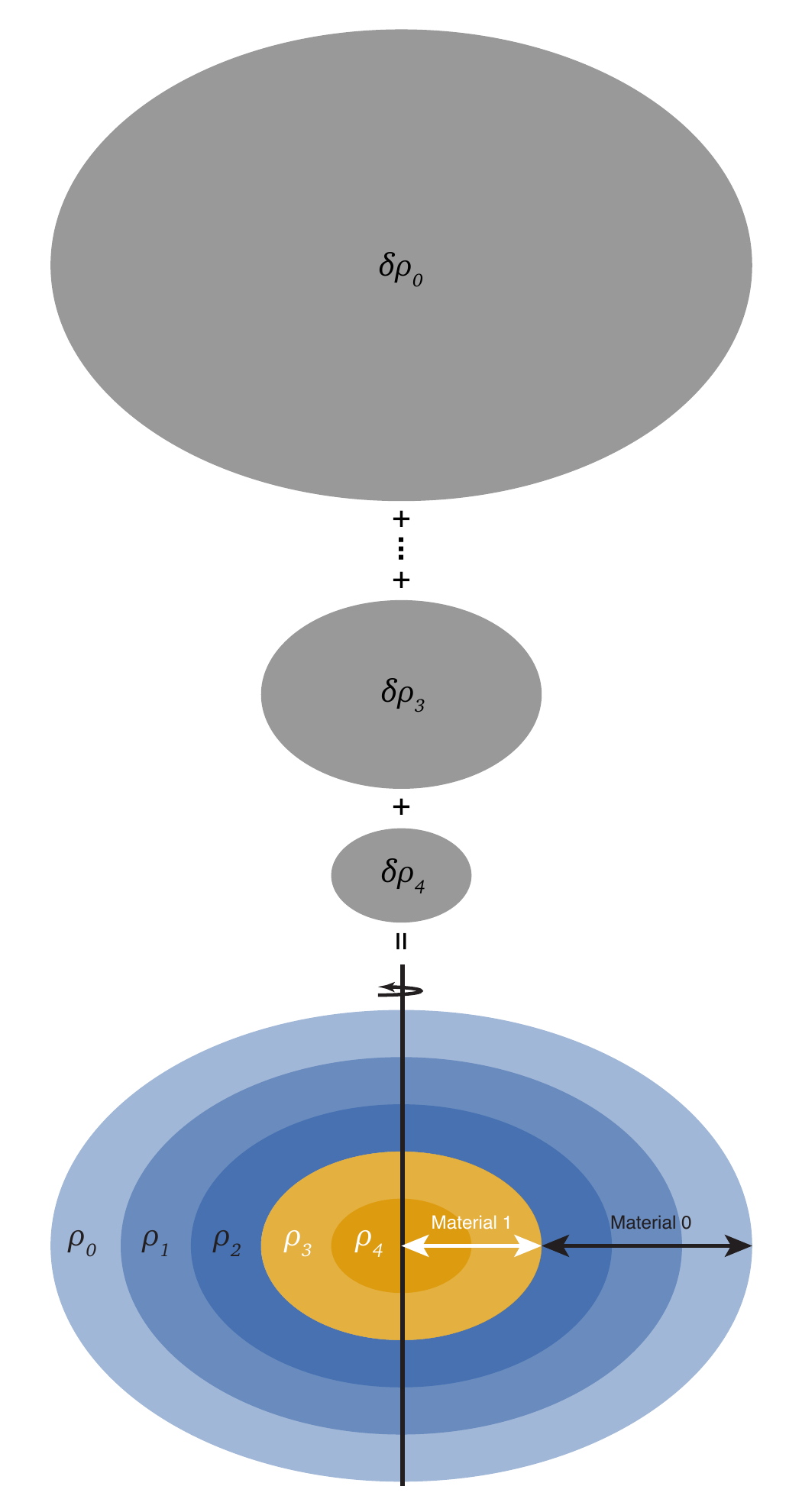}
\caption[]{Schematic of how an axisymmetric planetary structure is modeled in the HERCULES code. A body is described as a superposition of a number of constant-density spheroids of density $\delta\rho_{\rm i}$ (shown in gray). The superposition of the spheroids gives a body with increasing density with depth. The volumes between successive spheroids are called layers. Each layer has a constant density, $\rho_{\rm i}$, given by the sum of the densities of all the spheroids larger than the spheroid that defines the inner edge of the layer. Each layer belongs to a material layer which determines the relationship between pressure and total density in that layer by use of the material's equation of state. Two material layers are shown separately in blues and oranges. }
\label{fig:HERCULES}
\end{figure}

In order to be able to model rapidly rotating bodies, we built on the work of \cite{Kong2013} and \cite{Hubbard2012, Hubbard2013} to develop a new code. 
Here, we summarize the method, and full details are presented in \S\ref{sup:sec:HERCULES}.
The HERCULES ({\it Highly Eccentric Rotating Concentric U (potential) Layers Equilibrium Structure}) code models a body as a superposition of a number of overlapping, concentric, constant-density spheroids (Figure~\ref{fig:HERCULES}).
The surface of each spheroid is defined by an equipotential surface.
We refer to the region between two equipotential surfaces as a layer, and each layer is defined as being composed of a single material (e.g., silicate or iron). The density distribution in the planet is given by the superposition of the density of each of the spheroids. The total density of a layer in the body, $\rho_{\rm i}$, is given by the sum of the density of all the spheroids, $\delta \rho_{\rm i}$, that have an equatorial radius larger than the inner edge of the layer.
The code iteratively solves for the equilibrium structure and shape of the body while conserving the mass of each of the materials and the total AM.
During each iteration, the density of the spheroids is altered so that the total density is consistent with a hydrostatic pressure profile, the EOS of the materials, and an imposed thermal state for each of the material layers.
Each material's mass is conserved by altering the radius of each of the concentric layers. Although we refer to the volumes as spheroids, in our formulation the equipotential surfaces can be any rotationally symmetric surface whose radius decreases monotonically from equator to pole. 

The HERCULES code cannot reach the same level of precision as the CMS model.
The varying integration domains in the full solution to the Poisson's equation prevent the use of Gaussian quadrature, resulting in less precise numerical integration.
However, the HERCULES code is capable of efficiently calculating a variety of planetary structures, including rapidly rotating planets, making it a versatile tool to use in a wide range of problems in planetary science.

As in the SPH calculations, we modeled Earth-like bodies with 2/3 silicate and 1/3 iron by mass, and used the same equations of state derived from the M-ANEOS model.
A range of different thermal profiles were considered for the silicate (\S\ref{sec:profiles}).
Unless noted otherwise, 100 constant-density layers were used: 20 for the iron core and 80 for the silicate.
When the silicate was divided into two layers with different thermal states, 40 layers were used for each layer.
The equipotential surfaces were defined using $N_{\mu}$~$=$~1000 points, and terms up to spherical harmonic degree 12 were included in the iterative equation.
The HERCULES code requires a bounding pressure for the planet at the surface of the outermost layer.
Unless otherwise stated, we assumed a bounding pressure of $p_{\rm min}$~$=$~10~bar.

%xxxxxxxxxxxxxxxxxxxxxxxxxxxxxxxxxxxxxxxxxxxxxxxxxxxxxxxxxxxxxxxx
%xxxxxxxxxxxxxxxxxxxxxxxxxxxxxxxxxxxxxxxxxxxxxxxxxxxxxxxxxxxxxxxx
\subsection{Thermal profiles for isolated bodies}
\label{sec:profiles}

We considered isolated bodies with a range of thermal profiles. The variations in the thermal state of the core during accretion is not well understood. Here, we largely neglect these variations to focus on the thermal state of the silicate component. Each isolated body had an iron core of fixed specific entropy $S_{\rm core}$~$=$~1.5~kJ~K$^{-1}$~kg$^{-1}$. This core isentrope has a temperature of $\sim$3800~K at the pressure of the present-day core mantle boundary (CMB), similar to the present thermal state of Earth's core \citep[e.g.,][]{Anzellini2013}. The effect of varying core entropy is discussed in \S\ref{sec:HSSL}.

\begin{figure}
\centering
\includegraphics[scale=0.8333333]{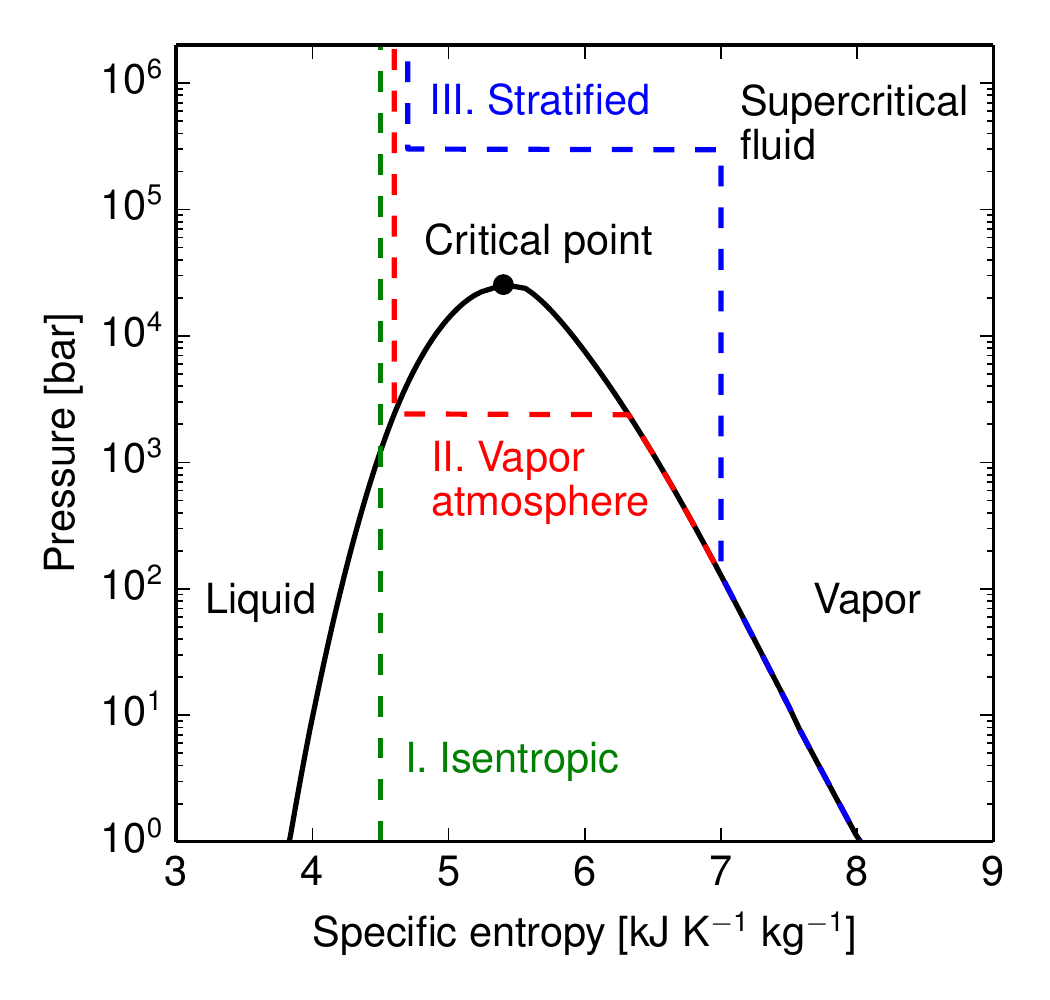}
\caption[]{Examples of the different silicate thermal profiles considered for the calculation of the structure of isolated bodies. The silicate portion of each body was modeled as a simple system with two condensed phases and vapor using an equation of state for forsterite. The colored dashed lines correspond to the different classes of thermal profiles described in \S\ref{sec:profiles}: (I) isentropic (green), (II) vapor atmosphere (red), and (III) stratified (blue). The solid black curve is the model liquid-vapor phase boundary for forsterite with the critical point ($S_{\rm crit}$~$=$~$5.40$~kJ~K$^{-1}$~kg$^{-1}$, $p_{\rm crit}$~$=$~$25.5$~kbar, $T_{\rm crit}$~$=$~$8810$~K, $\rho_{\rm crit}$~$=$~$1680$~kg~m$^{-3}$) shown by the black dot. Material in a thermal state plotting within the liquid-vapor phase boundary is a mixture of liquid and vapor with thermal states on each side of the boundary. Isentropes in pressure-temperature space are shown in Figure~\ref{sup:fig:profiles_PT}.}
\label{fig:profiles}
\end{figure}

For the silicate portion of bodies, we used three different classes of thermal profiles: (I) isentropic, (II) vapor atmosphere, and (III) stratified (Figure~\ref{fig:profiles}).
For isentropic profiles (class I, green line in Figure~\ref{fig:profiles}), the silicate portion of the body has a constant specific entropy, $S_{\rm lower}$, which is the simplest possible thermal state with which to make comparisons. Except for the coldest bodies considered in this study, the outer layers of these structures intersect the liquid-vapor phase boundary. In these cases, the low-pressure portions of the isentropic profile describe an ideal mixture of liquid and vapor, without any phase separation. 

Vapor atmosphere profiles (class II) are isentropic at high pressure with specific entropy, $S_{\rm lower}$, but if a profile intersects the liquid-vapor phase boundary, it follows the vapor side of the mixed phase region (red line in Figure~\ref{fig:profiles}). In this case, regions with pressures below the intersection point are assumed to be pure vapor on the phase boundary. The outer layers of such structures represent a saturated silicate vapor atmosphere and have a greater specific entropy compared to an isentropic thermal profile defined by the same $S_{\rm lower}$ (green line).
The vapor atmosphere profile approximates the scenario where condensate efficiently rains out from the low-pressure regions of a body and the vapor is stratified.
In a hot rocky body with a turbulent atmosphere, the rain out of condensates is unlikely to be wholly efficient, but the vapor atmosphere profile provides a useful end member case.

A stratified profile (class III) divides the silicate portion of a body into two separate material layers containing a specified mass fraction of the silicate. 
The lower layer is isentropic with specific entropy $S_{\rm lower}$.
The upper layer has the same thermal structure as a vapor atmosphere profile: isentropic at higher pressures with specific entropy $S_{\rm upper}$ and following the saturated vapor curve at lower pressures (blue dashed line in Figure~\ref{fig:profiles}).
The pressure at the boundary between the upper and lower layers varies between different bodies since the masses of the two silicate layers are dictated, not the pressure of transition.
Here, we considered stratified profiles with an upper layer that contains 50 or 25\% of the mass of the silicate.
These stratified profiles are intended to emulate the thermal structures that are typically produced in hydrodynamic simulations of giant impacts.
The impacting hemispheres of the colliding bodies are more highly shocked than the antipodal hemispheres.
After gravitational equilibration, the post-impact body is thermally stratified with lower entropy material from the antipodal hemispheres at higher pressures and higher entropy material from the impacted hemispheres at lower pressures.
Post-impact thermal profiles are much more variable than a simple two-layer structure, but the model stratified thermal profile allows us to examine the general effect of thermal stratification on planetary structure.

We have used specific entropy, rather than temperature, as the natural intensive variable to describe the thermal state for each layer within a planet. For liquid-vapor mixtures, the temperature is insufficient to determine the relative proportions of each phase at a given pressure. With specific entropy as the independent variable, the lever rule can be applied to determine the mass fraction of each phase using the material's liquid-vapor phase boundary, which is a dome in specific entropy--pressure space (Figure~\ref{fig:profiles}). In rocky planets, the thermal state of the mantle is often represented by the potential temperature, which is the temperature on the mantle isentrope at a reference pressure (usually 1 bar for terrestrial applications). The potential temperature is degenerate for thermal structures that intersect the liquid-vapor phase boundary (Figure~\ref{sup:fig:profiles_PT}), so it is not a useful parameter in this work. In addition, partially vaporized planets do not have a well-defined surface, and we must report the radius of the body at a specific pressure contour. 

We expect that the accuracy of the EOS model for silicates will be improved as new data are acquired on forsterite and other silicate chemical compositions \citep[e.g.,][]{Bolis2016,Sekine2016,Root2016,Davies2016,Davies2017}. In previous studies, the M-ANEOS model has been shown to underestimate the gain in entropy at high shock pressures \citep[e.g.,][]{Kraus2012}. The model shock temperatures for forsterite overlap with the error bars reported by \citet{Bolis2016} up to about $500$~GPa, which corresponds to a specific entropy slightly greater than the model critical point when shocked from standard pressure and temperature. As a result, the EOS model is inferred to underestimate the entropy gain at shock pressures $>\sim$500~GPa. Thus, the impact outcomes from the highest velocity events are likely to shift to a greater degree of vaporization with improving EOS models, but our general conclusions about changes in planetary structure relative to the degree of vaporization are robust. 

%xxxxxxxxxxxxxxxxxxxxxxxxxxxxxxxxxxxxxxxxxxxxxxxxxxxxxxxxxxxxxxxx
%xxxxxxxxxxxxxxxxxxxxxxxxxxxxxxxxxxxxxxxxxxxxxxxxxxxxxxxxxxxxxxxx
\subsection{Comparison of methods}

\begin{figure*}
\centering
\includegraphics[scale=0.8333333]{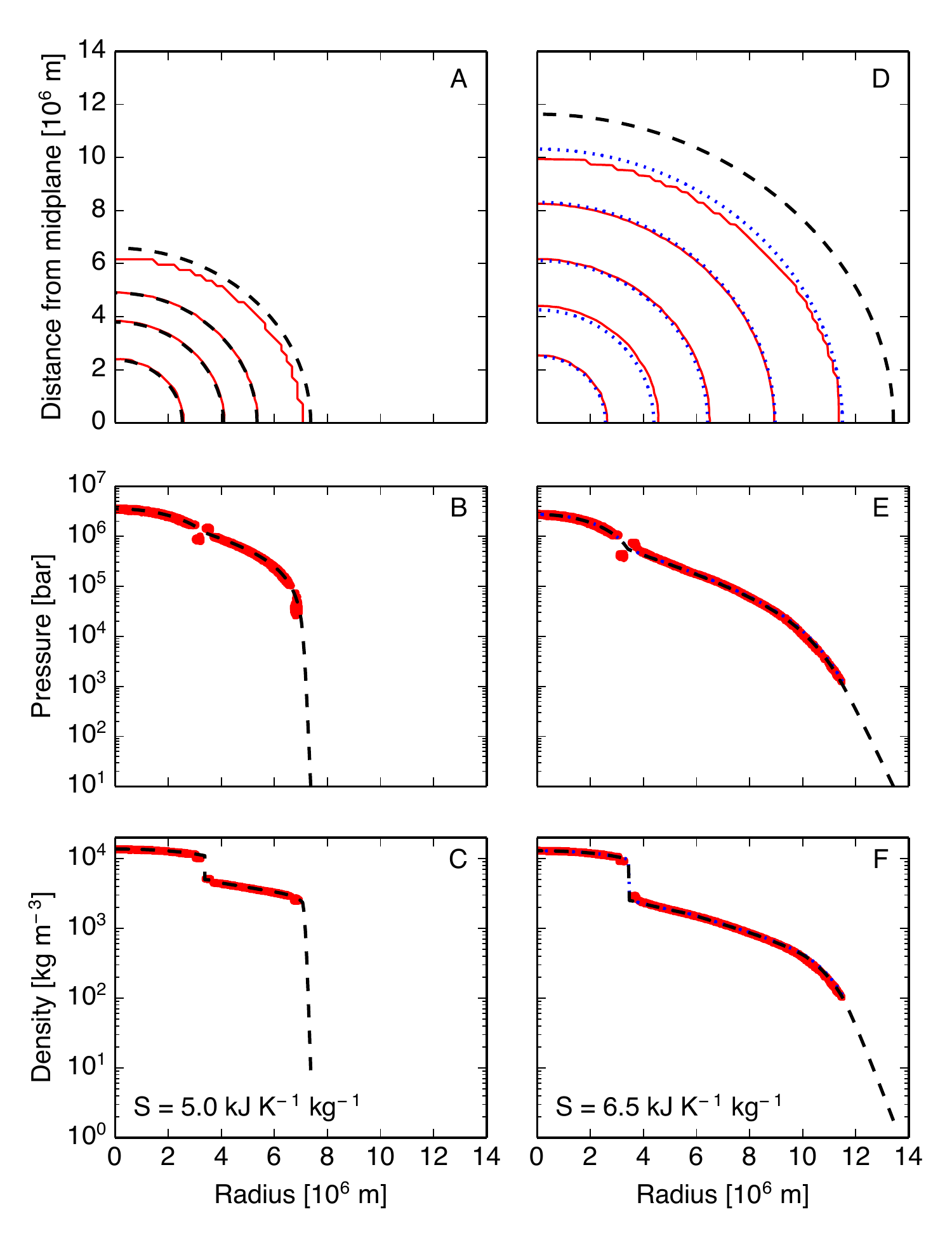}
\caption[]{A comparison of equilibrium planetary structures calculated using the GADGET-2 SPH code and the HERCULES code for a largely condensed body (A-C, silicate specific entropy $S_{\rm lower}$~$=$~$5.0$~kJ~K$^{-1}$~kg$^{-1}$) and a substantially vaporized body (D-F, silicate specific entropy $S_{\rm lower}$~$=$~$6.5$~kJ~K$^{-1}$~kg$^{-1}$). All bodies are Earth mass, have an angular momentum equal to that of the present-day Earth-Moon system ($L_{\rm EM}$) and an isentropic thermal profile (class I). SPH bodies are shown in red and HERCULES bodies with a bounding pressure, $p_{\rm min}$, of 10 bar are in black. In D-F, a HERCULES body with a bounding pressure of 1257~bar, the lowest pressure in the midplane of the corresponding SPH body, is also shown in blue. A,D show the axisymmetric SPH and HERCULES pressure contours on a quarter plane through the rotation axis. The pressure contours correspond to specific layers in the HERCULES code and are not the same pressures in A and D. The outer contour is always the bounding surface of the body. In D, only the surface of the outermost layer is shown for the HERCULES body with $p_{\rm min}$~$=$~$10$~bar. The lower panels display midplane pressure (B,E) and density (C,F) radial profiles.}
\label{fig:SPH_HERCULES}
\end{figure*}

\begin{table*}[t!]
{\renewcommand{\arraystretch}{1.3}
\centering
\caption{Comparison of key properties derived from HERCULES and SPH calculations of the two isolated planetary structures with different silicate thermal profiles shown in Figure~\ref{fig:SPH_HERCULES}. Each body has a constant specific entropy silicate layer  of $S_{\rm lower}$ (thermal profile class I).} 
\begin{tabular}{l c c c c c}
\label{tab:SPH_HERCULES}
& SPH & HERCULES &SPH & HERCULES & HERCULES \\ 
\cline{2-3} \cline{4-6}
&\multicolumn{2}{c}{$S_{\rm lower}$~$=$~5.0~kJ~K$^{-1}$~kg$^{-1}$} &\multicolumn{3}{c}{$S_{\rm lower}$~$=$~6.5~kJ~K$^{-1}$~kg$^{-1}$} \\ 
Minimum pressure [bar] & 26516 & 10 & 1257 & 1257 & 10 \\
Equatorial radius [10$^6$~km] & 6.93 & 7.37 & 11.47 & 11.50 & 13.42 \\
Aspect ratio (polar/equatorial radius) & 0.88 & 0.89 &0.86 & 0.90 & 0.87 \\
Angular velocity [10$^{-3}$ rad s$^{-1}$] & 0.372 & 0.370 &0.212 & 0.210 & 0.207 \\
\end{tabular}
}
\end{table*}

The GADGET-2 and HERCULES codes implement fundamentally different approaches for modeling the equilibrium structure of isolated bodies.
Nevertheless, the two techniques produce very similar structures for corotating bodies. Two examples of isentropic Earth-mass planets with a total angular momentum equal to the present-day Earth-Moon system ($L_{\rm EM}$~$=$~$3.5\times 10^{34}$ kg~m$^{-2}$~s$^{-1}$) are shown in Figure~\ref{fig:SPH_HERCULES}. 
The SPH calculation is shown in red and the HERCULES calculation, using a bounding pressure of $p_{\rm min}$~$=$~10~bar, is plotted in black.
The well-known issues with resolving boundaries with high density contrasts in SPH can be seen at the core-mantle boundary, with a layer of anomalous density particles on either side of the boundary.
However, the error at the boundary does not propagate to the rest of the structure.
Similarly, cooler SPH planets that have a sharp boundary with vacuum have an outer layer of particles with an increased radial separation compared to interior particles (Figure~\ref{fig:SPH_HERCULES}A-C with $S_{\rm lower}$~$=$~5~kJ~K$^{-1}$~kg$^{-1}$), but this layer captures the correct density and pressure at that radius. 

The limited resolution of the SPH simulations leads to an error in resolving the outer boundary of bodies, particularly for silicate specific entropies above the critical point. 
The SPH resolution is not sufficient to capture the mass of material in the low density, partially vaporized regions of hot bodies.
For the example planet with an isentropic profile below the critical point entropy (Figure~\ref{fig:profiles}A-C), the SPH calculation does not resolve pressures below about $2\times10^4$~bar. 
The scale height at the edge of a mostly condensed body is small, so the equatorial radius is only slightly smaller than that calculated with the HERCULES code with $p_{\rm min}$~$=$~10~bar, about 6\% in our example case.
However, for substantially vaporized bodies with specific entropies above the critical point value (e.g., Figure~\ref{fig:SPH_HERCULES}D-F with $S_{\rm lower}$~$=$~6.5~kJ~K$^{-1}$~kg$^{-1}$), the scale height of the low pressure regions is larger.
Although the minimum pressure of the SPH planet is lower, at 1257~bar in our example case, the difference in equatorial radius is greater, about 15\% smaller than that calculated with the HERCULES code with $p_{\rm min}=10$~bar.
To mimic the effect of the resolution limit in the SPH simulations, we also calculated a HERCULES body using a bounding pressure of $p_{\rm min}$~$=$~$1257$ bar (blue lines in Figure~\ref{fig:SPH_HERCULES}D-F), the lowest pressure in the midplane of the SPH planet.
In this case, the equatorial radius agreed with the SPH to within about 0.2\%.
The mass of the HERCULES body that lies outside of the surface of the SPH structure (outer red line in Figure~\ref{fig:SPH_HERCULES}D) and within the HERCULES 10-bar contour (black dashed line) is only about $10^{-3}$~$M_{\rm Earth}$. The mass of this unresolved outer, partially vaporized layer is well below the resolution of the SPH simulation, where each particle has a mass of about $10^{-5}$~$M_{\rm Earth}$.
For all the isolated, corotating planets considered here, the differences between HERCULES and SPH calculations typically involved less than the mass of 500 SPH particles ($<$~$0.5$~wt\%), resulting in errors in the SPH equatorial radii less than about 15\%.
Since there is little mass in the region not resolved by the SPH planets, the corotating angular velocity of the bodies calculated using SPH and HERCULES are not substantially different (Table~\ref{tab:SPH_HERCULES}).
SPH does not define the polar radius as well as the equatorial radius due to the difference in particle resolution along each axis of an oblate planet.
This leads to a small error in the aspect ratio, the ratio of the polar to equatorial radii (Table~\ref{tab:SPH_HERCULES}). 

In general, HERCULES is a more efficient and accurate method to model planetary structures compared to SPH. However, the current version of HERCULES can only calculate the structure of corotating bodies. A version of HERCULES that may solve for an imposed angular momentum structure is under development. As a result, we utilize SPH to study the structure of non-corotating bodies.

%xxxxxxxxxxxxxxxxxxxxxxxxxxxxxxxxxxxxxxxxxxxxxxxxxxxxxxxxxxxxxxxxxxxxxxxxxxxxxxxxxxxxxxxxxxxxxxxxxxxxxxxxxxxxxxxxxxxxxxxxxxxxxxxxxxxxxxxx
%xxxxxxxxxxxxxxxxxxxxxxxxxxxxxxxxxxxxxxxxxxxxxxxxxxxxxxxxxxxxxxxxxxxxxxxxxxxxxxxxxxxxxxxxxxxxxxxxxxxxxxxxxxxxxxxxxxxxxxxxxxxxxxxxxxxxxxxx
%xxxxxxxxxxxxxxxxxxxxxxxxxxxxxxxxxxxxxxxxxxxxxxxxxxxxxxxxxxxxxxxxxxxxxxxxxxxxxxxxxxxxxxxxxxxxxxxxxxxxxxxxxxxxxxxxxxxxxxxxxxxxxxxxxxxxxxxx

\begin{sidewaysfigure*}
\centering
\includegraphics[scale=0.83333333]{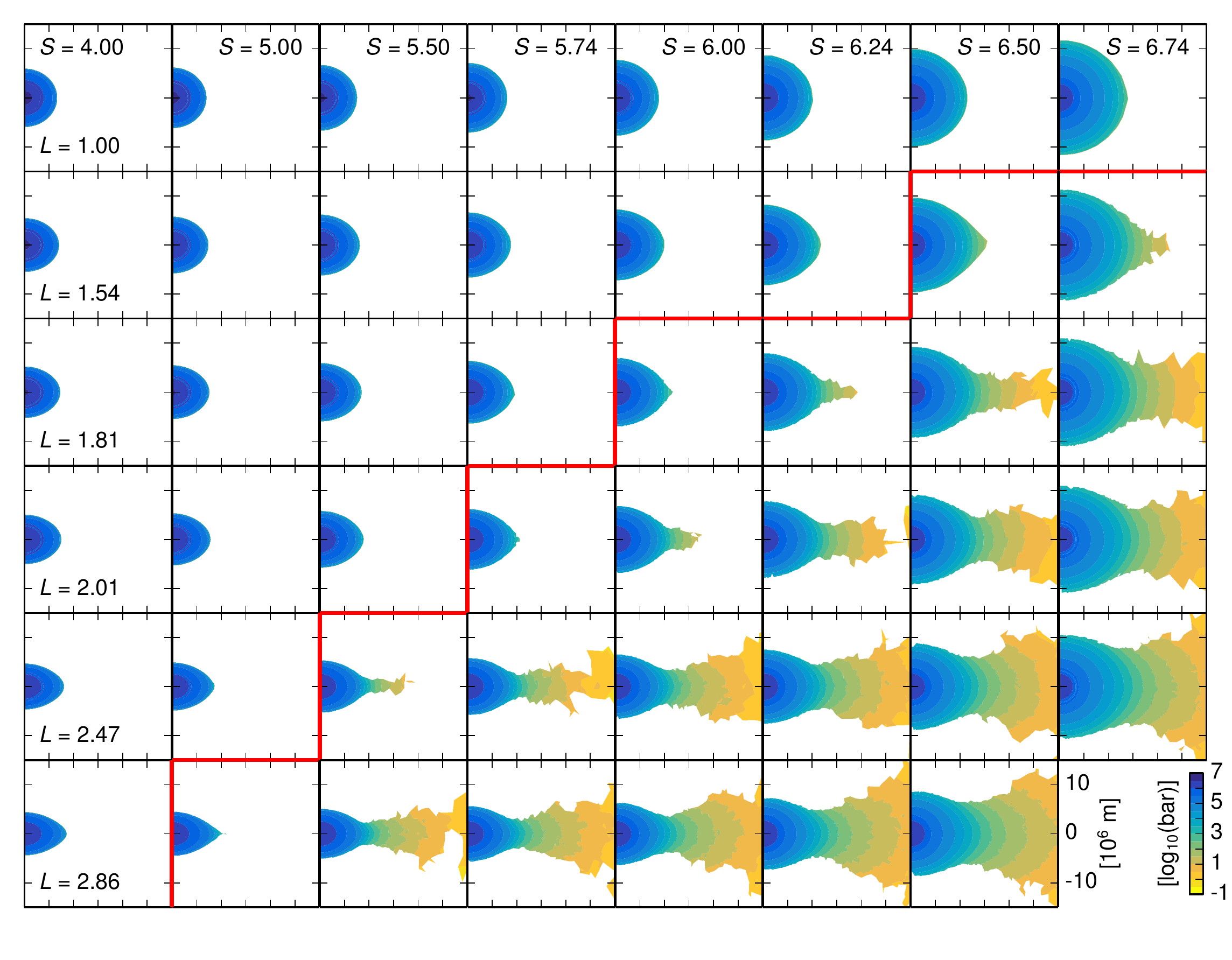}
\caption[]{The equilibrium structures of rocky planetary bodies are highly variable. Each panel displays a one-Earth-mass body with an isentropic thermal profile (class I). Colors denote pressure contours in a plane through the rotation axis for axisymmetric bodies, with different angular momenta (rows) and silicate specific entropies (columns), calculated using the GADGET-2 SPH code. Details of plotting technique are given in \S\ref{sec:structure}. Heating the body significantly inflates the radius, and increasing AM produces synestias, structures with connected disk-like regions. Bodies to the left of the red line are corotating, and bodies to the right of the red line are above the CoRoL and are synestias.}
\label{fig:shape_plot}
\end{sidewaysfigure*}

\begin{sidewaysfigure*}
\centering
\includegraphics[scale=0.8333333]{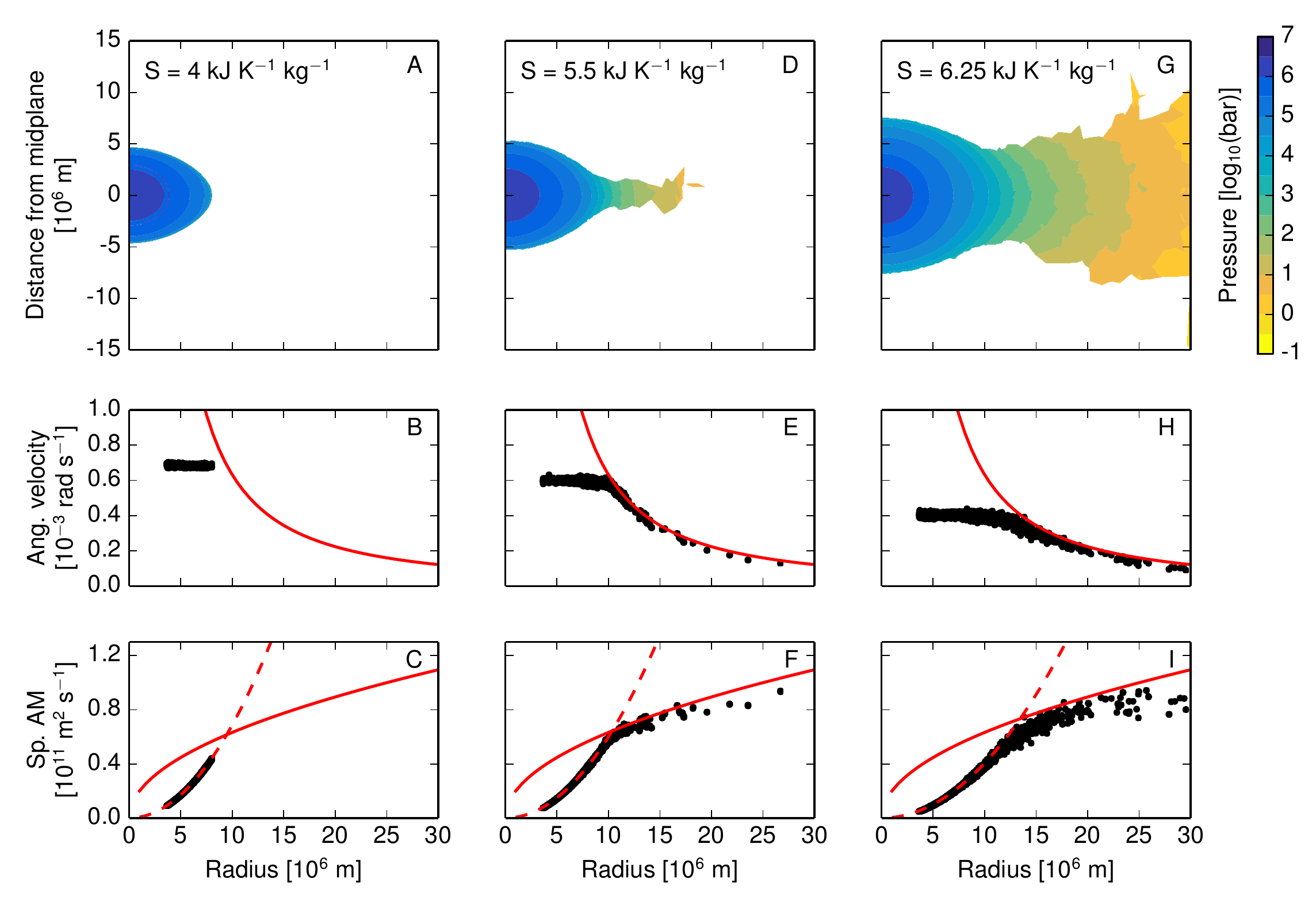}
\caption[]{The structures of rocky bodies with high angular momentum spans corotating, oblate bodies to a regime with a connected disk.  Each body has an Earth mass with total angular momentum of $2.45L_{\rm EM}$ and silicate specific entropies of 4.0 (A-C), 5.5 (D-F), and 6.25 (G-I)~kJ~K$^{-1}$~kg$^{-1}$. Structures were calculated using the GADGET-2 SPH code with isentropic thermal profiles (class I) imposed. Top row presents pressure contours in a plane through the rotation axis. Details of the plotting technique are described in \S\ref{sec:structure}. The angular velocities (middle row) and specific AM (bottom row) in the midplane of the silicate portion of the body are shown by black dots. Solid red lines denote the specific AM of a circular Keplerian orbit around a point mass and dashed red lines indicate the specific AM for corotation with the inner region of the structure.}
\label{fig:hot_planets}
\end{sidewaysfigure*}

\section{The structure of hot, rotating rocky bodies}
\label{sec:structure}

We calculated the structure of rocky bodies with varying thermal profiles and AM.
To illustrate the range of solutions, we determined the equilibrium structures for one Earth-mass bodies for a range of AM and silicate isentropes (thermal profile class I) using GADGET-2.
We find that there are a wide variety of possible structures, ranging from compact to highly extended, as shown by the pressure contours in Figure~\ref{fig:shape_plot}. Pressures were interpolated between SPH particles using a Delaunay triangulation. To avoid spurious contour lines in the extended regions with low particle density, the outermost few particles in each radial bin were not included when calculating the contours. The width of the radial bins and the number of particles excluded were varied depending on the structure being plotted. The gaps in some structures in Figure~\ref{fig:shape_plot} are due to the lack of resolution in those areas. In hot, high-AM structures the plotted pressure contours do not close as the scale height at the upper and lower surfaces of the structure is not resolved.
Increasing the specific entropy of the silicate leads to an increase in the fraction of vapor in the low-pressure regions of the body and a substantial increase in radius. 
At low AM, the structures are nearly spherical, oblate spheroids rotating with a single angular velocity, regardless of thermal state. 
However, bodies with higher AM do not always maintain a simple oblate spheroidal shape.
The equilibrium structure for an isolated body of a fixed mass and composition can attain a large range of morphologies and sizes, depending on thermal state and total AM.

To illustrate the dynamics of extended planetary structures, we compare the rotational profiles of bodies with different silicate specific entropies at a total AM of $2.45L_{\rm EM}$. In Figure~\ref{fig:hot_planets}, the pressure field and size of the structure are represented by the colored contours in the top row. Pressures were interpolated in the same manner as in Figure~\ref{fig:shape_plot}. The radial angular velocity and specific angular momentum in the equatorial plane are shown by the black dots in the middle and bottoms rows, respectively. High-AM bodies with low specific entropies have a constant angular velocity (i.e., are in solid body rotation) and consequently adopt very oblate equilibrium structures to conserve AM (Figure~\ref{fig:hot_planets}A-C). 
At higher specific entropies, the fraction of vapor in the outer regions is greater and the structure thermally expands. To maintain a fixed total AM, increasing the specific entropy is balanced by lower angular velocities. 
At a specific combination of AM and specific entropy, the angular velocity at the equator of the body intersects the Keplerian angular velocity. In Figure~\ref{fig:hot_planets}, the bodies in the left (A-C) and middle (D-F) columns are before and after this intersection. Beyond the intersection, it is not possible to attain an equilibrium corotating structure. The centripetal force required to remain bound and corotating is greater than the gravitational force, and a negative pressure gradient would be required to keep the equator at the same angular velocity as the interior of the body. A negative pressure gradient is non-physical and instead the outer edge of the body must adopt a Keplerian or sub-Keplerian angular velocity. With increasing specific entropy, the pressure gradient support of the structure increases, and the sub-Keplerian region of the structure expands substantially (Figure~\ref{fig:hot_planets}G-I).

We define the limit for planetary bodies with constant angular velocity as the {\it corotation limit} (CoRoL). The CoRoL is a surface that depends upon the mass, compositional layering, thermal profile, and AM of a body. Above the CoRoL, there is no solution for the planetary structure that is both hydrostatic and corotating.
For bodies below the CoRoL, there is a unique solution for a perfectly corotating structure.
However, above the CoRoL, there is no unique solution to the structure as it depends on the spatial distribution of mass and AM. 
The range of possible super-CoRoL structures depends on the mechanism that drove the body beyond the limit.

For the isolated SPH bodies shown in Figure~\ref{fig:shape_plot}, the silicate particles were incrementally heated in an attempt to obtain as small a structure as possible (see \S\ref{sec:method_SPH}). Incremental heating limited the mass and AM transported into the sub-Keplerian region of the structure during gravitational re-equilibration. Other processes that drive a planet beyond the CoRoL, such as giant impacts, can introduce more mass and AM farther out in the structure. 

We name structures beyond the CoRoL {\it synestias}, after the Greek {\it syn} for connected and {\it Hestia} for the goddess of architecture. The traditional definitions of mantle, atmosphere, and disk are not applicable in synestias
(Figure~\ref{fig:hot_planets}D-I). In this work, we refer to the corotating portion of both super-CoRoL and sub-CoRoL structures as the corotating region, and the Keplerian or sub-Keplerian portion of structures as the disk-like region or connected disk. The region between the corotating and disk-like regions we refer to as the transition region.
In typical synestias, there is a monotonic angular velocity profile connecting the corotating and disk-like regions (Figure~\ref{fig:hot_planets}E,H) and the transition region is very narrow. 
Synestias can be attained by a variety of celestial bodies and are not restricted to rocky bodies (\S\ref{sec:discussion_other}).

The shapes of synestias can vary from bodies with pinched equators to bodies with flared disk-like regions. Flaring typically only occurs at low pressures in the structure but can be substantial. The most extended structures in Figure~\ref{fig:shape_plot} have a vertical scale height in the low-pressure regions on the same order as the polar radius of the corotating region. The equatorial radius of synestias can extend beyond the Roche limit, which is the closest distance a satellite can withstand tidal forces from the planet ($\sim$$18 \times 10^6$~m for silicate satellites orbiting an Earth-mass body). 

%xxxxxxxxxxxxxxxxxxxxxxxxxxxxxxxxxxxxxxxxxxxxxxxxxxxxxxxxxxxxxxxxxxxxxxxxxxxxxxxxxxxxxxxxxxxxxxxxxxxxxxxxxxxxxxxxxxxxxxxxxxxxxxxxxxxxxxxx
%xxxxxxxxxxxxxxxxxxxxxxxxxxxxxxxxxxxxxxxxxxxxxxxxxxxxxxxxxxxxxxxxxxxxxxxxxxxxxxxxxxxxxxxxxxxxxxxxxxxxxxxxxxxxxxxxxxxxxxxxxxxxxxxxxxxxxxxx
%xxxxxxxxxxxxxxxxxxxxxxxxxxxxxxxxxxxxxxxxxxxxxxxxxxxxxxxxxxxxxxxxxxxxxxxxxxxxxxxxxxxxxxxxxxxxxxxxxxxxxxxxxxxxxxxxxxxxxxxxxxxxxxxxxxxxxxxx
\section{The corotation limit (CoRoL) for rocky planets}
\label{sec:HSSL}

We have determined the CoRoL as a function of mass, AM, and thermal state for Earth-like planets using the HERCULES code. As the current version of HERCULES cannot calculate the structure of non-corotating bodies, including synestias, we find the CoRoL by extrapolation from corotating bodies. In AM increments of 0.05 or 0.01~$L_{\rm EM}$, we calculated the equilibrium structure of corotating bodies with a given thermal profile, compositional layering, and mass. Above a certain AM, close to the CoRoL, HERCULES is no longer able to find a physical solution to the structure as, during at least one iteration, the total potential gradient at the edge of the body changes sign. This causes the equipotential surfaces to cross at higher latitudes, breaking the assumptions of the model. In order to find the CoRoL, we linearly extrapolate to higher AM to find the point at which the corotating and Keplerian angular velocities at the edge of the structure intersect (Figure~\ref{sup:fig:CoRoLfind}). This point is the CoRoL and is generally close to the highest AM structure found by HERCULES.

\begin{figure}
\centering
\includegraphics[scale=0.8333333]{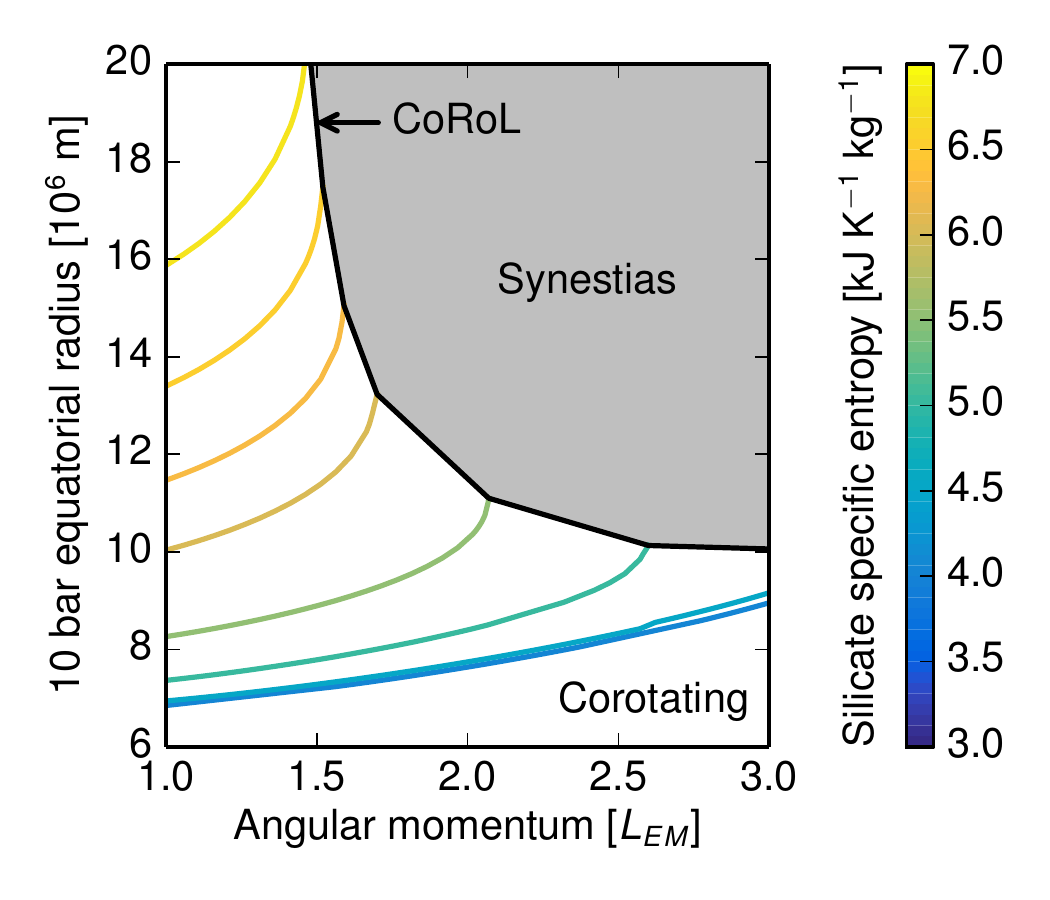}
\caption[]{The radius of Earth-like bodies varies strongly with specific entropy and angular momentum. Each colored line is the range of equatorial radii of bodies with isentropic thermal profiles (class I) of fixed silicate specific entropy and varying angular momenta, calculated using the HERCULES code. The black line denotes the limit for equilibrium corotating structures (CoRoL). Above the CoRoL, structures are synestias and a portion of the body must form a disk-like region.}
\label{fig:HSSL_R-L}
\end{figure}

The equatorial radius, and hence the CoRoL, is very sensitive to the specific entropy of the outer regions of the body and the total AM. Figure~\ref{fig:HSSL_R-L} presents the calculated equatorial radius for Earth-mass bodies with isentropic silicate layers (thermal profile class I), where each line indicates the radius for a constant specific entropy and varying total AM. 
Each line terminates at the limit of corotating equilibrium structures, and the locus of such points defines the CoRoL (black line) for this specific planetary mass, compositional layering, and thermal profile. 
Hotter bodies have significantly expanded equatorial radii and cross the CoRoL at lower AM.
Even below the CoRoL, the equatorial radius varies by a factor of three depending on thermal state and AM.

\begin{figure}
\centering
\includegraphics[scale=0.8333333]{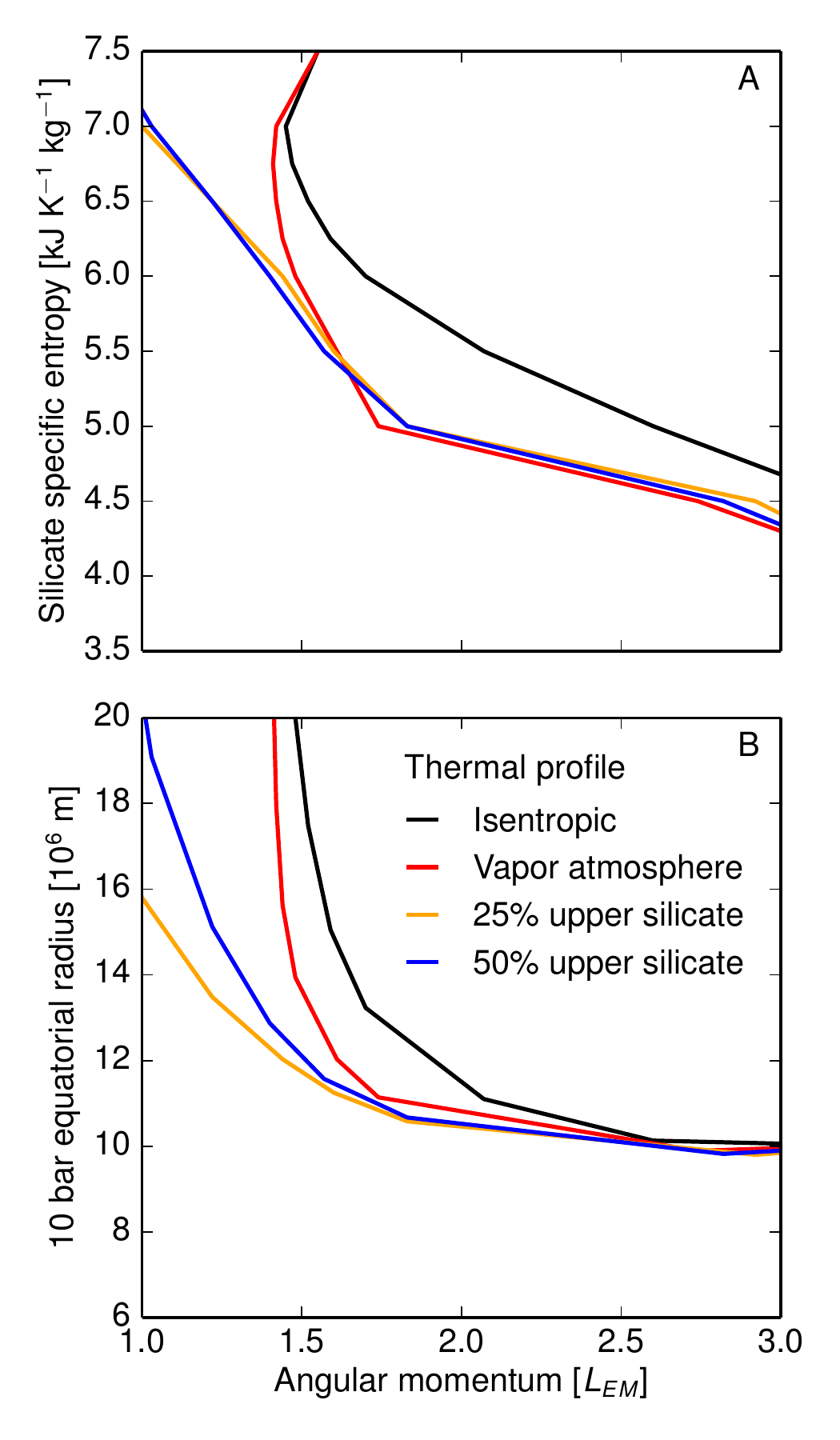}
\caption[]{The CoRoL depends on the thermal structure of a body. Using the HERCULES code, we calculated the CoRoL for bodies with isentropic (black line, class I), vapor atmosphere (red line, class II), and stratified silicate (class III) thermal profiles. We present the CoRoL in terms of (A) the silicate specific entropy used to define the thermal profile and (B) the equatorial radius as a function of angular momentum. For stratified profiles only the specific entropy of the upper silicate layer was varied, and the lower layer was held constant with $S_{\rm lower}$~$=$~$4$~kJ~K$^{-1}$~kg$^{-1}$. Stratified profiles (\S\ref{sec:profiles}) with 25\% (orange line) and 50\% (blue line) by mass upper silicate layers are shown.}
\label{fig:HSSL_thermal}
\end{figure}

The CoRoL is also sensitive to the thermal profile within the body (\S\ref{sec:profiles}).Figure~\ref{fig:HSSL_thermal} presents the CoRoL in terms of (A) the specific entropy of the silicate layer (upper layer for stratified profiles (class III)) and (B) the 10-bar equatorial radius, as a function of AM.
The AM required to exceed the CoRoL with an isentropic profile (class I) is offset from the other thermal profiles because the density profile assumes an ideal mixture of liquid and vapor in the mixed phase region. 
Vapor atmosphere profiles (class II), with pure gas densities in the lowest pressure layers, produce bodies with larger equatorial radii and reach the CoRoL at lower AM for the same value for $S_{\rm lower}$.
For low specific entropies, the CoRoL for bodies with isentropic (class I) and vapor atmosphere (class II) thermal profiles tend towards each other because the mixed phase region of the structure is absent or negligible.
At high specific entropies, the CoRoL for bodies with isentropic (class I) and vapor atmosphere (class II) thermal profiles converge as the mixed phase regions of isentropic bodies become increasingly dominated by vapor. At these high specific entropies, the CoRoL turns over as the slower rotation rate of the more extended structures begins to overcome the effect of an increased equatorial radius.

We also considered bodies with stratified thermal profiles (class III) with a colder lower silicate layer ($S_{\rm lower}$) and hotter upper silicate layer ($S_{\rm upper}$). For $S_{\rm upper}$~$>\sim$~ 5.5~kJ~K$^{-1}$~kg$^{-1}$, such structures cross the CoRoL at a lower AM compared to bodies with the vapor atmosphere profile (class II) with a silicate specific entropy equal to $S_{\rm upper}$. 
The cold lower silicate layer is more compact and closer to the rotation axis than the equivalent mass fraction of bodies with vapor atmosphere profiles or isentropic profiles with high specific entropies.
Accordingly, the structure has a lower moment of inertia, and the cold lower silicate layer of a stratified body does not accommodate as much AM for a given angular velocity.
Thus, stratified bodies must rotate faster, compared to an isentropic body with the same AM, and the equator of the planet intersects the Keplerian orbit at a lower total AM.
This effect increases with a larger difference in specific entropy between the upper and lower silicate layers. The CoRoL does not turn over for thermally stratified structures in the range of specific entropies considered because the cold dense cores of such bodies have a comparatively low moment of inertia.
Because the isentropic case requires a higher specific entropy to reach the CoRoL compared to stratified structures, the suite of isentropic structures in Figure~\ref{fig:shape_plot} has fewer synestias over this range of AM and specific entropy than would be achieved by stratified bodies.
The fact that thermal stratification causes bodies to cross the CoRoL at lower AM is significant because giant impacts create stratified structures (\S\ref{sec:impacts}).

The different thermal profiles have substantially different equatorial radii at the CoRoL (Figure~\ref{fig:HSSL_thermal}B) because of the strong sensitivity of the radius to the thermal state of the outermost layers. However, in terms of specific entropy, the CoRoL is similar for the different stratified thermal profiles (Figure~\ref{fig:HSSL_thermal}A). Consequently, the total angular momentum and specific entropy of the outer silicate layers are reliable metrics to determine if a particular stratified body is above the CoRoL. 

For the expected range of values for Earth, the specific entropy of the iron core has little effect on the CoRoL for Earth-like bodies. 
For the iron EOS model used here, the present thermal state of Earth's core corresponds to a specific entropy of $\sim$1.5~kJ~K$^{-1}$~kg$^{-1}$. 
The thermal history of the core is debated, so we considered a range of specific entropies, $S_{\rm core}$~$=$~1 to 2~kJ~K$^{-1}$~kg$^{-1}$, corresponding to temperatures of 700 to 11,000~K at the pressure of the present-day core-mantle boundary.
This temperature range covers estimates for the early Earth's core that were calculated by requiring a core dynamo of its present strength for all of Earth history \citep[e.g.,][]{ORourke2016}.
A higher specific entropy core slightly increases the AM required to exceed the CoRoL because the increased moment of inertia of the extended core reduces the corotating angular velocity at a given AM.
We find that, for the range of core entropies considered, the equatorial radius of the structure and the AM required to exceed the CoRoL vary by only a few percent due to the low thermal expansivity of iron at high pressures.

\begin{figure}
\centering
\includegraphics[scale=0.8333333]{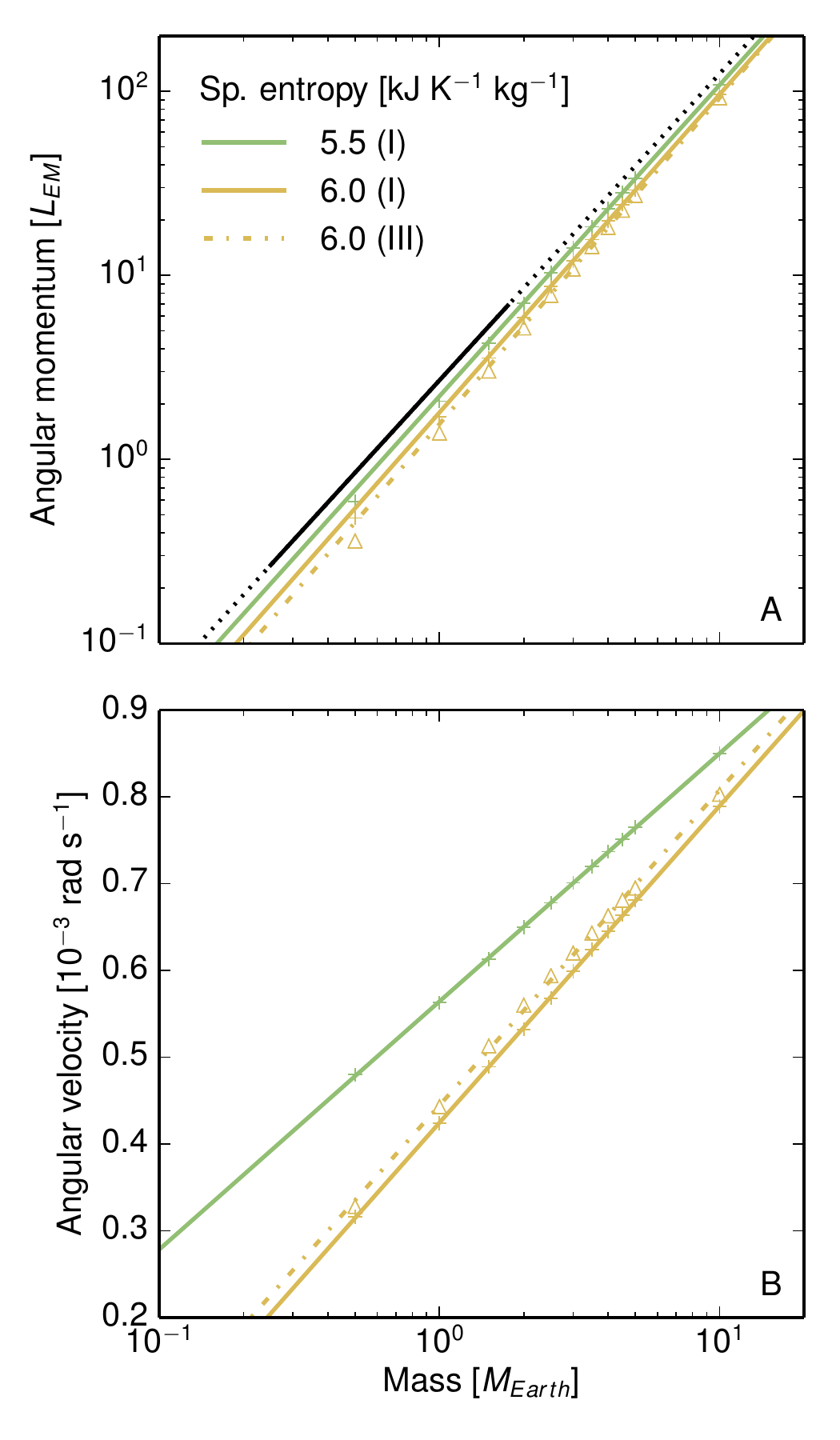}
\caption[]{The angular momentum required to exceed the CoRoL (A) and the angular velocity of bodies at the CoRoL (B) for Earth-like bodies of varying masses calculated using the HERCULES code. Symbols are for bodies with isentropic (class I, $+$) and stratified (class III, $\bigtriangleup$) silicate thermal profiles. Stratified profiles with 50\% by mass cold lower silicate layer $S_{\rm lower}$~$=$~$3$~kJ~K$^{-1}$~kg$^{-1}$ were used. The lines are power law (A) or log-linear (B) fits to the calculated points. Colors indicate the specific entropy of the silicates or upper silicate layer for the isentropic and stratified profiles, respectively. The black dashed line in A indicates the average AM expected for terrestrial bodies, extrapolated from the results of \citet{Kokubo2010}. The solid black line indicates the range of planetary masses formed in the simulations of \citet{Kokubo2010}.}
\label{fig:HSSL_M-L}
\end{figure}

The CoRoL is also dependent on the mass of the body (Figure~\ref{fig:HSSL_M-L}).
We modeled the relationship between mass and the AM required to exceed the CoRoL for Earth-like bodies by a simple power law of the form $L_{\rm CoRoL}\propto M^{\gamma}$, where $M$ is the mass of the body.
The value of the exponent, $\gamma$, only varies slightly between different thermal structures.
We fit the exponent for the three example thermal profiles shown in Figure~\ref{fig:HSSL_R-L}. For the 5.5 and 6.0~kJ~K$^{-1}$~kg$^{-1}$ isentropic profiles (class I, solid lines), $\gamma$~$=$~$1.69$ and 1.73, respectively. For the example stratified planet (class III) with 50~wt\% at 6.0~kJ~K$^{-1}$~kg$^{-1}$ (dot-dashed line), $\gamma$~$=$~$1.77$. These fits are good for larger mass planets but miscalculate the AM to exceed the CoRoL by 10 to 25\% for 0.5~$M_{\rm Earth}$ bodies and produce a misfit on the order of a few percent for $M_{\rm Earth}$ bodies.
The angular velocity of a body at the CoRoL also increases with mass and varies significantly for different thermal structures. The increase in angular velocity scales linearly with the logarithm of the mass of the body for the whole range of mass we considered.

Table~\ref{sup:tab:HSSL} presents a summary of the CoRoL for different rocky bodies calculated with the HERCULES code.

% XXX could mention future work to look at core entropies from impacts in the discussion

%xxxxxxxxxxxxxxxxxxxxxxxxxxxxxxxxxxxxxxxxxxxxxxxxxxxxxxxxxxxxxxxxxxxxxxxxxxxxxxxxxxxxxxxxxxxxxxxxxxxxxxxxxxxxxxxxxxxxxxxxxxxxxxxxxxxxxxxx
%xxxxxxxxxxxxxxxxxxxxxxxxxxxxxxxxxxxxxxxxxxxxxxxxxxxxxxxxxxxxxxxxxxxxxxxxxxxxxxxxxxxxxxxxxxxxxxxxxxxxxxxxxxxxxxxxxxxxxxxxxxxxxxxxxxxxxxxx
%xxxxxxxxxxxxxxxxxxxxxxxxxxxxxxxxxxxxxxxxxxxxxxxxxxxxxxxxxxxxxxxxxxxxxxxxxxxxxxxxxxxxxxxxxxxxxxxxxxxxxxxxxxxxxxxxxxxxxxxxxxxxxxxxxxxxxxxx

\begin{sidewaysfigure*}
\centering
\includegraphics[scale=0.83333333333]{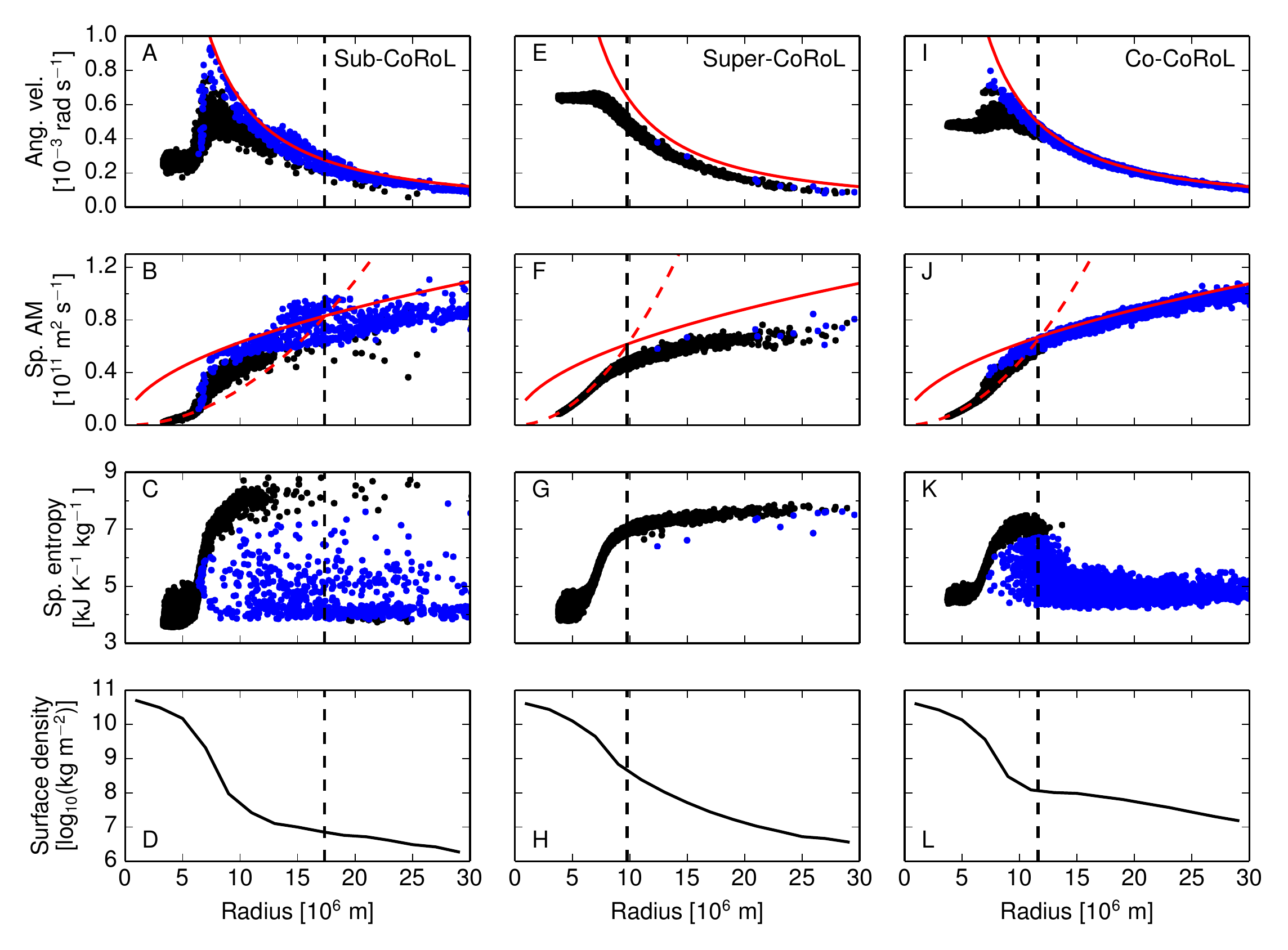}
\caption[]{Giant impacts produce a range of post-impact structures, with varying dynamic and thermodynamic properties (Table~\ref{sup:tab:impacts}). Dots represent silicate SPH particles in the midplane for three example impact scenarios (one in each column) that produce substantial disk-like regions at 48~hrs after first contact. The dynamical structures identified as: sub-CoRoL (1st column), super-CoRoL (2nd column), and co-CoRoL (3rd column) structures. Solid red lines denote the angular velocity or specific AM of a circular Keplerian orbit around a point mass. Dashed red line indicates the specific AM of material corotating with the inner region. The black particles are either at pressures above the critical point (25.5~kbar) or are pure vapor, and the blue particles are mixtures of liquid and vapor. The super-Keplerian particles in A and B are the remnants of a disrupting moonlet.
}
\label{fig:MAD}
\end{sidewaysfigure*}

\begin{sidewaysfigure*}
\centering
\includegraphics[scale=0.83333333333]{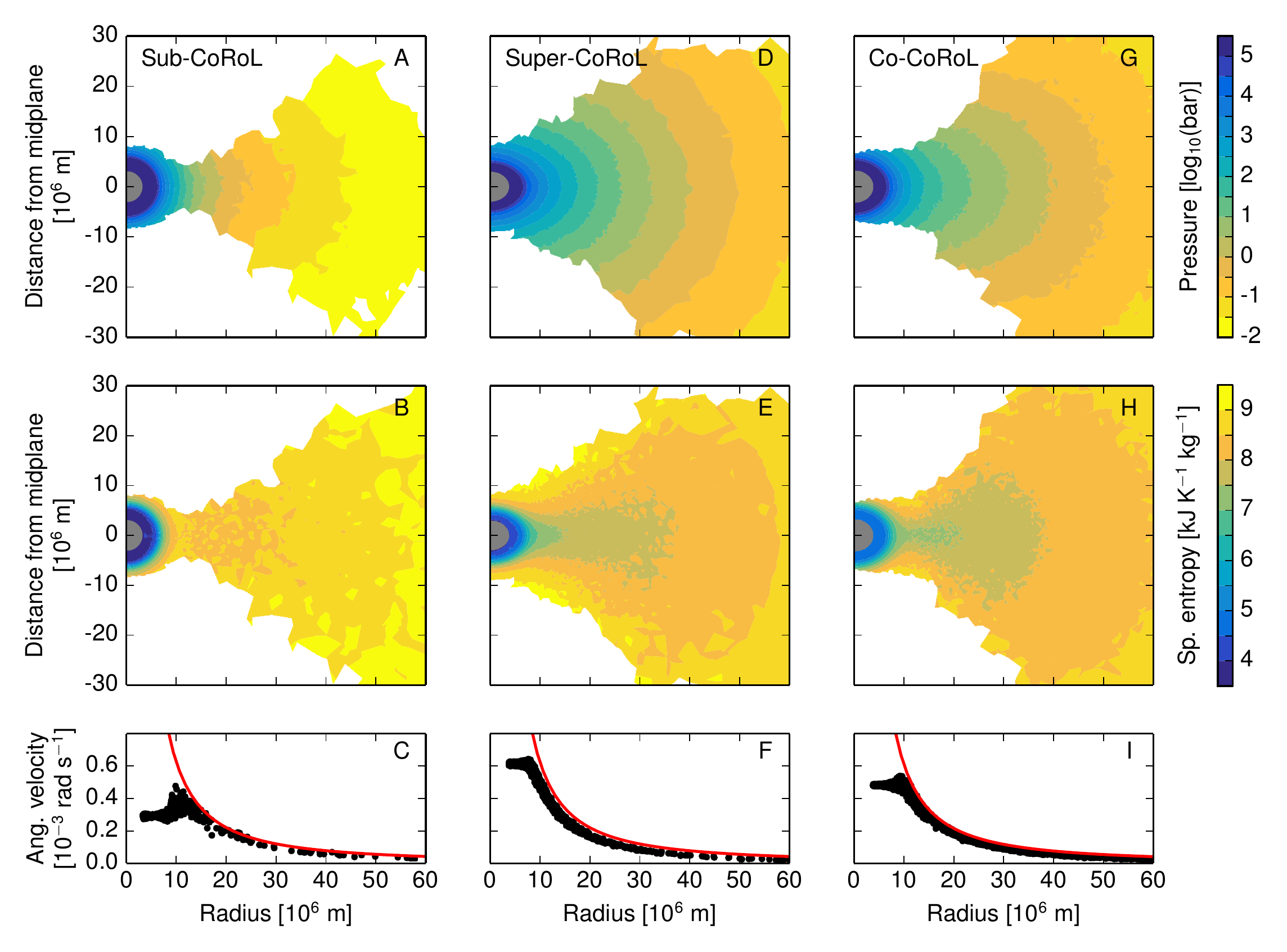}
\caption[]{The approximate vapor structure for the three example post-impact structures shown in Figure~\ref{fig:MAD}. The post-impact states were processed by removing condensed material and re-equilibrating the vapor structure as described in \S\ref{sec:impacts_structure}. Pressure (A,D,G) and entropy (B,E,H) contours were plotted as described in \S\ref{sec:structure}. Lower panels show the angular velocity (C,F,I) for the silicate SPH particles in the midplane. Solid red lines denote the angular velocity of a circular Keplerian orbit around a point mass. Pressure and entropy scale bars in this figure differ from elsewhere in the paper.}
\label{fig:PI_pressure}
\end{sidewaysfigure*}

\section{Post-impact structures}
\label{sec:impacts}
\subsection{The structure of post-impact states}
\label{sec:impacts_structure}

Rocky bodies are naturally forced into hot, rotating states by giant impacts during the later stages of accretion. We seek to understand the significance of hot, rotating planetary structures by assessing the range of possible structures that could be attained during planet formation and the frequency of occurrence. First we examine and attempt to classify the dynamical and thermal structures of post-giant impact states. In order to do this, we calculated the outcome of impacts between Earth-like bodies with a range of collision parameters using SPH, as described in \S\ref{sec:method_SPH}, and combined our results with the simulations from \citet{Cuk2012}. We examined the instantaneous structures of post-impact structures at 24 to 48 hrs after first contact, when the structure had reached a quasi-steady mass and AM distribution. 
This suite of impacts covers a range of impact energies but focuses on those that have high enough specific impact energy to be described as giant impacts by \citet{Quintana2016} (\S\ref{sec:impacts:Nbody}). The full set of impacts are summarized in Table~\ref{sup:tab:impacts}. 

We find that there is a broad range of post-impact structures, in terms of both thermal and dynamical state. Three example post-impact structures with substantial disk-like regions are shown in Figure~\ref{fig:MAD}. All structures have a corotating region close to the rotational axis, and a Keplerian or sub-Keplerian disk-like region farther out. The transition region varies substantially between different structures. A large fraction of our suite of post-impact states have very narrow transition regions and the corotating region grades smoothly into the disk-like region, as is the case for isolated synestias (Figure~\ref{fig:hot_planets}). But, in some cases there is a shear boundary (an increasing angular velocity with radius) in the transition region which varies in magnitude and width (Figure~\ref{fig:MAD}A and I). The prevalence of structures with narrow transition regions is likely biased by the many high-AM impacts with fast, small impactors in our suite of impacts. The nature of the transition region is determined both by the AM and thermal state of the structure, as we discuss below.

Impact-induced heating is very heterogeneous and post-impact structures are far from isentropic. The inner regions of the structure are thermally stratified with a monotonically increasing specific entropy with radius (Figure~\ref{fig:MAD}C,G,K). This general profile is produced by buoyancy and rotational forces during the impact acting to impose a monotonically decreasing density profile. Such gravitationally-equilibrated profiles neglect any effects from chemistry, thermal equilibration, or shear strength that may act on the dynamical timescale of hours. For impacts between initially condensed bodies, the post-impact lowermost mantle is a high-pressure liquid or a liquid-solid mixture \citep[see discussions in][]{Stewart2015,Nakajima2015} with specific entropies typically between 3 and 5~kJ~K$^{-1}$~kg$^{-1}$. 
The specific entropy of the silicates in the inner regions increases substantially with radius (Figure~\ref{fig:MAD}C,G,K), typically far exceeding the specific entropy of the critical point ($S_{\rm crit}$~$=$~$5.40$~kJ~K$^{-1}$~kg$^{-1}$).
In most cases we considered, the corotating region does not intersect the liquid-vapor phase boundary. At high pressure in the body, the silicate grades smoothly from a liquid into a supercritical fluid and then vapor. In the cases where the thermal profile does intersect the liquid-vapor phase boundary for a range of pressures within the corotating region, the condensate fraction is usually small.
%%The high specific entropy of the outer layers of the corotating region result in the radius of the corotating region being much larger than the radius of a fully condensed planet with the same mass and AM.

The thermal state of the silicates farther out in the structure is more variable. 
Some structures have disk-like regions that have high specific entropy and are mostly vaporized even to beyond the Roche limit (Figure~\ref{fig:MAD}G), while others have disk-like regions that are much colder and have a large mass fraction that is condensed (Figure~\ref{fig:MAD}C,K).
In the latter case, the transition between the high-entropy inner regions and the colder disk-like regions typically occurs in the dynamical transition region. 
The parcels of material represented by SPH particles cannot thermally equilibrate with each other.
As a result, colder, more condensed particles can coexist with hotter, mostly vapor particles in the transition region.
The colder particles are less pressure supported than the vaporized particles and rotate at a lower angular velocity at the same radius.
These colder particles complicate the dynamics in the transition region for the SPH post-impact structures (as shown in Figure~\ref{fig:MAD}).
In reality, the material represented by the colder and hotter SPH particles would thermally equilibrate, and in most cases in our suite of post-impact states, the transition region would evolve to be dominantly vapor. In all cases, there may be some (comparatively colder) material in orbit at larger distances (e.g., a distal debris disk) that is not continuous with the central structure.

We examined whether post-impact structures are below or above the CoRoL. 
Sub-CoRoL structures (e.g., Figure~\ref{fig:MAD}A-D) do not have sufficient AM for the hot, inner regions of the structure to extend out to the intersection between the corotating and Keplerian angular velocities (black dashed line in Figure~\ref{fig:MAD}).
Consequently, the vapor at the edge of the hot inner regions is sheared as the angular velocity profile transitions from corotating to sub-Keplerian. 
The disk-like regions of sub-CoRoL structures tend to have a low vapor fraction in the midplane. 
In SPH post-impact, sub-CoRoL structures there are colder particles in the transition region that contribute to the shear, but the angular velocity profile is still dominated by the shear in the vapor.
Conversely, in super-CoRoL structures (synestias), the hot inner regions extend out to the intersection between the corotating and Keplerian angular velocities.
There is a smooth transition in the angular velocity of the vapor from the corotating inner region to the sub-Keplerian disk-like region. 
In some super-CoRoL structures, there are colder particles in the transition region that have different angular velocities than the vapor, leading to an apparent shear. However, this shear does not significantly affect the overall smooth angular velocity profile.
In a few cases, it is difficult to determine whether the structure is above or below the CoRoL. 
An example of such a structure is shown in Figure~\ref{fig:MAD}I-L.
The vapor of the inner region only just extends to the intersection between the corotating and Keplerian angular velocities, and there is a substantial mass of colder particles in the transition region.
As a result, the mass averaged angular velocity profile has a small shear boundary. 
It is not clear from the SPH post-impact structure alone whether such structures are above the CoRoL nor whether they would remain so following thermal equilibration.
We classify such structures as co-CoRoL as they are at the border between sub-CoRoL and super-CoRoL states.

We would like to compare the shape of post-impact structures to those of the isolated synestias; however, the inability of impact codes to correctly model phase separation makes it difficult to calculate the pressure field that would be expected immediately after an impact.
To approximate the vapor pressure field in each of our example cases, we have modified the post-impact states produced by SPH to remove the condensed material.
We removed unbound SPH particles and the condensed mass fraction of bound SPH particles in the disk-like region from the simulation at 24~hrs after the impact. The structure was then re-equilibrated for a further 24~hrs with the mass of any condensate removed at each time step. The resulting structure approximates the continuous vapor pressure field at several dynamical times after the impact if the condensate was fully decoupled from the vapor and the effects of cooling were neglected. 
The calculated pressure fields, and corresponding specific entropy fields, for the three examples cases are given in Figure~\ref{fig:PI_pressure}.
The pressure fields and angular velocity profiles for the super-CoRoL and co-CoRoL cases are similar to those found for isolated synestias (Figures~\ref{fig:shape_plot} and \ref{fig:hot_planets}).
Impacts emplace significant amounts of mass and AM far from the rotation axis and the vapor structure can extend beyond the Roche limit, with significant vapor pressure (10's of bar) at several Earth radii.
The larger masses in the disk-like regions and the increase of specific entropy with decreasing pressure in the post-impact states (Figure~\ref{fig:PI_pressure}B,E,H) leads to more exaggerated flaring than in the isentropic isolated bodies in Figures~\ref{fig:shape_plot} and \ref{fig:hot_planets}. The concentration of cold material in the midplane is due to buoyancy driven settling and is partly a result of the lack of thermal equilibration in the SPH code. For the particular example of a co-CoRoL case in Figure~\ref{fig:MAD}, the structure after condensate extraction is above the CoRoL and the resulting angular velocity profile structure becomes monotonically decreasing with radius.
The sub-CoRoL example also shows a flared pressure field similar to the other cases but with lower pressures.
The angular velocity profile in the sub-CoRoL case still has a shear boundary after removing the condensate, and the profile is similar to the expected solution for a shearing fluid \citep{Chandrasekhar1969,Desch2013}.
%XXXXX SJL I don't think we need a section explaining this redistribution. It is pretty self explanatory. 

\begin{sidewaysfigure*}
\centering
\includegraphics[scale=0.8333333]{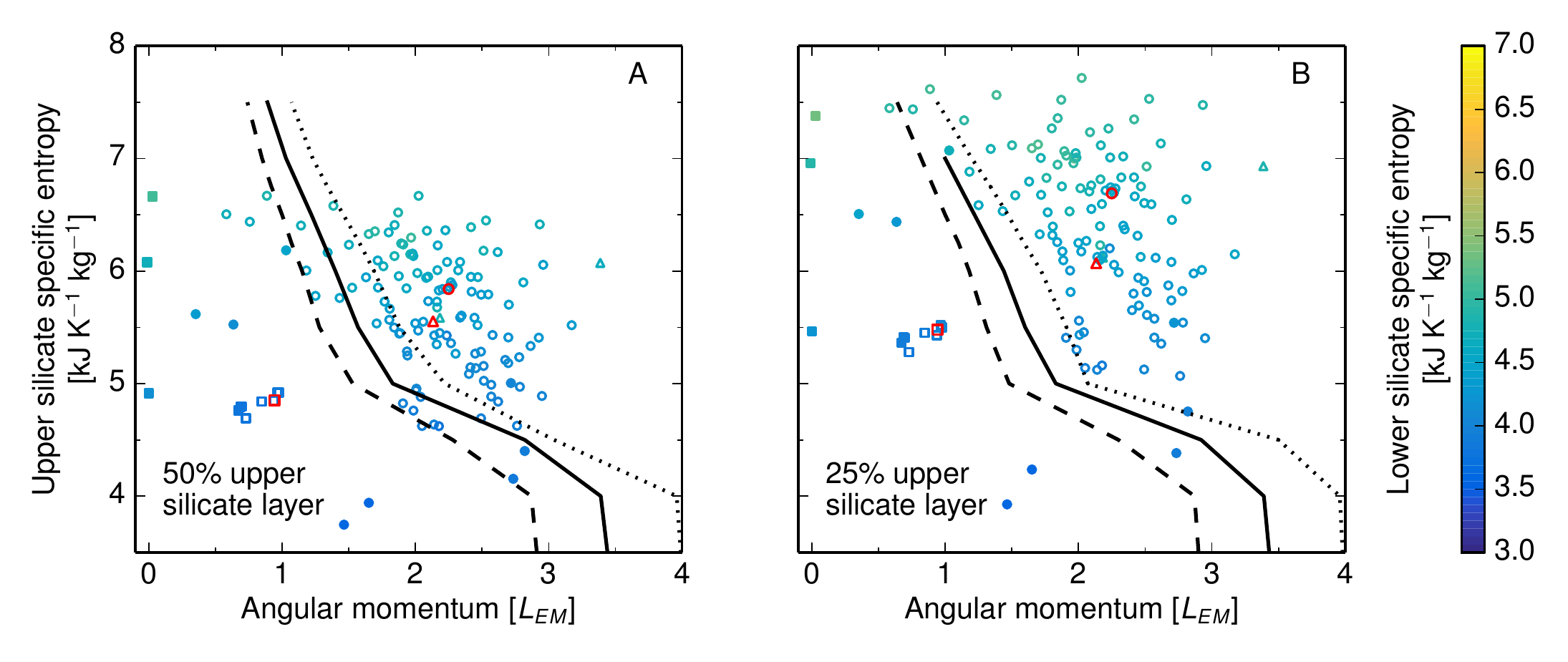}
\caption[]{Giant impacts can produce structures that are above the CoRoL. The characteristics of the post-impact structures can be assessed by their total angular momentum and average specific entropy of the outer 50\% (A) or 25\% (B) by mass of silicates. 
The symbols indicate the dynamical group of the structure: sub-CoRoL ($\Box$), super-CoRoL ($\bigcirc$) and co-CoRoL ($\bigtriangleup$). 
The colors indicate the average specific entropy of the corresponding lower 50\% (A) or 75\% (B) by mass of silicates. 
Filled symbols indicate structures with little or no mass in the disk-like region of the structure (see text for discussion of such cases).
Only impacts that have a total bound mass of the post-impact structures between 0.9 and 1.1 $M_{\rm Earth}$ are plotted, and the lines denote the CoRoL for Earth-like composition bodies of 0.9 (dashed), 1.0 (solid) and 1.1 (dotted) $M_{\rm Earth}$. The CoRoL lines indicate the upper silicate layer specific entropy and AM limit for stratified thermal profiles ($S_{\rm lower}$~$=$~$4$~kJ~K$^{-1}$~kg$^{-1}$) with 50~wt\% (A) or 25~wt\% (B) upper silicate layers. Red symbols correspond to the example cases plotted in Figure~\ref{fig:MAD}.}
\label{fig:HSSL_impact}
\end{sidewaysfigure*}

In Figure~\ref{fig:HSSL_impact}, we compare the post-impact structures to the CoRoL for stratified thermal profiles. As shown in Figure~\ref{fig:HSSL_thermal}, the CoRoL is defined by similar AM and outer specific entropy for a variety of stratified thermal profiles.
The majority of our post-impact states that we identified as super-CoRoL or co-CoRoL exceed the CoRoL calculated for stratified bodies when considering the specific entropy of the outer 50 or 25~wt\% of the body.
All our sub-CoRoL structures fall below the CoRoL for stratified bodies. There are some structures that we identified as super-CoRoL structures that fall below the CoRoL plotted in Figure~\ref{fig:HSSL_impact}.
These structures typically have thermal profiles where the top few percent of the mass has very high specific entropies. As a consequence, the structures intersect the CoRoL but with very little mass in the disk-like region (filled circles in Figure~\ref{fig:HSSL_impact}). The existence of very hot outer regions is not captured in the average entropy of the structure and comparing such structures with the CoRoL for stratified planets with 50 or 25~wt\% hot outer layers is not a good indicator of whether the structure is above the CoRoL.

Next, we consider the structure of post-impact bodies in the context of proposed Moon-formation scenarios.
The example cases shown in Figures~\ref{fig:MAD} are examples of proposed Moon-forming giant impacts. A-D show a typical structure formed by an impact in the style of the canonical Moon-forming impact \citep[e.g.,][]{Canup2001,Canup2004,Canup2008}: sub-CoRoL structures with relatively cold disk-like regions. E-H and I-L show example high-AM, high-energy Moon-forming impacts from \citet{Cuk2012} and \citet{Canup2012}, respectively. These high-AM impacts tend to produce super-CoRoL or co-CoRoL structures with substantially vaporized disk-like regions. Note that both the high-AM studies proposed a range of impact energies and impact parameters as potential Moon-forming events. Previous studies of post-impact states after the canonical impact \citep[e.g.,][]{Canup2001,Canup2004,Canup2008} did not observe synestias as they only considered bodies with an AM similar to the present-day Earth-Moon system. Such bodies would require very high specific entropies in their outer layers to exceed the CoRoL (Figure~\ref{fig:HSSL_impact}). Such hot thermal states are not reached in canonical style impacts.

A significant implication of this analysis is that some post-impact structures cannot be analyzed as a separate planet and disk, as is common in studies of post-impact states. Instead, the post-impact structures are synestias, which must be treated as a single extended structure. The impact-generated synestias have a large variation in the mass of the disk-like region. The proportion of mass and AM in the disk-like region is determined by the impact conditions \citep[with consistent results between different types of codes, e.g.,][]{Canup2013}. Because these structures exceed the CoRoL, any process that transfers mass from the disk-like region to the corotating region (and preserves the mean specific entropy and AM), would not be able to transfer all the mass to the corotating region. Or, more colloquially, the entire disk-like region cannot fall down. Because the structure is expected to cool by radiation (perhaps mitigated by internal dynamics such as viscous heating and continuing accretion), an impact-generated synestia will evolve to a state below the CoRoL, but the timescale will depend on the evolutionary path for that particular body. This investigation of post-impact states is a snapshot of the (near) initial structure after the event.

%XXXXXXXXXXXXXXXXXXXXXXXXXXXXXXXXXXXX
\subsection{The likelihood of hot, rapidly rotating post-impact states}
\label{sec:impacts:Nbody}

Next, we examine the likelihood of generating hot, rapidly rotating post-impact states during the giant impact stage of terrestrial planet formation. First, we examine the specific entropy of post-impact structures. We find that the specific entropy of the outer silicate portions of post-impact structures scales well with a modified specific impact energy, $Q_{\rm S}$. $Q_{\rm S}$ is a variation of the parameter developed in \citet{Leinhardt2012} that adjusts for the geometry of collisions between similarly sized bodies to estimate the relative deposition of impact energy into the post-impact body for different collision scenarios (supporting information \S\ref{sup:sec:Qs}).

\begin{figure}
\centering
\includegraphics[scale=0.8333333]{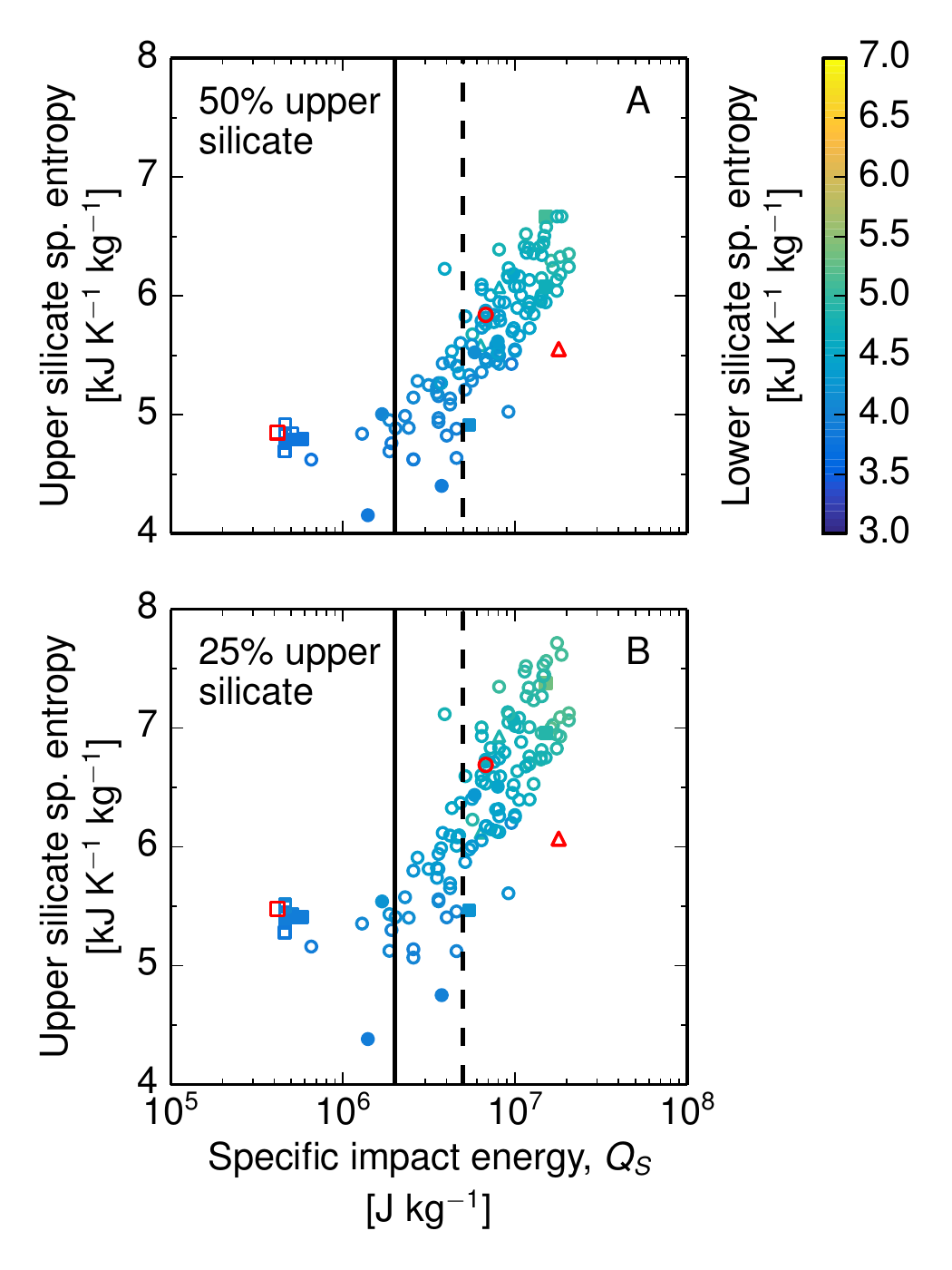}
\caption[]{The specific entropy of the upper silicate layers of post-impact structures scales well with the specific impact energy, $Q_{\rm S}$. The scaling for both the specific entropy of the upper 50~wt\% (A) and 25~wt\% (B) is shown.
Symbols indicate the dynamical group of the structure: sub-CoRoL ($\Box$), super-CoRoL ($\bigcirc$) and co-CoRoL ($\bigtriangleup$). The colors indicate the average specific entropy of the corresponding lower 50~wt\% (A) or 75~wt\% (B) of the silicates. Filled symbols indicate structures with little or no mass in the disk-like region of the structure. Only post-impact structures with bound masses between 0.9 and 1.1~$M_{\rm Earth}$ are shown. Red symbols correspond to the example cases plotted in Figure~\ref{fig:MAD}.}
\label{fig:QS}
\end{figure}

We find that the specific entropy of the outer silicate layers scales linearly with the logarithm of specific impact energy for post-impact bodies with masses between 0.9 and 1.1 $M_{\rm Earth}$ and $Q_{\rm S}$~$>\sim$$10^6$~J~kg$^{-1}$ (Figure~\ref{fig:QS}).
$Q_{\rm S}$ does not account for any entropy increase caused by secondary impacts in graze and merge collisions or by reimpacting debris.
Such effects can be significant, particularly for low energy impacts, and are likely responsible for the apparent plateau in entropies at low $Q_{\rm S}$ in our suite of impacts as these low energies are dominated by graze and merge, canonical-style Moon-forming impacts.
Impacts with $Q_{\rm S}$~$>$~$2\times10^6$~J~kg$^{-1}$ generally deposit sufficient energy such that the upper 25~wt\% of the silicates have an average specific entropy that exceeds the critical point value for the equation of state, $S_{\rm crit}$~$=$~$5.4$~kJ~K$^{-1}$~kg$^{-1}$ (Figure~\ref{fig:QS}B, solid line).
For impacts with $Q_{\rm S}$~$>$~$5\times10^6$~J~kg$^{-1}$, the upper 50~wt\% of silicates typically attain mean specific entropies above the critical point (Figure~\ref{fig:QS}A, dashed line).
Such hot post-impact structures are dominantly vapor at low pressures. They have much larger radii compared to entirely condensed planets of the same mass and AM. The correlation between post-impact thermal state and specific impact energy provides a straightforward means to determine which collisions during accretion lead to highly vaporized structures. 

\begin{figure}
\centering
\includegraphics[scale=0.8333333]{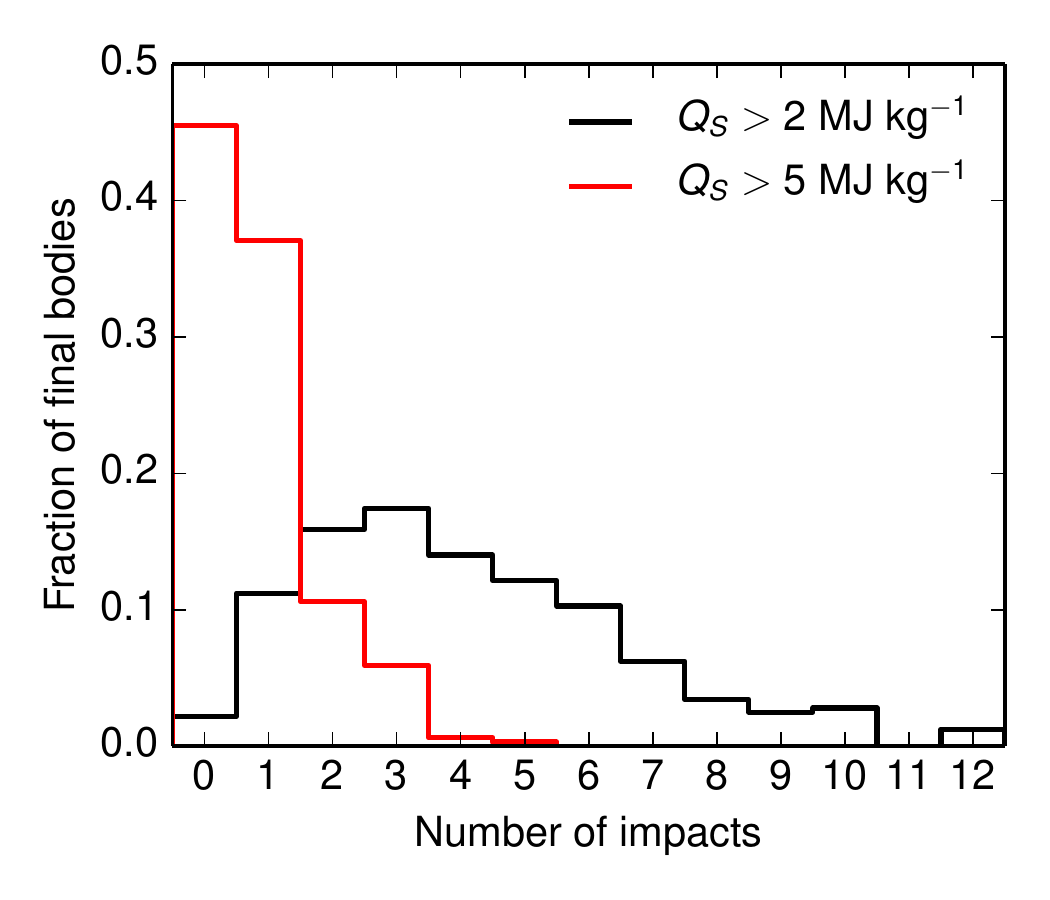}
\caption[]{High specific energy impacts are common during terrestrial planet formation. Histograms present the fraction of planets with final masses above 0.5 $M_{\rm Earth}$ that experienced a given number of impacts with specific impact energies, $Q_{\rm S}$, greater than $2$~$\times$~$10^6$~J~kg$^{-1}$ (black) or $5$~$\times$~$10^6$~J~kg$^{-1}$ (red). Such planets generally experience several giant impacts with $Q_{\rm S}$~$>$~$2$~$\times$~$10^6$~J~kg$^{-1}$ and about half suffer an event with $Q_{\rm S}$~$>$~$5$~$\times$~$10^6$~J~kg$^{-1}$.}
\label{fig:impact_likelihood}
\end{figure}

$N$-body simulations of terrestrial planet formation have recently incorporated more realistic collision outcome models including fragmentation \citep[e.g.,][]{Chambers2013,Carter2015,Quintana2016}. \citet{Quintana2016} used the same modified specific impact energy, $Q_{\rm S}$, to evaluate the collisions in their simulations and graciously provided us with the collision histories for the final planets. We calculated the number of high energy impacts that near Earth-mass planets experience as they accrete. We find that most near Earth-mass planets experienced several giant impacts with $Q_{\rm S}$~$>$~$2\times10^6$~J~kg$^{-1}$ during accretion and over half also experienced at least one impact with $Q_{\rm S}$~$>$~$5\times10^6$~J~kg$^{-1}$ (Figure~\ref{fig:impact_likelihood}). Assuming that the scaling of thermal state with impact energy holds for smaller bodies (Figure~\ref{fig:QS}), we expect rocky bodies to form highly vaporized structures multiple times during accretion. Other $N$-body studies without fragmentation and with different initial conditions and alternative configurations of the giant planets \citep[e.g.,][]{Obrien2006,Hansen2009,Raymond2009,Walsh2011,Levison2015a} all produce high-energy impacts in the final stages of accretion although the frequency of different energy impacts may change between these different scenarios. 

The rotational states of terrestrial bodies during accretion is difficult to track, especially because partitioning of AM between the growing bodies, collision ejecta, and a planetesimal population is poorly understood. $N$-body studies have found that each giant impact can dramatically alter the spin state of a growing planet \citep[e.g.,][]{Agnor1999}. \citet{Kokubo2010} have conducted the best assessment to date of the spin state of rocky planets during accretion. 
They used an $N$-body simulation of planet formation with bimodal impact outcomes, either perfect merging or hit-and-run, and tracked the AM of each of the bodies in the simulation. 
\citet{Kokubo2010} found that the mean AM of rocky planets is large, e.g., 2.69~$L_{\rm EM}$ for Earth-mass planets, and that the distribution of AM is wide. The mean angular velocity of final planets was similar for the entire mass range in their study, covering planetary masses between about 0.25 and 1.75 $M_{\rm Earth}$. We infer that that the AM of terrestrial bodies roughly scales as $L$~$=$~$2.69 L_{\rm EM} (M/M_{\rm Earth})^{5/3}$ (black line in Figure~\ref{fig:HSSL_M-L}A).
These results indicate that the AM of planets during accretion is expected to be large and that the present-day AM of planets have been modified by satellite and solar tides.

The prevalence of both hot and high-AM states for terrestrial bodies during accretion makes the formation of synestias highly likely. In Figure~\ref{fig:HSSL_impact}, we find that thermal states with dominantly vapor outer layers (e.g., $S_{\rm upper}$~$>$~$S_{\rm crit}$) reach the CoRoL for AM greater than about $1.5 L_{\rm EM}$. \citet{Kokubo2010} found that a majority of Earth-mass planets exceed this AM at the end of accretion. Our examination of impact energies during accretion find that nearly all Earth-mass planets experience one or more collisions that produce dominantly vapor outer layers. Therefore, we conclude that it is common for Earth-mass bodies to exceed the CoRoL and form synestias one or more times during accretion. 
%Can we be more precise than majority? XXXXX 1.5 L_EM is 1.3 /hr for The fig 3 of Kokubo and Genda 2010. That is close to the end of the 1 sigma error bars so it is >~84% have L>1.5LEM. But I don't want to try and throw out a number based on their graph. We could ask them for their data. 

%xxxxxxxxxxxxxxxxxxxxxxxxxxxxxxxxxxxxxxxxxxxxxxxxxxxxxxxxxxxxxxxxxxxxxxxxxxxxxxxxxxxxxxxxxxxxxxxxxxxxxxxxxxxxxxxxxxxxxxxxxxxxxxxxxxxxxxxx
%xxxxxxxxxxxxxxxxxxxxxxxxxxxxxxxxxxxxxxxxxxxxxxxxxxxxxxxxxxxxxxxxxxxxxxxxxxxxxxxxxxxxxxxxxxxxxxxxxxxxxxxxxxxxxxxxxxxxxxxxxxxxxxxxxxxxxxxx
%xxxxxxxxxxxxxxxxxxxxxxxxxxxxxxxxxxxxxxxxxxxxxxxxxxxxxxxxxxxxxxxxxxxxxxxxxxxxxxxxxxxxxxxxxxxxxxxxxxxxxxxxxxxxxxxxxxxxxxxxxxxxxxxxxxxxxxxx
\section{Discussion}
\label{sec:discussion}

%xxxxxxxxxxxxxxxxxxxxxxxxxxxxxxxxxxxxxxxxxxxxxxxxxxxxxxxxxxxxxxxx
%xxxxxxxxxxxxxxxxxxxxxxxxxxxxxxxxxxxxxxxxxxxxxxxxxxxxxxxxxxxxxxxx

\subsection{Dynamic stability of hot, rotating planetary structures}

%%% XXX STSM what about post-impact structure stability?

Rapidly rotating, axisymmetric bodies can be susceptible to dynamic instabilities, such as the bar instability that causes the breakdown of axial symmetry and drives reorganization of the structure \citep[e.g.,][]{Chandrasekhar1969}.
Constant-density, rotating spheroids are also known to transition from axisymmetric Maclaurin spheroids to non-axisymmetric Jacobi spheroids above a critical degree of rotational flattening, because Jacobi spheroids are lower energy states \citep[][]{Chandrasekhar1969}. 
If rapidly rotating axisymmetric planetary structures were susceptible to such instabilities, the structure could rearrange or break up.
Here, we examine the stability of the rapidly rotating planetary structures considered in this work. 

The stability of rapidly rotating fluid bodies is typically evaluated by the ratio of the rotational kinetic energy, $T$, to the gravitational potential energy, $W$.
Rotating bodies are unstable to non-axisymmetric modes if the ratio $T/|W|$ is above a critical value, which varies depending on the body in question. Constant-density Maclaurin spheroids \citep{Chandrasekhar1969}, stellar systems \citep{Ostriker1973a}, rapidly rotating white dwarfs \citep{Ostriker1969}, and polytropic stars \citep{Ostriker1973} have critical values in the range 0.14 to 0.27. 
All the pre- and super-CoRoL structures considered in this work have an energy ratio $T/|W|$ below this range of critical values. 
For example, the highest value for the structures shown in Figure~\ref{fig:shape_plot} is $0.07$. For the post-impact structures we studied for this paper (Table~\ref{sup:tab:impacts}) the maximum value is $0.085$ and the average value is $0.042$.
We also did not observe any instabilities or triaxial structures in our isolated body SPH calculations, where the equilibrium shapes were calculated over many dynamical timescales. 
Therefore, based on the energy stability criteria and our empirical observations, the rapidly rotating planetary structures presented in this work are likely to be dynamically stable. 

%xxxxxxxxxxxxxxxxxxxxxxxxxxxxxxxxxxxxxxxxxxxxxxxxxxxxxxxxxxxxxxxx
%xxxxxxxxxxxxxxxxxxxxxxxxxxxxxxxxxxxxxxxxxxxxxxxxxxxxxxxxxxxxxxxx

\subsection{Lifetimes of hot planetary structures}

Although hot, extended planetary structures are dynamically stable, they evolve by radiating energy. In some cases, AM may be reduced by tides or resonant interactions.
For synestias, a sufficient decrease in AM or thermal energy will cause the body to fall below the CoRoL and adopt a more compact, corotating structure.
Bodies that are far enough away from their host stars will eventually cool to a silicate magma ocean overlain by a volatile dominated atmosphere. 
Calculating the timescales for the collapse of synestias and the cooling of post-impact states to magma oceans is key to understanding the influence of post-impact states on the evolution of terrestrial planets.

Synestias formed by impacts are substantially vapor in their disk-like regions. 
Unless the body is very close to its parent star, the outer regions of the structure are expected to cool quickly by radiation.
When the photosphere is controlled by silicate condensation, the radiative temperature is high \citep[$\sim$2300~K,][]{Lock2016LPSC}, but the energy that needs to be radiated to fall below the CoRoL is substantial.
Condensing the vapor requires significant energy loss, both to cool to the phase boundary and to release the latent heat of vaporization of silicates. The contraction of the structure upon cooling also releases a comparable amount of potential energy.
The amount of energy that needs to be radiated to fall below the CoRoL depends on how the internal energy is redistributed during cooling because the thermal structure affects the location of the CoRoL.
We can estimate a lower limit on the timescale to cool below the CoRoL by considering just the potential energy difference between a post-impact structure and a corotating body of the same AM, with a thermal state slightly below the CoRoL.
This approach neglects the loss of energy required to reduce the specific entropy of the outer layer of silicates (for a fixed AM) to below the CoRoL, which could be substantial.
We assume a radiative temperature of $\sim$2300~K for silicate vapor. 
However, the surface area of the structure will change with time, and the difference in radiative surface area between the post-impact state and the sub-CoRoL structure is an order of magnitude.
As a result, there is a substantial uncertainty in cooling time.
Considering the full range of possible radiative surfaces, typical Earth-mass post-impact synestias would require a minimum time of order $10^1$ to $10^3$~yrs to cool to a state below the CoRoL. 

Alternatively, a synestia could be brought below the CoRoL by removing AM from the structure by tides, interactions with the host star \citep{Murray1999}, or three body interactions with the host star and a moon \citep{Goldreich1966,Touma1994,Cuk2012,Wisdom2015,Cuk2016,Tian2017}.
A significant reduction of AM by tides takes 10$^3$~-~10$^{9}$~yrs, depending on the mechanism, which is generally greater than the radiative cooling timescale.
AM may also be redistributed by viscous spreading, but the whole structure would remain above the CoRoL (for a fixed thermal state). 
Thus, an impact-generated Earth-mass synestia would be stable for greater than 10's to 1000's of years.

The timescale for a post-impact structure to cool to a fully condensed magma ocean is longer. 
To estimate the timescale, we calculate the difference in potential, kinetic and internal energy between post-impact states and corotating, isentropic condensed planets of the same AM (left column in Figure~\ref{fig:shape_plot}).
We again assume a radiative temperature of $\sim$2300~K and consider radiative surface areas ranging from the post-impact state to a fully condensed body.
For highly extended post-impact states that are above the CoRoL, the cooling time to a magma ocean state is on the order of $10^2$ to $10^3$~yrs.
For post-impact states produced by less energetic impacts that do not begin above the CoRoL, the timescale is shorter, but still on the order of $10^2$ to $10^3$~yrs.

For all partially vaporized states, the cooling time may be substantially extended if the chemistry of the system produces a lower-temperature photosphere.
For example, refractory components in the photosphere may form relatively low-temperature hazes.
The chemistry of the silicate atmosphere will also evolve as it cools, potentially changing the effective radiative temperature. 
In addition, the energy input from continued accretion will increase the cooling timescale of post-giant impact structures, especially if the accreting bodies were small and primarily deposited their energy in the high-entropy outer layers of the body.
The lifetime of hot, extended structures affects the accretion and evolution of terrestrial planets (\S\ref{sec:discussion_formation}).

%xxxxxxxxxxxxxxxxxxxxxxxxxxxxxxxxxxxxxxxxxxxxxxxxxxxxxxxxxxxxxxxx
%xxxxxxxxxxxxxxxxxxxxxxxxxxxxxxxxxxxxxxxxxxxxxxxxxxxxxxxxxxxxxxxx
\subsection{CoRoL vs.\ solid body rotational breakup}

The rotational breakup of small rocky bodies has been extensively studied due to its importance in bounding the possible rotational states of minor planets and in understanding the origin of multiple asteroid systems \citep[e.g.,][]{Harris1996,Pravec2007,Richardson2006,Holsapple2004,Richardson2005,Cuk2007,Walsh2008} \citep[see review by][]{Walsh2015}.
Solid bodies with a range of material strength properties have been considered. 
Cohesionless asteroids share similarities with fluid bodies, and the spin stability limit for self-gravitating solid bodies with deformation has been studied in detail \citep[e.g.,][]{Holsapple2004}. 

The CoRoL for the larger rocky bodies studied in this work is different than the critical spin limit for small bodies. 
The typical assumption of incompressibility used in the small body literature does not hold for planet-sized bodies.
In addition, planetary bodies with hot thermal states intersect the liquid-vapor phase boundary, leading to a significant decrease in average density.
Accordingly, the CoRoL is typically reached at a lower AM compared to the critical spin limit for a rigid, condensed body.

The outcome of exceeding spin stability is also different for asteroids and hot planets. Upon exceeding the critical spin limit, an asteroid will shear into two or more bodies with relative motions that conserve AM \citep{Cuk2007,Walsh2008,Walsh2015}. In contrast, a hot planetary body that exceeds the CoRoL can remain a single structure with the excess AM forming a disk-like region. Therefore, the CoRoL is a new dynamical transition that is fundamentally different from the well-studied critical spin limit for smaller solid bodies.

%xxxxxxxxxxxxxxxxxxxxxxxxxxxxxxxxxxxxxxxxxxxxxxxxxxxxxxxxxxxxxxxx
%xxxxxxxxxxxxxxxxxxxxxxxxxxxxxxxxxxxxxxxxxxxxxxxxxxxxxxxxxxxxxxxx
\subsection{Analysis of post-impact structures}
\label{sec:discussion_analysis}

The structure and evolution of post-impact states can have a substantial impact on planet formation and the final properties of terrestrial planets (\S\ref{sec:discussion_satellites}).
In particular, the properties of the disk-like regions of post-impact structures control the mechanisms and efficiency of satellite formation (\S\ref{sec:discussion_formation}).
The numerical methods used to model giant impacts cannot be used to directly model satellite formation.
Hence, dedicated disk structure and evolution models have been used to study the formation of moons from the disk-like regions of post-impact structures, with initial conditions based on the results of impact simulations.
Informed by work on astrophysical disks, most studies treat post-impact structures as consisting of a distinct planet and disk, both for modeling the disk evolution and for analyzing post-impact states. 
However, as discussed in \S\ref{sec:impacts} and shown in Figure~\ref{fig:MAD}, above the CoRoL the entire post-impact structure can be continuous between the corotating and disk-like regions.
Even below the CoRoL, there may be significant vapor pressure in the transition region between the corotating and disk-like regions (e.g., Figure~\ref{fig:PI_pressure}A), and the disk-like region cannot be treated in isolation from the rest of the structure.
Here, we summarize previous work on the structure of circumterrestrial disks formed by giant impacts and identify some issues with common approaches taken in these studies.

Published studies of giant impacts have made different assumptions on how to divide post-impact structures into a planet and a disk. 
Since most impact studies have utilized SPH, in this section we refer to a parcel of material as a particle. 
The majority of studies \citep[e.g.,][]{Canup2001a,Canup2001,Canup2004,Canup2008,Canup2012,Nakajima2014,Nakajima2015} have used an iterative routine to divide the structure, e.g., as described in \citet{Canup2001a}. %%% probably the first use is much earlier from Cameron's work SJL What do you want to do about this. I can't find a description of what they did to identify disk particles in the Cameron papers i am aware of.
First, an initial guess is made to estimate the mass and equatorial radius of the planet. 
Particles are defined as being part of the planet or disk based on their AM.
A particle is classified as being in the disk if it has sufficient AM such that the semi-major axis of a circular Keplerian orbit with the same AM is greater than the planet's equatorial radius. Bound particles that are not classified as being in the disk are considered to be part of the planet. 
The equatorial radius of the planet is recalculated based on the total mass and AM of all the particles classified as being in the planet, assuming the planet has a similar bulk density to the present-day Earth and limited rotational flattening.
The particles are then reclassified based on the new mass and radius of the planet and the procedure is repeated until convergence.
Alternatively, \citet{Cuk2012} used a density contour of $1000$~kg~m$^{-3}$ to define the planet's equatorial radius due to the large rotational flattening of the planet. 
As in the iterative routine, material with sufficient AM to be in a circular Keplerian orbit above the equator of the planet was considered to be part of the disk.
Low density material that did not meet the AM requirement to be in the disk was considered to be part of a vapor atmosphere around the planet.
The combined mass of the planet and the atmosphere was used for calculating the orbit of particles.
In both methods, once the structure is divided into a planet and disk, the surface density of the disk is calculated by moving the mass of each disk particle to the semi-major axis of the equivalent circular Keplerian orbit for the AM of that particle.
Although the details differ slightly, previous work applied the same principle of dividing the structure based on AM, and we refer to both methods as the conventional analysis.

The conventional analysis neglects the connection between the disk-like region and the corotating region of the structure through the vapor phase.
Also, the radius of the planet defined in the conventional analysis does not account for the large mass of high specific entropy, low-density material in the corotating and transition regions of the structure because the planet is assumed to have a bulk density corresponding to condensed material (of order 10$^{3}$~kg~m$^{-3}$).
The assumptions made in analyzing post-impact states are important as the properties of the post-impact structure (e.g., the disk mass, AM and surface density) calculated using the conventional analysis and reported in impact studies are used to inform the initial conditions of disk evolution models, either directly or indirectly.

Most studies of disks in the aftermath of giant impacts have focused on the Moon-forming event \citep[e.g.,][]{Thompson1988,Ida1997,Kokubo2000,Ward2012,Ward2014,Ward2017,Salmon2012,Salmon2014,Nakajima2014,Canup2015,Charnoz2015}.
In the canonical giant impact, $\sim$20~wt\% of the disk would have been initially vapor \citep{Canup2001,Canup2004}, but in the recently proposed high-AM models, the disk would have been largely vaporized \citep{Cuk2012,Canup2012,Nakajima2014}. 
Modeling the structure and evolution of partially vaporized disks is challenging since the material in the disk is comprised of multiple phases and multiple chemical components.
It is necessary to make a number of simplifying assumptions to  make the problem tractable. 
The simplest disk models are $N$-body simulations that neglect the vapor in the disk and treat the disk material as fully condensed particles \citep[e.g.,][]{Ida1997,Kokubo2000}.
As the disk evolves under gravitational forces, material with an orbit that intersects the planet is assumed to be lost from the disk and is removed from the simulation.

More advanced disk models have made approximations of the multiphase physics \citep[e.g.,][]{Thompson1988,Ward2012,Ward2014,Ward2017,Salmon2012,Salmon2014,Nakajima2014,Canup2015,Charnoz2015}.
Typically, particles of condensate are assumed to experience mutual collisions that damp the eccentricity and inclination of their orbits.
Outside the Roche limit, the vapor is assumed to condense quickly and the condensate rapidly collides together to form moonlets.
Inside the Roche limit, the condensate separates from the gas and forms a liquid layer \citep{Ward2012} or froth \citep{Thompson1988} in the midplane of the disk, surrounded by a vapor atmosphere.
Viscosity in the disk causes the inner disk to spread and the surface density distribution to evolve.
As in $N$-body simulations, material that spreads inside the radius of the planet is assumed lost from the disk, although some preliminary work has considered a more complex inner boundary \citep{Desch2013,Charnoz2015}.
Although the above assumptions are the most important for our discussion here, a number of other assumptions are made in different models of the inner disk, including: a constant surface density  \citep[e.g.,][]{Salmon2012,Salmon2014}, neglecting the radial pressure gradient \citep[e.g.,][]{Charnoz2015} and using a single component equation of state \citep[e.g.,][]{Ward2012}.
All the disk models are initialized with conditions that are based on the conventional analysis described above and evolve under the assumption of unfettered mass flow to the planet. 

Our work identifies several issues with the assumptions made in previous studies of post-impact structures and circumterrestrial disks.
First, post-impact structures cannot always be divided into a planet and disk that can be modeled separately.
For synestias, the corotating region and the disk-like region are continuous, as demonstrated by the angular velocity and thermodynamic profiles of the structure shown in Figure~\ref{fig:MAD}E-H. 
In such cases, calculating the structure or evolution of a region of the structure without considering the whole will lead to substantial errors.
Even for sub-CoRoL structures, there can be significant pressures ($\sim$1000's bar in the transition region between the corotating and disk-like regions, e.g., Figure~\ref{fig:PI_pressure}A), due to the high specific entropy of the material. In such cases, the structure and evolution of the disk-like region cannot be considered in isolation from the rest of the structure because the influence of the vapor in the corotating and transition regions is not negligible. 
Due to the hot thermal state of the corotating and transition regions, the definition of the planet in the conventional analysis does not correspond to the inner edge of the disk-like region in the post-impact structure.
For example, for the structure shown in Figure~\ref{fig:MAD}A-D, in the conventional analysis the mass of the planet is 0.98$M_{\rm Earth}$ using the iterative routine of \cite{Canup2001a} and 0.95$M_{\rm Earth}$ using the approach of \cite{Cuk2012}. The combined mass of the planet and atmosphere using the \citep{Cuk2012} approach is 0.99$M_{\rm Earth}$.
In contrast, the mass of the corotating region in the SPH post-impact structure is only 0.88$M_{\rm Earth}$ but the radius of the corotating region extends out to $\sim$6500~km, similar to the radius of the planet defined in the conventional analysis, 6532~km or 6691~km. The hot vapor in the transition region extends out to beyond 8000~km, much greater than the conventional radius of the planet. Thus, the inner edge of the disk-like region is much farther out than assumed in the conventional analysis. The thickness of the transition region may be dependent on the viscosity of the vapor \citep{Desch2013} and the thickness may not be captured accurately in SPH. However, even if the transition region was much thinner, the large radius of the hot inner portion of the structure cannot be ignored.

Second, mass and AM cannot simply be lost from the disk-like regions of a structure to the planet.
The pressure at the interface between the corotating and disk-like regions in most structures is high (10$^3$-10$^4$~bar) and these regions tend to be majority vapor. Vapor cannot simply flow across the boundary because of the pressure gradient.
The fate of condensed mass in the disk that falls or spreads inwards is more complicated. 
Falling condensate would encounter the high specific entropy of the corotating and transition regions, thermally equilibrate and vaporize.
Vaporizing condensates could increase the pressure support in the corotating region and, in turn, the disk-like region. 
Increasing the vapor pressure could add vapor mass to the disk-like region, particularly in the case of synestias. 
The magnitude of the effect of the infalling condensate depends on the balance between the potential energy released by the falling condensates and the redistribution of thermal energy.
Crucially, mass cannot flow irretrievably from the disk-like region to the corotating region, and previous studies have artificially depleted the mass in the disk-like region.

Third, the redistribution of the disk mass onto Keplerian orbits in the conventional analysis produces unrealistically compact surface density distributions for multiphase disks. 
In post-impact structures the disk-like regions are typically a variable mixture of condensate and vapor (for example, see the range of specific entropies in Figure~\ref{fig:MAD}C,G,K).
The condensates are not supported by the pressure gradient in the vapor.
If the condensates decouple from the vapor then they would fall inwards via buoyancy forces.
The codes that are typically used in giant impact studies do not include the separation of vapor and condensed phases.
This inability of the impact code to model multiphase behavior is the rationale behind the redistribution of mass in the conventional analysis.
The conventional analysis assumes that the mass fraction of vapor is small and so the effect of vapor pressure is neglected.
However, for disk-like regions that are substantially vaporized, the vapor pressure cannot be neglected.
Even assuming condensate is decoupled from the vapor, the pressure gradient can support the fraction of mass that is vapor on orbits at greater semi-major axes than purely Keplerian orbits.
If the condensates are coupled to the vapor, then the whole mass of the disk can be supported by pressure gradient forces.
If there is a substantial fraction of vapor, the conventional analysis produces a surface density distribution that is more compact than a pressure-supported disk.
In impacts where the majority of the disk-like region is vaporized (e.g., the example in Figure~\ref{fig:MAD}E-H), the structure of the disk-like region is correctly modeled by the impact code, given sufficient resolution and time for the structure to reach quasi-equilibrium (e.g., Figures~\ref{fig:MAD}E-H and \ref{fig:PI_pressure}D-F). 
In these cases, no post-simulation redistribution of mass is necessary.
Even so, there is a technical difficulty in calculating the post-impact structure for a particular impact even in completely vapor structures. 
The numerical viscosity in the SPH code leads to some redistribution of mass and AM on the dynamical timescales required for the structure to gravitationally equilibrate. 
For the example in Figure~\ref{fig:PI_pressure}D-F, the mass outside of Roche increased by a few tenths of a lunar mass over days of simulation time.
Thus, the equilibrium structure evolves over the calculation time.
Given that the real viscosity in the structure is poorly constrained, there is uncertainty in the predicted post-impact structure for a specific impact scenario.
In a mixed phase disk, more careful analysis is needed to correctly model the dynamics and thermodynamics of liquid and vapor both during the impact and in the immediate aftermath. 

\begin{figure}
\centering
\includegraphics[scale=0.8333333]{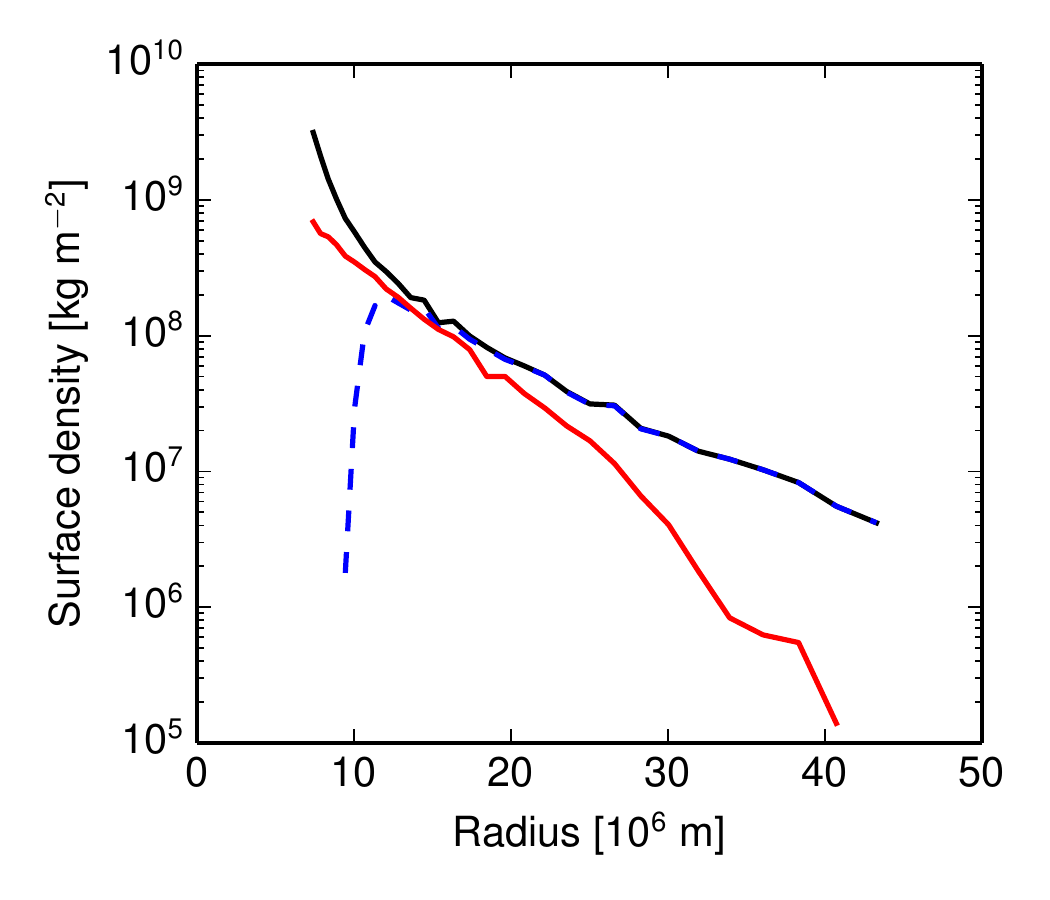}
\caption[]{The conventional analysis for processing the output of giant impact simulations artificially depletes the mass in the disk-like regions for substantially vaporized structures. For example, the surface density in the disk-like region of the structure shown in Figure~\ref{fig:MAD}E-H (black) is significantly higher than the surface density calculated using the conventional analysis in the style of \citet{Cuk2012} (red). The disk mass is moved inwards from its original position (blue dashed).
}
\label{fig:surface_density}
\end{figure}

It is possible to quantify the error that is introduced in analyzing the output of an impact simulation using the conventional analysis in the case that the disk-like region is majority vapor.
There is initially no phase separation and the code used to simulate the impact can accurately model the hydrostatic post-impact structure.
For cases similar to the Moon-forming impacts proposed by \cite{Cuk2012}, the conventional analysis artificially decreases the mass of the disk-like region by several lunar masses.
The surface density of the disk-like region is correspondingly reduced as shown in Figure~\ref{fig:surface_density}.
The black line in Figure~\ref{fig:surface_density} shows the true surface density of the structure and the surface density of the disk in the conventional analysis is given in red. The blue dashed line shows the original position of the disk material in the conventional analysis.
The mass depletion of the disk-like region is strongest close to the planet and far from the rotational axis.
Thus, the mass in the disk-like regions can be substantially reduced by the conventional analysis.
For the example in Figure~\ref{fig:surface_density}, the mass beyond Roche is reduced from 1.7 to 0.5 lunar masses.
Note that the surface density of the disk-like regions within a synestia (black line in Figure~\ref{fig:surface_density}) is not constant with radius as assumed in some disk evolution models. 
In most impact cases, some mass is injected into orbits far from the central mass. 
This mass will cool rapidly and can be modeled as pure condensate.

\cite{Nakajima2014} have previously noted the significance of pressure gradients in the structure of highly vaporized post-impact structures.
They analyzed the disk structures generated by specific examples of canonical, high-energy and high-AM, and intermediate giant impacts.
They found that, in the disk-like regions of high-energy, high-AM structures, the specific AM was nearly constant with radius, due to strong pressure support. 
Such structures are likely to be unstable by the Rayleigh criteria. 
As a result, \cite{Nakajima2014} redistributed the mass in the high-energy, high-AM cases assuming an arbitrary stable surface density profile that conserved the mass and AM of the disk.
However, they calculated the disk mass and AM to be conserved using the conventional analysis. The disk mass was significantly depleted by the use of the conventional analysis, and the calculated disks have less mass than found in the disk-like region of synestias formed under the same impact conditions.

Given the significant errors in the structure calculated using the conventional approach, it is important that better techniques are developed to analyze the results of giant impact simulations. For vapor dominated synestias, we suggest that removing the escaping mass from an SPH simulation and calculating the hydrostatic structure of the bound mass provides a reasonable estimate of the post-impact structure, as shown in Figure~\ref{fig:PI_pressure}. In addition, the treatment of the interface between the corotating and disk-like regions in disk evolution models requires significant technical developments \citep[as discussed in][]{Charnoz2015}. 

%XXXXmaybe add in cooling analogy -SJL what do you mean?

%xxxxxxxxxxxxxxxxxxxxxxxxxxxxxxxxxxxxxxxxxxxxxxxxxxxxxxxxxxxxxxxx
%xxxxxxxxxxxxxxxxxxxxxxxxxxxxxxxxxxxxxxxxxxxxxxxxxxxxxxxxxxxxxxxx
\subsection{Synestias and satellite formation}
\label{sec:discussion_satellites}

In the canonical disk model (e.g., Figure~\ref{fig:MAD}A-D), recent work suggests that satellite accretion occurs in a multi-stage process \citep{Salmon2012}.
The material outside the Roche limit condenses and accretes quickly (on a timescale of weeks) to form a proto-moon. The proto-moon temporarily confines the edge of the Roche-interior liquid-vapor disk via resonant interactions.
As the Roche-interior fluid disk cools and viscously spreads both inwards and outwards, moonlets whose orbits are raised beyond the Roche limit are accreted onto the proto-moon on a timescale of 100's of years \citep{Machida2004, Salmon2012, Charnoz2015}. The multiphase dynamics of a canonical circumterrestrial disk are challenging to model, and our understanding of the physical processes in the disk is incomplete. For example, most studies of the canonical disk evolution make an assumption of energy balance between viscous spreading of the Roche-interior disk and radiative cooling; however, \citet{Charnoz2015} have shown that radiative cooling dominates the system. Furthermore, this work demonstrates that canonical disk models must include the vapor pressure support from the corotating region. 

Satellite accretion from the various post-impact structures described here can be fundamentally different than accretion from a canonical circumterrestrial disk.
Post-impact synestias can have a much higher surface density in the disk-like region (e.g., Figure~\ref{fig:MAD}). 
Consequently, the cooling time of vapor beyond the Roche limit reaches timescales relevant for satellite formation. 
For example, \citet{Lock2016LPSC} estimate that substantial vapor pressure ($>$~bars) can persist at the Roche limit for 10's of years after the impact for the synestia example in Figure~\ref{fig:MAD}E-H. 
Synestias do not have a shear boundary between the corotating region and disk-like region, and therefore, the whole structure can mix more easily than in the sub-CoRoL structures formed in canonical giant impacts. 
Condensates that decouple from the vapor in the lowest surface density regions beyond the Roche limit have angular momenta such that they orbit within the vapor structure. 
In some cases, the proto-moon may form beyond the Roche limit and grow to near its final mass while remaining within the Earth-composition vapor of the synestia.
\citet{Lock2016LPSC} proposed that accretion within an impact-generated synestia can explain the isotopic and chemical properties of our Moon. A future paper will investigate the evolution of synestias and the formation of our Moon.

%xxxxxxxxxxxxxxxxxxxxxxxxxxxxxxxxxxxxxxxxxxxxxxxxxxxxxxxxxxxxxxxx
%xxxxxxxxxxxxxxxxxxxxxxxxxxxxxxxxxxxxxxxxxxxxxxxxxxxxxxxxxxxxxxxx
\subsection{Planet formation with hot rocky bodies}
\label{sec:discussion_formation}

\begin{figure}
\centering
\includegraphics[scale=0.83333333]{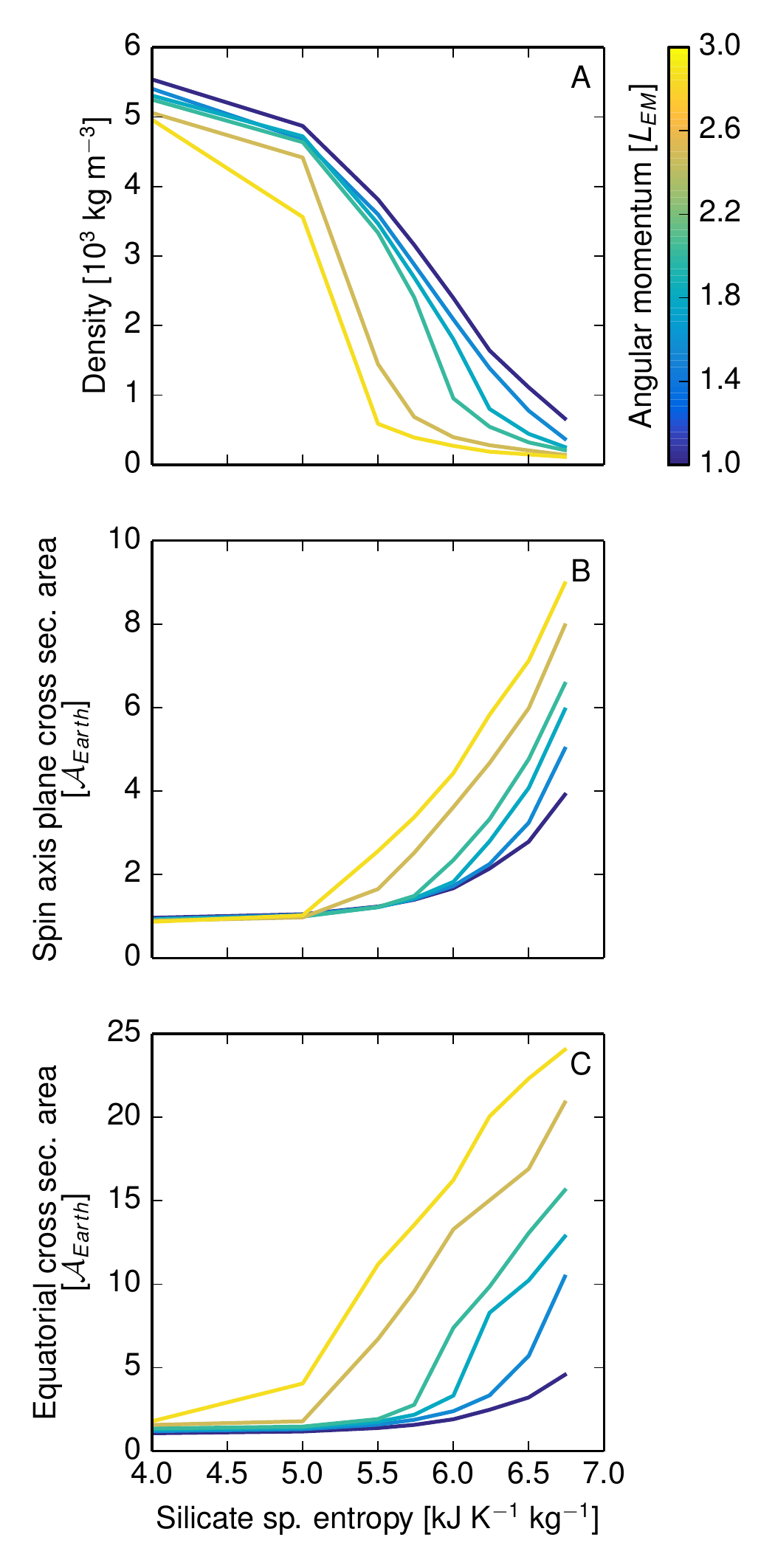}
\caption[]{The extended structures of hot, rotating rocky bodies lead to very low bulk densities, comparable to cool planets with substantial gaseous envelopes. Curves represent the bulk densities (A) and cross-sectional areas in a plane perpendicular to the rotation axis (B) and in the equatorial plane (C) for the Earth-mass bodies shown in Figure~\ref{fig:shape_plot} as a function of silicate specific entropy (thermal profile class I). Cross-sectional areas are given in units of the Earth's present equatorial cross-sectional area, $A_{\rm Earth}$. Colored lines show bodies of constant angular momentum.}
\label{fig:hot_density}
\end{figure}

We have shown that partially vaporized post-impact bodies are common during accretion (\S\ref{sec:impacts}); however, the implications of the highly variable thermal and physical structures of such bodies on the formation of terrestrial planets have not been considered. 
Here we comment on a few key areas of planet formation affected by the possible range of planetary structures: accretion efficiency, core formation, and chemical evolution.

At present, models of the accretion of terrestrial planets do not account for changes in thermal state. The physical size of planetesimals and planets are usually calculated by assuming a fixed bulk density over the entire calculation \citep[typically 3~g~cm$^{-3}$ in $N$-body codes, e.g.,][]{Quintana2016}. For rotating rocky bodies with the highest specific entropy cases considered here, the bulk density could be an order of magnitude lower. Figure~\ref{fig:hot_density} presents the bulk density and areal cross sections for the suite of Earth-mass bodies shown in Figure~\ref{fig:shape_plot}. Partial vaporization is accompanied by a significant decrease in bulk density for all rotational states. Post-impact states can be significantly larger than isolated bodies with even lower density and larger cross sections. All giant impacts results in some inflation of the radius of a body. The larger sizes of hot rocky planets (Figure~\ref{fig:shape_plot}) translate to larger collisional cross sections and an increased probability of impacts, and hence more efficient accretion (Figure~\ref{fig:hot_density} B,C).
In particular, the expanded cross section of a body after an impact will increase the fraction of impact debris reaccreted in the early time period before the debris can be perturbed from crossing orbits.
In proposed Moon-forming giant impacts, typically a few to several lunar masses of material is ejected \citep{Canup2004, Cuk2012, Canup2012}, which is accreted onto the Earth and nearby planets within $\sim$10~Myr under current assumptions \citep{Jackson2012,Bottke2015}.
It has been suggested that the warm debris disks produced by giant impacts would be observable around other stars \citep{Jackson2012}. But recently, \citet{Kenyon2016} noted that the observed occurrence rate of warm debris disks around young stars is much lower than what would be expected from the prevalence of extra-solar rocky planets around mature stars and current planet formation models.
\citet{Kenyon2016} inferred that terrestrial planet formation must be quick and neat, with efficient reaccretion of impact-produced debris. 
Realistic capture cross sections of planets could be a factor in explaining such observations and should be included in planet formation studies.

The final stages of core formation in rocky planets is expected to occur during, or in the aftermath of, giant impacts \citep[e.g.,][]{Rubie2007a,Rubie2015}. The pressure and temperature profiles with depth depend on the thermal and rotational state of the post-impact body.
The partitioning of elements between the mantle and core is controlled by the pressure and temperature at which core and mantle materials equilibrate.
Most studies of Earth's accretion assume that core material equilibrates at temperatures between the liquidus and solidus \citep[e.g.,][]{Rubie2007a,Rubie2015}. However, in the impact simulations summarized in Figures~\ref{fig:HSSL_impact} and \ref{fig:QS}, much of the post-impact body has temperatures far above the liquidus, with some sections potentially above the solvus for silicate-iron mixtures \citep{Wahl2015}.
In contrast, the lowermost mantle of post-impact states is much colder and potentially partially solid \citep{Solomatov2000,Stewart2015,Nakajima2015}.
The superheated post-impact structure is important for core formation because the time for the whole structure to cool below the liquidus is much longer than the settling time for iron.
Metal-silicate equilibration in the aftermath of giant impacts can occur over a wide range of pressure and temperature conditions, which affects partitioning of elements between the mantle and the core.
For example, \citet{Badro2016} recently argued that core formation at very high mantle temperatures would result in the partitioning of magnesium into the core which could provide a substantial power source for the terrestrial dynamo \citep{ORourke2016,ORourke2016a}.
The thermal and rotational states of bodies after giant impacts is therefore critical to understanding the conditions of core formation and interpreting the observed chemical tracers.

Upon cooling, the partially vaporized structures formed by giant impacts reach a thermal state associated with a conventional magma ocean planet, where the silicates are liquid and overlain by an atmosphere comprised of volatile compounds (e.g., H$_2$O, CO/CO$_2$, N$_2$, Ar, etc.).
However, the stage of evolution between the post-impact state and a magma ocean has been largely ignored in previous studies of planet formation.
During this period the structure is evolving rapidly, leading to changing pressure and temperature conditions in the body.
The rapid cooling of the body is likely to drive convection and potentially mix large fractions of the body, establishing the thermal and chemical structure of the magma ocean. 
Understanding this process is critical to understanding the possible survival of distinct geochemical signatures that predate the Moon-forming giant impact \citep{Rizo2016a,Mukhopadhyay2012,Peto2013,Tucker2012,Parai2012}.
Furthermore, the period after an impact is also the time of highest impactor flux due to the reaccretion of impact debris.
It has been suggested that bombardment from a population of small bodies \citep[e.g.,][]{Schlichting2015} or from single large bodies \citep{Genda2005,Stewart2014} could remove significant amounts of atmosphere, and hence volatile elements, from terrestrial planets that are cold enough to have a liquid or solid surface and atmosphere.
Impact erosion of planetary atmospheres has also been invoked as a possible process to fractionate Earth's budget of volatile elements \citep{Tucker2014}. 
%%% will always intersect at very low pressures for silicates (vs. H2) SJL What do you mean?
Temporarily, typical hot post-impact structures do not have a distinct boundary between a vapor atmosphere and a liquid or solid surface. The corotating region of the structure grades smoothly from a silicate vapor to a supercritical fluid. 
Removal of volatiles by impacts is less efficient when the silicate potion of a body is partially vaporized as the impact energy is not efficiently deposited in a single shock at the planets surface but instead the impactor is gradually slowed by the continuous density gradient in the structure. 
Furthermore, during the period with substantial silicate vapor, the silicate components are more abundant than the volatile components in the atmosphere, and material ejected by small impactors is likely to be dominated by silicate components rather than volatiles. Thus, volatile loss by impacts may be inefficient during the hot post-impact period.
More work is needed to understand the key phase in terrestrial planet evolution between giant impacts and magma oceans. 

%xxxxxxxxxxxxxxxxxxxxxxxxxxxxxxxxxxxxxxxxxxxxxxxxxxxxxxxxxxxxxxxx
%xxxxxxxxxxxxxxxxxxxxxxxxxxxxxxxxxxxxxxxxxxxxxxxxxxxxxxxxxxxxxxxx
\subsection{Exoplanets}
Around other stars, super-Earth size planets are common \citep{Morton2016}. For the {\it Kepler} mission, super-Earths were defined by radii between 1.25 and 2$R_{\rm Earth}$ \citep{Borucki2011}. The composition of super-Earths appear to span cool rocky bodies to partially gaseous bodies, with a possible composition transition around 1.5$R_{\rm Earth}$ \citep{Weiss2014,Rogers2015}. However, the rocky portions of such planets have been assumed to have a high bulk density, based on the mass-radius relationships for cold bodies \citep[e.g.,][]{Zharkov1978,Valencia2006, Swift2012, Hubbard2013, Zeng2013, Zeng2016,Unterborn2016}. The equilibrium surface temperatures of exoplanets are generally higher than the terrestrial bodies in our solar system, and a few known cases are even hot enough to consider a surface with partially vaporized silicates (e.g., Kepler-78b \citep{Sanchis-Ojeda2013} and 55 Cancri e \citep{Demory2016}). 

Planets with a bulk density less than cold rock or ice are assumed to have a substantial volatile gas mass fraction. Our calculation of the bulk density of high specific entropy rocky bodies (Figure~\ref{fig:hot_density}A) overlaps those typically associated with ice giant and gas giant planets (e.g., at densities below $\sim$$3000$~kg~m$^{-3}$). 

Furthermore, the rotational distortion of exoplanets can be significant, producing highly asymmetric bodies (Figure~\ref{fig:hot_density}B,C). 
Significantly, this effect is largest for hot rocky planets where substantial rotational flattening can be achieved even at modest spin periods of several hours. 
Most of the rocky exoplanets discovered so far are close to their host stars and expected to be tidally locked or in spin-orbit resonances \citep[e.g.,][]{Correia2010}. However, the rotational states of exoplanets further from their host stars are unknown and a range of rotation rates must be considered for the formation and evolution of exoplanets.
If, in calculating the planetary size from transit measurements, the areal cross section of significantly flattened exoplanets are interpreted as spherical, the viewing angle could lead to a significant error in the inferred volume of the planet. The structures of hot and/or rotating rocky bodies should be considered when inferring the possible compositions of exoplanets. Future work will use the HERCULES code to calculate mass-radius relationships for rocky bodies that include thermal and rotational effects and the structures of planets with hot, rocky interiors and varying molecular atmospheres.

Super-Earths are also expected to form via giant impacts in a manner similar to the accretion of our terrestrial planets \citep{Chambers2010}. 
As noted in \S\ref{sec:impacts}, \cite{Kokubo2010} found that 0.25 to 1.75~$M_{\rm Earth}$ planets had a similar average angular velocity at the end of accretion. 
Accounting for the planetary structure assumptions made by \cite{Kokubo2010}, the average AM of rocky planets scales as $\sim$$M^{5/3}$, where $M$ is the planetary mass. 
If this result holds for systems that form more massive planets, the mean AM of hot super-Earths is expected to exceed the CoRoL, which scales with mass by a similar power law (Figure~\ref{fig:HSSL_M-L}).
Therefore, we expect that synestias also form during the accretion of super-Earths. Because some exoplanets are much closer to their star, the cooling time of synestias may be much longer than for planets in our solar system. The large population of exoplanets is used to test general planet formation theories. If the physical properties of hot rocky planets are an important factor during general planet formation, the large population of super-Earths may be the most promising target for detailed investigations.

%xxxxxxxxxxxxxxxxxxxxxxxxxxxxxxxxxxxxxxxxxxxxxxxxxxxxxxxxxxxxxxxx
%xxxxxxxxxxxxxxxxxxxxxxxxxxxxxxxxxxxxxxxxxxxxxxxxxxxxxxxxxxxxxxxx
\subsection{Giant planets and stars}
\label{sec:discussion_other}
%XXX{\bf this section needs to be vetted by some experts}

Angular momentum and disk formation are central topics in the formation of stars \citep[e.g.,][]{Bodenheimer1995} and giant planets \citep[e.g.,][]{Peale2015}. In most cases, disk evolution is studied without explicitly calculating the structure of the central object, as has been the case with the lunar disk. However, a few studies have calculated structures for stars and giant planets that are similar to the synestias found in this work. Here, we discuss examples of such structures and compare them to terrestrial synestias. 

\citet{Ostriker1968} calculated theoretical structures for massive white dwarfs. They found solutions for rapidly rotating structures with imposed AM profiles, including cases that resulted in a central quasi-corotating region and differentially rotating outer region. They argued that the structures, which exceed the corotation limit calculated by \citet{James1964}, are stable. The structures are morphologically similar to the synestias calculated here, e.g., Figure~4 in \citet{Ostriker1968}. These differentially rotating structures are consistent with the inferred angular velocities of white dwarfs.

\citet{Bodenheimer1971} extended the previous work to massive main sequence stars. By systematically increasing the total AM, with different imposed AM profiles, he found that the equilibrium structures of the stars varied significantly. Again, solutions with differential rotation have morphological similarities to synestias (see his Figure~4). These structures are expected to be transient while main sequence stars shed AM.

Early studies of the formation of gas giant planets considered the contraction of a gas cloud. \citet{Bodenheimer1977} found that it was possible to contract into a configuration with a central planet connected to a circumplanetary disk (e.g., his Figure~3). These structures, with different imposed AM profiles, are morphologically similar to the synestias generated here for terrestrial bodies. These early models of gas giant planet formation in isolation have been superseded by studies that include transfer of material from the protoplanetary disk \citep[see review by][]{Peale2015}. In isolated contraction models, a so-called spin-out disk forms because the contracting body cannot absorb all of the original AM in the cloud. \citet{Ward2010} proposed that although the early stages of the formation of circumplanetary disks around gas giants may include a spin-out stage, the system must transition to an accretion disk fed by the protoplanetary disk. The details of the formation of gas giants and their satellites are not settled, and the connection between the central planetary structure and satellite-forming disk has not been fully incorporated into existing models.

In these examples of other rapidly rotating astrophysical objects, the structures share many similarities to the terrestrial synestias. They are continuous in density, monotonic in the angular velocity profile and typically have extended or flared outer regions.
However, unlike the terrestrial synestias, the angular velocity profiles for the outer regions of these other astrophysical bodies are not strictly Keplerian because of the imposed AM profile.
In our impact-generated synestias, the AM profiles were not imposed. The distribution of mass and AM between the corotating region and disk-like region was determined by the impact conditions. As for the case of terrestrial bodies, if gas giant planets and stars form synestias during their formation and evolution, a full understanding of the evolution of the system may require direct consideration of the physical structure of the central object.

%xxxxxxxxxxxxxxxxxxxxxxxxxxxxxxxxxxxxxxxxxxxxxxxxxxxxxxxxxxxxxxxxxxxxxxxxxxxxxxxxxxxxxxxxxxxxxxxxxxxxxxxxxxxxxxxxxxxxxxxxxxxxxxxxxxxxxxxx
%xxxxxxxxxxxxxxxxxxxxxxxxxxxxxxxxxxxxxxxxxxxxxxxxxxxxxxxxxxxxxxxxxxxxxxxxxxxxxxxxxxxxxxxxxxxxxxxxxxxxxxxxxxxxxxxxxxxxxxxxxxxxxxxxxxxxxxxx
%xxxxxxxxxxxxxxxxxxxxxxxxxxxxxxxxxxxxxxxxxxxxxxxxxxxxxxxxxxxxxxxxxxxxxxxxxxxxxxxxxxxxxxxxxxxxxxxxxxxxxxxxxxxxxxxxxxxxxxxxxxxxxxxxxxxxxxxx
\section{Conclusions}
\label{sec:conclusions}

The physical structure of planetary bodies provides the essential framework needed to investigate planet formation and evolution. Assumptions about structure influence the inquiry of physical and chemical processes. The models for planetary structure that have been used to date have typically been simple, such as the constant-density bodies in $N$-body simulations of accretion. 
However, planetary structure can vary significantly during formation and evolution. In particular, during accretion terrestrial bodies can be forced into hot, rapidly rotating states by giant impacts, and there is a need for a better understanding of such structures.

Calculating the structure of hot, rapidly rotating rocky bodies is particularly challenging because of the need to include substantial flattening and multiple phases in the equation of state.
To overcome these difficulties, we have developed a flexible new code, HERCULES, that can calculate the physical structure of planetary bodies with varying thermal and rotational states.
HERCULES solves for the axisymmetric shape and internal profile of a planet, given the mass, compositional layering, specific entropy profile, and AM. 

In this work, we investigated the structure of Earth-like bodies over a wide range of thermal and rotational states. 
We found a that the size and shape of Earth-like bodies varies considerably over the expected range of internal energy and AM during accretion. 
We showed that there is a corotation limit for the structure of terrestrial bodies that depends on mass, compositional layering, thermal state and AM.
We have named super-CoRoL structures synestias. Synestias typically consist of an inner corotating region connected to an outer disk-like region.
By analyzing the results of $N$-body simulations of planet formation, we found that high-entropy, highly vaporized post-impact states are common during terrestrial planet accretion. 
Given the estimated range of planetary AM during the giant impact stage, we find that many post-impact structures are likely to be synestias.

The occurrence of partially vaporized post-impact states during accretion significantly affects various aspects of planet formation.
The inclusion of realistic post-impact states into models of growing bodies is needed to understand the history of terrestrial planets.
In particular, the dynamics and evolution of synestias are fundamentally different than traditional models of planets surrounded by a dynamically distinct disk.
Previous analyses of the structure of the Earth and a circumterrestrial disk made assumptions that are violated when a synestia is created. 

The large population of exoplanets, with larger rocky bodies and higher stellar flux compared to our solar system, offers an opportunity to apply our new understanding of hot, rotating rocky planetary structures. Future work will include the calculation of mass-radius curves for a range of rocky bodies with varying thermal and rotational states. The HERCULES code is an efficient tool for calculating planetary structures and may be integrated into studies of many aspects of planet formation and evolution that have previously been neglected.

%xxxxxxxxxxxxxxxxxxxxxxxxxxxxxxxxxxxxxxxxxxxxxxxxxxxxxxxxxxxxxxxxxxxxxxxxxxxxxxxxxxxxxxxxxxxxxxxxxxxxxxxxxxxxxxxxxxxxxxxxxxxxxxxxxxxxxxxx
%xxxxxxxxxxxxxxxxxxxxxxxxxxxxxxxxxxxxxxxxxxxxxxxxxxxxxxxxxxxxxxxxxxxxxxxxxxxxxxxxxxxxxxxxxxxxxxxxxxxxxxxxxxxxxxxxxxxxxxxxxxxxxxxxxxxxxxxx
%xxxxxxxxxxxxxxxxxxxxxxxxxxxxxxxxxxxxxxxxxxxxxxxxxxxxxxxxxxxxxxxxxxxxxxxxxxxxxxxxxxxxxxxxxxxxxxxxxxxxxxxxxxxxxxxxxxxxxxxxxxxxxxxxxxxxxxxx
\section*{Acknowledgments}
This work was supported by NESSF grant NNX13AO67H, NASA Origins grant NNX11AK93G, NASA solar system workings grant NNX15AH54G and DOE-NNSA grants DE-NA0001804 and DE-NA0002937. 
We thank Sean Raymond and Elisa Quintana for providing N-body simulation data. 
We also thank Don Korycansky, Phil Carter, Dave Stevenson and Miki Nakajima for useful discussions, and Francis Nimmo and two anonymous reviewers for comments that greatly improved this work.
The modified version of GADGET-2 and the EOS tables are contained in the supplement of \cite{Cuk2012}.
The HERCULES code is included in the supporting information for this paper. A user guide will be available upon the completion of SJL's PhD thesis.

%xxxxxxxxxxxxxxxxxxxxxxxxxxxxxxxxxxxxxxxxxxxxxxxxxxxxxxxxxxxxxxxxxxxxxxxxxxxxxxxxxxxxxxxxxxxxxxxxxxxxxxxxxxxxxxxxxxxxxxxxxxxxxxxxxxxxxxxx
%xxxxxxxxxxxxxxxxxxxxxxxxxxxxxxxxxxxxxxxxxxxxxxxxxxxxxxxxxxxxxxxxxxxxxxxxxxxxxxxxxxxxxxxxxxxxxxxxxxxxxxxxxxxxxxxxxxxxxxxxxxxxxxxxxxxxxxxx
%xxxxxxxxxxxxxxxxxxxxxxxxxxxxxxxxxxxxxxxxxxxxxxxxxxxxxxxxxxxxxxxxxxxxxxxxxxxxxxxxxxxxxxxxxxxxxxxxxxxxxxxxxxxxxxxxxxxxxxxxxxxxxxxxxxxxxxxx
%\section*{References}
\bibliographystyle{elsarticle-harv} 
\bibliography{References}

%This top part might not be allowed but they shouldn't care too much
%make all the sections labelled with an S for supp
\renewcommand{\thepage}{S\arabic{page}}  
\renewcommand{\thesection}{S\arabic{section}}   
\renewcommand{\thetable}{S\arabic{table}}   
\renewcommand{\thefigure}{S\arabic{figure}}
\renewcommand{\theequation}{S\arabic{equation}}

%reset page count and section count
\setcounter{page}{1}
\setcounter{section}{0}
\setcounter{figure}{0}
\setcounter{table}{0}
\setcounter{equation}{0}
\setcounter{tnote}{0}
\setcounter{fnote}{0}
\setcounter{footnote}{0}
\setcounter{cnote}{0}
\setcounter{author}{0}
\setcounter{affn}{0}

\renewenvironment{abstract}{\global\setbox\absbox=\vbox\bgroup
  \hsize=\textwidth\def\baselinestretch{1}%
  \noindent\unskip\textbf{Contents}
 \par\medskip\noindent\unskip\ignorespaces}
 {\egroup}

\begin{frontmatter}

\title{Supplementary materials for "The structure of terrestrial bodies: Impact heating, corotation limits and synestias"}

\begin{abstract}
%%%Remove or add items as needed%%%
\begin{enumerate}
\item Text S1 to S4
\item Figures S1 to S8
\item Tables S1 to S3
\item Captions for large Tables S4 and S5
\item HERCULES code
\end{enumerate}
\end{abstract}
\end{frontmatter}

%xxxxxxxxxxxxxxxxxxxxxxxxxxxxxxxxxxxxxxxxxxxxxxxxxxxxxxxxxxxxxxxxxxxxxxxxxxxxxxxxxxxxxxxxxxxxxxxxxxxxxxxxxxxxxxxxxxxxxxxxxxxxxxxxxxxxxxxx
%xxxxxxxxxxxxxxxxxxxxxxxxxxxxxxxxxxxxxxxxxxxxxxxxxxxxxxxxxxxxxxxxxxxxxxxxxxxxxxxxxxxxxxxxxxxxxxxxxxxxxxxxxxxxxxxxxxxxxxxxxxxxxxxxxxxxxxxx
%xxxxxxxxxxxxxxxxxxxxxxxxxxxxxxxxxxxxxxxxxxxxxxxxxxxxxxxxxxxxxxxxxxxxxxxxxxxxxxxxxxxxxxxxxxxxxxxxxxxxxxxxxxxxxxxxxxxxxxxxxxxxxxxxxxxxxxxx
\section{HERCULES code}
\label{sup:sec:HERCULES}

We have developed a new code for calculating the equilibrium structure of rapidly rotating, fluid bodies.
The HERCULES ({\it Highly Eccentric Rotating Concentric U (potential) Layers Equilibrium Structure}) code improves on the concentric Maclaurin spheroid \citep{Hubbard2013} by allowing the calculation of the structure of rapidly rotating bodies. Additionally, HERCULES can also solve for the structure of a body that is consistent with material equations of state (EOS) and conserves the mass and AM of a body.

In HERCULES, a body is modeled as a series of nested, constant density spheroids (Figure~\ref{fig:HERCULES}), with the surface of each spheroid an equipotential surface. The structure of the body is found by iteratively solving for the shape, density, and equatorial radius of each of the spheroids, and the rotational velocity of the body. Fundamental to finding the equilibrium structure is the calculation of the gravitational potential due to each of the individual spheroids, in order to determine the equipotential surfaces. In \S\ref{sup:sec:single_spheroid}, we summarize the calculation of the potential for a single constant density spheroid. We then describe the HERCULES code in \S\ref{sup:sec:concentric_layer}.

%xxxxxxxxxxxxxxxxxxxxxxxxxxxxxxxxxxxxxxxxxxxxxxxxxxxxxxxxxxxxxxxx
%xxxxxxxxxxxxxxxxxxxxxxxxxxxxxxxxxxxxxxxxxxxxxxxxxxxxxxxxxxxxxxxx
\subsection{Gravitational potential due to a single spheroid}
\label{sup:sec:single_spheroid}

In order to find the equilibrium structure of a body in HERCULES we need an expression for the gravitational potential due to a constant density spheroid, both inside and outside the spheroid. Conventionally, the potential due to a spheroid has been calculated by using a small parameter expansion \citep[e.g.][]{Zharkov1978}, but for rapidly rotating bodies such methods either breakdown or require the inclusion of a large number of terms. Alternatively, \cite{Hubbard2012} formulated an expression for the gravitational potential due to a single, constant-density spheroid as an expansion in Legendre polynomials. However, as identified by \cite{Kong2013}, the expression formulated by \cite{Hubbard2012} does not converge for bodies that are more oblate than a certain limit if the radius of the point for which the potential is being calculated is between the equatorial and polar radius of the body, i.e., if $b$~$<$~$r$~$<$~$a$ where $a$ and $b$ are the equatorial and polar radii of the spheroid respectively.
\cite{Kong2013} provided a solution to the potential due to a Maclaurin spheroid as an expansion in Legendre polynomials that converges for significantly oblate bodies in the case that $b$~$<$~$r$~$<$~$a$.
\cite{Hubbard2013} also found an expression for the potential at a point where $r\leq b$. 
Here, we combine the results of these previous studies and consider the potential due to a single spheroid in all possible regimes, using consistent notation and a derivation in the style of \cite{Kong2013}. 

The gravitational potential at a point due to a body is given by
\begin{linenomath*}
\begin{align}
V(r, \mu, \phi) = G \int_{\mathcal{V}} \frac{\rho(r', \mu', \phi')}{ | \vec{r} - \vec{r'} |} d\mathcal{V}' \;\;.
\end{align}
\end{linenomath*}
$\vec{r}$~$=$~$(r, \mu, \phi)$ is the position vector at the point at which the potential is being calculated, the evaluation point, with $r$ the radius, $\mu$~$=$~$\cos \theta$ where $\theta$ is the angle from vertical, and $\phi$ is the azimuthal angle. $\vec{r}'$~$=$~$(r', \mu', \phi')$ is the position vector of the mass described by the density distribution $\rho(r', \mu', \phi')$. $G$ is the gravitational constant and $\mathcal{V}$ is the volume in which there is mass. The denominator can be expanded in terms of spherical harmonics, with the expansion depending on the relative magnitude of $r$ and $r'$ (see Equations 9 and 10 in \cite{Kong2013}).
Here we will only consider axisymmetric bodies for which 
\begin{linenomath*}
\begin{equation}
 \frac{1}{ | \vec{r} - \vec{r'} |} = \begin{dcases*}
        \sum_{l=0}^{\infty}  \frac{ \left(r' \right )^l}{r^{l+1}} P_l(\mu) P_l(\mu')   & when  $\: r > r' $\\
	\sum_{l=0}^{\infty}  \frac{ r^l}{\left ( r' \right )^{l+1}} P_l(\mu) P_l(\mu')   &  when $\: r < r' $	
        \end{dcases*} \;\;,
\end{equation}
\end{linenomath*}
where $P_l(\mu)$ is the Legendre polynomial of degree $l$.
The expression for the potential therefore depends on the position of the evaluation point, relative to the mass distribution. 

\begin{figure*}
   \centering
   \includegraphics[scale=0.8333333]{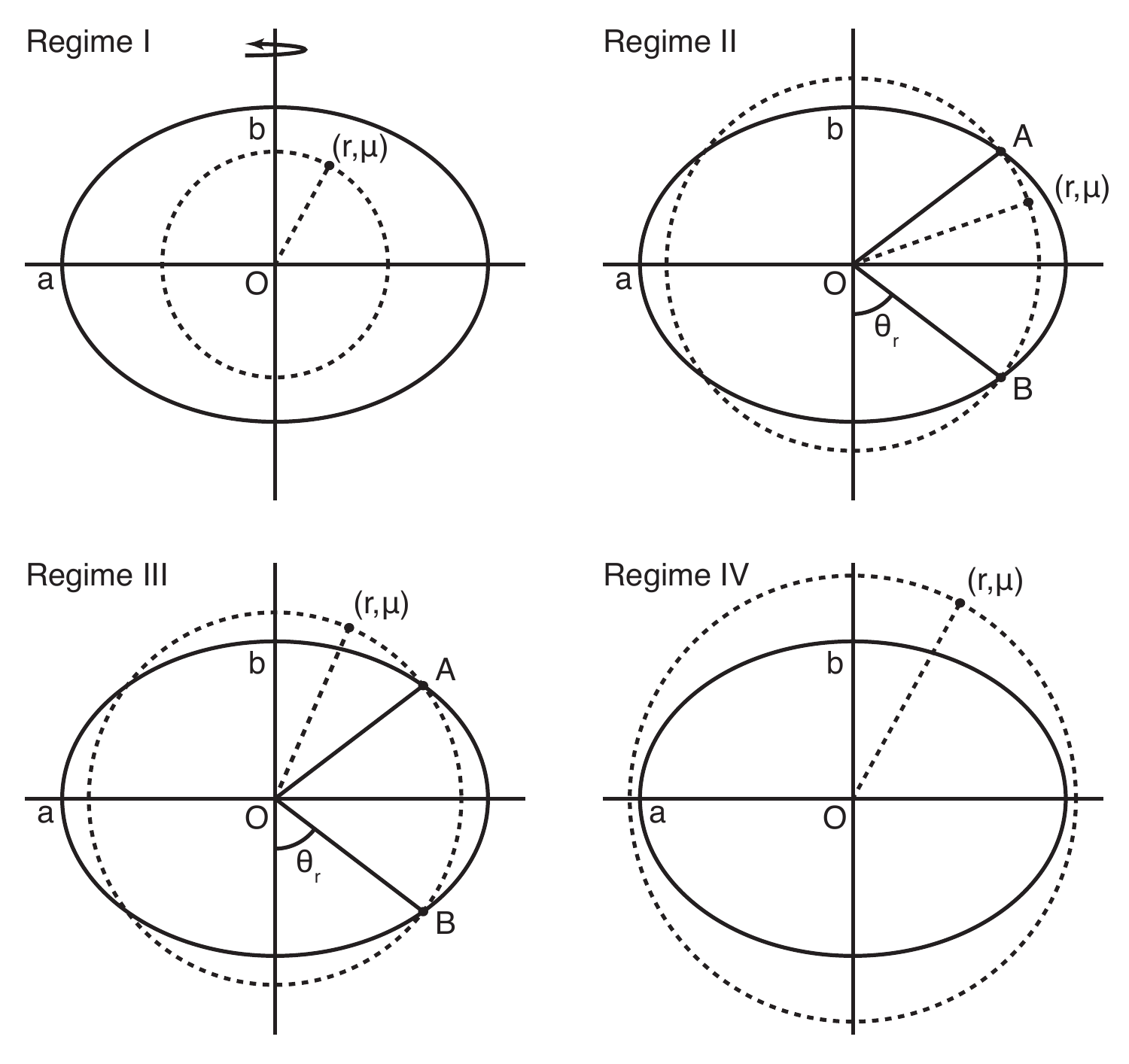} 
   \caption{Schematic of the four regimes for calculating the potential of a constant density spheroid, see text for details. Black solid line shows a cross section through the spheroid, with the rotation axis vertical. The equatorial radius, $a$, and polar radius, $b$, are labeled. The point at which the potential is being evaluated, $(r,\mu)$, is marked by the point at the end of the dashed line. The dashed circle indicates the locus of all other points with a radius $r$ from the origin, O. The points on the spheroid where the radius of the surface is equal to $r$, i.e. $\tilde{r}(\mu)$ where $\mu$~$=$~$\pm \mu_r$, are labelled A and B.}
   \label{sup:fig:Regimes}
\end{figure*}

We consider four regimes for calculating the potential: $r$~$\leq$~$b$; $b$~$<$~$r$~$<$~$a$ with the evaluation point within the body; $b$~$<$~$r$~$<$~$a$ with the evaluation point outside the body; and $r$~$\geq$~$a$ (see Figure~\ref{sup:fig:Regimes}).
Note that these regimes are not the same as the domains used by \cite{Kong2013}. We consider each of the regimes for calculating the potential for a body of uniform density, $\rho$, in turn.
This body is assumed to be rotationally symmetric and symmetric across the equatorial plane.
Additionally, the expressions given here assume that the radius of the surface of the body is monotonically decreasing from the equator to the pole, but there is no requirement for the body to be a Maclaurin spheroid as in \cite{Kong2013}.

%%%%%%%%%%%%%%%%%%%%%%%%%%%%%%%%%
\subsubsection{Regime I: $r$~$\leq$~$b$}
\label{sup:sec:R1}

The first regime is that for calculating the potential at a point within the spheroid, where the volume interior to the radius $r$ is all within the spheroid. 
This case was considered by \cite{Hubbard2013} but we rederive it here.
In this regime the potential is a sum of two terms; the first due to the mass within a sphere with radius $r$, and the second an integral over the mass in the ellipsoidal shell exterior to $r$.
Given the symmetry of the body, the potential function can be written as
\begin{linenomath*}
\begin{align}
\begin{aligned}
{} & V^I(r,\mu)= \frac{4\pi}{3} \rho G r^2 \\
& + 2\pi \rho G \int_{-1}^{1} \int_{r}^{\tilde{r}(\mu')} \left [ \sum_{l=0}^{\infty} \frac{r^{l}}{(r')^{l+1}} P_{l}(\mu)P_{l}(\mu') (r')^2 dr' \right ] d\mu' \;\;,
\end{aligned}
\end{align}
\end{linenomath*}
where $\rho$ is the density of the spheroid and $\tilde{r}(\mu)$ is the radius of the surface of the body at $\mu$.
Using the symmetry properties of the integrals, the expression simplifies to
\begin{linenomath*}
\begin{align}
\begin{aligned}
{} & V^I(r,\mu)=  4\pi \rho G \left \{ \frac{r^2}{3} \right. \\
& \left. + \int_{0}^{1} \int_{r}^{\tilde{r}(\mu')} \left [ \sum_{k=0}^{\infty} \frac{r^{2k}}{(r')^{2k+1}} P_{2k}(\mu)P_{2k}(\mu') (r')^2 dr' \right ] d\mu' \right \} \;\;.
\end{aligned}
\end{align}
\end{linenomath*}
Now we integrate over the radial direction, considering the $k$~$=$~0,1 terms separately
\begin{linenomath*}
\begin{align}
\begin{aligned}
{} & V^I(r,\mu)=  4\pi \rho G \left \{ \frac{r^2}{3} \right . \\
& \left . + \frac{1}{2} \int_{0}^{1} \left [ (\tilde{r}(\mu'))^2 - r^2 \right ] d\mu' \right . \\
& \left . + r^2 P_{2}(\mu)  \int_{0}^{1} \left [ \ln \left ( \frac{\tilde{r}(\mu')}{r}\right ) P_{2}(\mu') \right ] d\mu' \right . \\
& \left . + \sum_{k=2}^{\infty} \frac{r^{2k} P_{2k}(\mu)}{2-2k}  \int_{0}^{1} \left [ (\tilde{r}(\mu'))^{\, 2-2k} - r^{\, 2-2k} ) P_{2k}(\mu')  \right ] d\mu'  \right \} \;\;.
\end{aligned}
\end{align}
\end{linenomath*}
This is similar in form to the expression found in \cite{Kong2013} for Regimes II and III. 
Therefore we use notation similar to that of \cite{Kong2013} and rewrite
\begin{linenomath*}
\begin{align}
\begin{aligned}
V^I(r,\mu) {} &= \frac{4\pi}{3} \rho G r^2 + \frac{G M}{r} \left [ \left ( \frac{r}{a} \right ) N_0(\xi, 1) \right. \\
& \left. -  \left ( \frac{r}{a} \right )^3 N_2(\xi,1) P_2(\mu) \right . \\
& \left . - \sum_{k=2}^{\infty}  \left ( \frac{r}{a} \right )^{2k+1} N_{2k}(\xi,1) P_{2k}(\mu) \right ] \;\;,
\end{aligned}
\end{align}
\end{linenomath*}
where
\begin{linenomath*}
\begin{align}
\begin{aligned}
\label{eqn:1lay_M} 
M= \frac{4 \pi \rho}{3}  \int_{0}^{1} \left [ \tilde{r}(\mu') \right ]^3 d\mu'  = \frac{4 \pi \rho}{3} a^3  \int_{0}^{1} \left [ \tilde{\xi}(\mu') \right ]^3 d\mu'
\end{aligned}
\end{align}
\end{linenomath*}
is the mass of the spheroid,
\begin{linenomath*}
\begin{align}
& \begin{aligned}
\label{eqn:1lay_N0}
N_0(\xi,\mu_r)  {} &=  \left (  \frac{3}{2} \right ) \frac{ \int_{0}^{\mu_r} \left [ (\tilde{\xi}(\mu'))^2 - \xi^2 \right ] d\mu' }{\int_{0}^{1} \left [ \tilde{\xi}(\mu') \right ]^3 d\mu' } \;\;, 
\end{aligned} \\[4ex]
& \begin{aligned}
\label{eqn:1lay_N1}
N_2(\xi,\mu_r) {} &= - \frac{ 3 \int_{0}^{\mu_r} \left [ \ln \left ( \tilde{\xi}(\mu') / \xi \right ) P_2(\mu') \right ] d\mu' }{\int_{0}^{1} \left [ \tilde{\xi}(\mu') \right ]^3 d\mu' } \;\;,
\end{aligned}
\end{align}
\end{linenomath*}
and
\begin{linenomath*}
\begin{align}
& \begin{aligned}
\label{eqn:1lay_N2k}
N_{2k}(\xi,\mu_r) {} & =  -\left ( \frac{3}{2k-2} \right ) \\
 &  \times \frac{ \int_{0}^{\mu_r} \left [ \left ( \xi^{\, 2-2k} - (\tilde{\xi}(\mu'))^{\, 2-2k} \right ) P_{2k}(\mu') \right ] d\mu' }{\int_{0}^{1} \left [ \tilde{\xi}(\mu') \right ]^3 d\mu' }, \\
&  \; \;\;\;   \textrm{for} \; k \ge 2 \;\;.
\end{aligned}
\end{align}
\end{linenomath*}
$\tilde{\xi}(\mu)$~$=$~$\tilde{r}(\mu)/a$ is the normalized radius on the surface of the body, $\xi$~$=$~$r/a$ is the normalized radius at which we are calculating the potential, and $a$ is the equatorial radius of the spheroid. We define $N_2$ with the opposite sign from that used in \cite{Kong2013} to be consistent with the definition of $N_{2k}$. This definition makes the expression for regimes II and III simpler.

The notation used here is similar to \cite{Kong2013}, but not exactly the same. \cite{Kong2013} use $\xi_0$~$=$~$r/a$ (which we denote $\xi$) and use $\xi(\mu)$ to denote the normalized radius at a point on the surface for which we have used $\tilde{\xi}(\mu')$.
We feel that the notation used in \cite{Kong2013} is confusing when considering multiple spheroids and so have adopted this new notation for clarity.

%%%%%%%%%%%%%%%%%%%%%%%%%%%%%%%%%
\subsubsection{Regime II: interior point with $b$~$<$~$r$~$<a$}
\label{sup:sec:R2}

This case is the same as that considered by \cite{Kong2013}, but with the point at which the potential is being calculated within the body.
However, the location of the evaluation point relative to the surface does not change the expression for the potential in this regime and
\begin{linenomath*}
\begin{align}
\begin{aligned}
{}  & V^{II}(r,\mu) = \\
& 2\pi \rho G \int_{-\mu_r}^{+\mu_r} \int_0^r \left [ \sum_{l=0}^{\infty} \frac{(r')^l}{r^{l+1}} P_l(\mu)P_l(\mu') (r')^2 dr' \right ] d\mu' \\
& + 2\pi \rho G \int_{-\mu_r}^{+\mu_r} \int_{r}^{\tilde{r}(\mu')} \left [ \sum_{l=0}^{\infty} \frac{r^l}{(r')^{l+1}} P_l(\mu)P_l(\mu') (r')^2 dr' \right ] d\mu' \\
& +2\pi \rho G \int_{\mu_r}^{1} \int_0^{\tilde{r}(\mu')} \left [ \sum_{l=0}^{\infty} \frac{(r')^l}{r^{l+1}} P_l(\mu)P_l(\mu') (r')^2 dr' \right ] d\mu' \\
& +2\pi \rho G \int_{-1}^{-\mu_r} \int_0^{\tilde{r}(\mu')} \left [ \sum_{l=0}^{\infty} \frac{(r')^l}{r^{l+1}} P_l(\mu)P_l(\mu') (r')^2 dr' \right ] d\mu' \;\;. 
\label{eqn:Kong15}
\end{aligned}
\end{align}
\end{linenomath*}
$\mu_r$~$=$~$cos\, \theta_r$ (Figure~\ref{sup:fig:Regimes}) is the value of $\mu$ for which the observation radius and the surface intersect, i.e. $\tilde{r}(\mu_r)$~$=$~$r$, in the upper hemisphere.
This can be integrated to give
\begin{linenomath*}
\begin{align}
\begin{aligned}
V^{II}(r,\mu) {} &= \frac{G M}{r} \left \{ \left [ 1 - K_0(\xi, \mu_r) + \left ( \frac{r}{a} \right ) N_0 (\xi, \mu_r) \right ] \right .\\
& \left . - \sum_{k=1}^{\infty} \left [   \left (\frac{r}{a} \right )^{ \, 2 k + 1} N_{2k}(\xi, \mu_r) \right . \right. \\
& \left.\left.  + \left (\frac{a}{r} \right )^{2 k} (J_{2k}- K_{2k}(\xi, \mu_r) )\right ] P_{2k}(\mu) \right \} \;\;,
\label{eqn:RegII}
\end{aligned}
\end{align}
\end{linenomath*}
where
\begin{linenomath*}
\begin{align}
& \begin{aligned}
\label{eqn:1lay_K0} 
K_0(\xi,\mu_r)  {} &=   \frac{ \int_{0}^{\mu_r} \left [ (\tilde{\xi}(\mu'))^3 - \xi^3 \right ] d\mu' }{\int_{0}^{1} \left [ \tilde{\xi}(\mu') \right ]^3 d\mu' } \;\;,
\end{aligned} \\[4ex]
& \begin{aligned}
\label{eqn:1lay_K2k} 
K_{2k}(\xi,\mu_r) {} &= - \left ( \frac{3}{2k+3} \right ) \\
&  \times \frac{ \int_{0}^{\mu_r} \left [ \left ( (\tilde{\xi}(\mu'))^{\, 2k+3} - \xi^{\, 2k+3} \right ) P_{2k}(\mu') \right ] d\mu' }{\int_{0}^{1} \left [ \tilde{\xi}(\mu') \right ]^3 d\mu' }, \\
&  \; \;\;\;  \textrm{for} \; \ge 1 \;\; ,
\end{aligned} 
\end{align}
\end{linenomath*}
and
\begin{linenomath*}
\begin{align}
\label{eqn:1lay_J2k} 
J_{2k}  {} &=  - \left ( \frac{3}{2k+3} \right ) \frac{ \int_{0}^{1} \left [ (\tilde{\xi}(\mu'))^{\, 2k+3}  P_{2k}(\mu') \right ] d\mu' }{\int_{0}^{1} \left [ \tilde{\xi}(\mu') \right ]^3 d\mu' } \;\;.
\end{align}
\end{linenomath*}
$J_{2k}$ are the gravitational moments and are not dependent on $r$.
This is the same expression found by \citet{Kong2013}.

In dealing with single Maclaurin spheroids, \citet{Kong2013} could assume that 
\begin{linenomath*}
\begin{align}
\mu_r=\frac{\sqrt{a^2-r^2}}{r \sqrt{\left (\frac{a}{b} \right )^2 -1}} \;\;.
\end{align}
\end{linenomath*}
For a general surface this is not the case and we find $\mu_r$ for each potential surface numerically.

%%%%%%%%%%%%%%%%%%%%%%%%%%%%%%%%%
\subsubsection{Regime III: exterior point with $b$~$<$~$r$~$<$~$a$}
\label{sup:sec:R3}

This is the exact case that was considered by \cite{Kong2013}.
The expression in Equation \ref{eqn:RegII} for the potential applies and $V^{III}(r,\mu)$~$=$~$V^{II}(r,\mu)$.

%%%%%%%%%%%%%%%%%%%%%%%%%%%%%%%%%
\subsubsection{Regime IV: $r$~$\geq$~$a$}
\label{sup:sec:R4}

This regime was considered previously by \cite{Hubbard2012} and \cite{Kong2013}.
Since all the mass of the spheroid is interior to the evaluation point we need only consider a single term
\begin{linenomath*}
\begin{align}
\label{sup:eqn:V4_1}
 V^{IV} {} & (r,\mu)= \\
 & 2\pi \rho G \int_{-1}^{1} \int_0^{\tilde{r}} \left [ \sum_{l=0}^{\infty} \frac{(r')^l}{r^{l+1}} P_l(\mu)P_l(\mu') (r')^2 dr' \right ] d\mu' \;\;.
\end{align}
\end{linenomath*}
Note that this expression in Equation 13 of \cite{Kong2013} is missing the $2\pi$ prefactor.
Integrating Equation \ref{sup:eqn:V4_1} radially we get the expression
\begin{linenomath*}
\begin{align}
V^{IV}(r,\mu) {} &= \frac{G M}{r} \left [ 1-\sum_{k=1}^{\infty} \left ( \frac{a}{r} \right )^{2k} J_{2k} P_{2k}(\mu) \right ] \;\;,
\end{align}
\end{linenomath*}
which is the general expression for the potential outside of an axisymmetric body.

%xxxxxxxxxxxxxxxxxxxxxxxxxxxxxxxxxxxxxxxxxxxxxxxxxxxxxxxxxxxxxxxx
%xxxxxxxxxxxxxxxxxxxxxxxxxxxxxxxxxxxxxxxxxxxxxxxxxxxxxxxxxxxxxxxx
\subsection{Concentric layer model}
\label{sup:sec:concentric_layer}

%%%%%%%%%%%%%%%%%%%%%%%%%%%%%%%%%
\subsubsection{Model overview}
\label{sup:sec:model_overview}

In the HERCULES code, the body is modeled as a series of $N_{\rm lay}$ nested spheroids with uniform density, $\delta \rho_i$. Although we refer to the volumes as spheroids, in our formulation, the equipotential surfaces can be any rotationally symmetric surface whose radius decreases monotonically from equator to pole. Spheroids are numbered so that the 0th spheroid defines the surface of the body with numbers increasing inwards (Figure~\ref{fig:HERCULES}). The surface of each spheroid is an equipotential surface, with the same numbering as the spheroids. The radius of equipotential surface $i$ at $\mu$ is denoted $\tilde{r}_i (\mu)$ Normalized radii for each surface, $\tilde{\xi}_i(\mu)$, are normalized by the equatorial radius of that surface, i.e. $\tilde{\xi}_i(\mu)$~$=$~$\tilde{r}_i(\mu)/a_i$. Each equipotential surface is defined by $N_{\mu}$ linearly spaced $\mu$ points. The equatorial and polar radii of an equipotential surface, or equivalently spheroid, $i$ are $a_i$ and $b_i$ respectively.

The regions of the body between the equipotential surfaces are referred to as layers. The layers are numbered by the equipotential surface at the top of the layer. The mass distribution in the planet is given by the superposition of the mass of each of the constant density spheroids. All mass located between the two equipotential surfaces defining a layer is considered to be part of that layer.
The total density, $\rho_i$, of a given layer is the sum of the densities of all the spheroids that are larger than the spheroid that defines the inner edge of that layer, i.e.,
\begin{linenomath*}
\begin{align}
\rho_j=\sum_{i\leq j} \delta\rho_i \;\;,
\end{align}
\end{linenomath*}
with $\rho_0$~$=$~$\delta \rho_0$. Layers are divided into $N_{\rm mat}$ material layers, e.g. silicate and iron, with material layers numbered from the outside in, beginning from zero. The material a layer is made of dictates the equation of state (EOS) used in relating pressure to the density of the layer. The mass of material layers are also conserved separately (see \S\ref{sup:sec:Mconc}). The body is rotating at an angular velocity, $\omega_{\rm rot}$.

The equilibrium structure of a body is found using a multi-staged iteration. Firstly the shape of the equipotential surfaces is found, based on the mass distribution from the previous iteration. The shape of the constant density spheroids is then adjusted to match the equipotential surfaces. A separate conservation iterative routine alters the radii of each layer to conserve the mass of each of the material layers and the angular velocity of the body is recalculated to conserve the total AM (see \S\ref{sup:sec:Mconc} and \S\ref{sup:sec:Lconc}).
Within the conservation iteration, the density of the spheroids is altered to be consistent with the hydrostatic pressure profile and the given material EOS (\S\ref{sup:sec:press_calc}). The shape and conservation iterations are then repeated until the convergence criteria are met (see \S\ref{sup:sec:convergence}).

%%%%%%%%%%%%%%%%%%%%%%%%%%%%%%%%%
\subsubsection{Shape iterative equation}
\label{sup:sec:itterative_eqn}

For the shape iteration we solve for the equipotential surfaces within the structure, including both the gravitational and centrifugal potential. 
The gravitational potential at any point is given by the sum of the potential from each of the spheroids, accounting for the different regimes discussed in \S\ref{sup:sec:single_spheroid},
\begin{linenomath*}
\begin{align}
\begin{aligned}
V (r, \mu) {}& = \sum_{i=0}^{i_{I}} V^{I}_i (r, \mu) + \sum_{i=i_{I}+1}^{i_{IV}-1} V^{II}_i (r, \mu) \\
& +  \sum_{i=i_{IV}}^{N_{\rm lay}-1} V^{IV}_i (r, \mu) \;\;.
\end{aligned}
\end{align}
\end{linenomath*}
$i_{I}$ is the lowermost spheroid for which $r$~$<$~$b_i $ and $i_{IV}$ is the uppermost spheroid for which $r$~$>$~$a_i $. Note that as regime II and III have the same expression for potential we have combined both into regime II here. The centrifugal potential, $Q$, for a corotating body is given by
\begin{linenomath*}
\begin{align}
\begin{aligned}
 Q (r,\mu) =\frac{1}{3} r^2 \omega_{\rm rot}^2 \left [ 1 - P_2(\mu) \right ] \;\;,
 \end{aligned}
\end{align}
\end{linenomath*}
where $\omega_{\rm rot}$ is the corotating angular velocity. The total potential $U$ is given by summing the gravitational and centrifugal potentials i.e.,  $U$~$=$~$V+Q$.

In each shape iteration, to find the shape of the $j$th equipotential surface corresponding to the new mass distribution, we find the locus of points at which the total potential in the structure equals the potential at a radius $a_{\rm j}$ in the midplane. In other words, we solve 
\begin{linenomath*}
\begin{align}
U (\tilde{r}_j(\mu), \mu) -U(a_j,0) = 0 \;\;.
\end{align}
\end{linenomath*}
for $\tilde{r}_{\rm j} (\mu)$ for each $\mu$ point that describes the surface.
Combining our previous results, the equation that must be solved iteratively to find $\tilde{r}_j (\mu)$ is therefore
\begin{linenomath*}
\begin{align}
\begin{aligned}
\label{eqn:all_start}
 V(\tilde{r}_j(\mu), \mu) + Q(\tilde{r}_j(\mu), \mu)-V(a_j,0) - Q(a_j,0) = 0 \;\;,
\end{aligned}
\end{align}
\end{linenomath*}
with
\begin{linenomath*}
\begin{align}
& \begin{aligned}
V (r, \mu) & = \sum_{i=0}^{i_{I}} V^{I}_i (r, \mu) \\
& + \sum_{i=i_{I}+1}^{i_{IV}-1} V^{II}_i (r, \mu) +  \sum_{i=i_{IV}}^{N_{\rm lay}-1} V^{IV}_i (r, \mu) \;\;,
\end{aligned} \\[4ex]
& \begin{aligned}
V_i^I(r,\mu){} & =  \frac{4\pi}{3} \delta\rho_i G r^2 + \frac{G M_i}{r} \left [ \left ( \frac{r}{a_i} \right ) N_{i,0}(\xi_i, 1) \right . \\
& \left . - \sum_{k=1}^{\infty}  \left ( \frac{r}{a_i} \right )^{2k+1} N_{i,2k}(\xi_i,1) P_{2k}(\mu) \right ] \;\;,
\end{aligned} \\[4ex]
& \begin{aligned}
V^{II}_i(r,\mu) {} &= \frac{G M_i}{r} \\
& \left \{ \left [ 1 - K_{i,0}(\xi_i, \mu_{r,i}) + \left ( \frac{r}{a_i} \right ) N_{i,0} (\xi_i, \mu_{r,i}) \right ] \right .\\
& \left . - \sum_{k=1}^{\infty} \left [   \left (\frac{r}{a_i} \right )^{ \, 2 k + 1} N_{i,2k}(\xi_i, \mu_{r, i}) \right. \right. \\
& \left.  \left. + \left (\frac{a_i}{r} \right )^{2 k} (J_{i,2k}-K_{i,2k}(\xi_i, \mu_{r,i}) )\right ] P_{2k}(\mu) \right \} \;\;,
\end{aligned}
\end{align}
\begin{align}
& \begin{aligned}
V^{IV}_i(r,\mu) {} &= \frac{G M_i}{r} \left [ 1-\sum_{k=1}^{\infty} \left ( \frac{a_i}{r} \right )^{2k} J_{i,2k} P_{2k}(\mu) \right ] \;\;,
\end{aligned} \\[4ex]
& \begin{aligned}
Q (r,\mu) =\frac{1}{3} r^2 \omega_{\rm rot}^2 \left [ 1 - P_2(\mu) \right ] \;\;,
\end{aligned} \\[4ex]
& \begin{aligned}
\label{eqn:Ni0}
N_{i,0}(\xi_i,\mu_r)  {} & = \left (  \frac{3}{2} \right ) \frac{ \int_{0}^{\mu_r} \left [ (\tilde{\xi_i}(\mu'))^2 - \xi_i^2 \right ] d\mu' }{\int_{0}^{1} \left [ \tilde{\xi_i}(\mu') \right ]^3 d\mu' } \;\;,
\end{aligned} \\[4ex]
& \begin{aligned}
\label{eqn:Ni2}
N_{i,2}(\xi_i,\mu_r) {} &= - \frac{ 3 \int_{0}^{\mu_r} \left [ \ln \left ( \tilde{\xi_i}(\mu') / \xi_i \right ) P_2(\mu') \right ] d\mu' }{\int_{0}^{1} \left [ \tilde{\xi_i}(\mu') \right ]^3 d\mu' } \;\;,
\end{aligned} \\[4ex]
& \begin{aligned}
\label{eqn:Ni2k}
N_{i,2k}(\xi_i,\mu_r) {} &=  -\left ( \frac{3}{2k-2} \right ) \\ 
&\frac{ \int_{0}^{\mu_r} \left [ \left ( \xi_i^{\, 2-2k} - (\tilde{\xi_i}(\mu'))^{\, 2-2k} \right ) P_{2k}(\mu') \right ] d\mu' }{\int_{0}^{1} \left [ \tilde{\xi_i}(\mu') \right ]^3 d\mu' }, \\ 
& \textrm{for} \; k \ge 2 \;\;,
\end{aligned} \\[4ex]
& \begin{aligned}
\label{eqn:Ki0}
K_{i,0}(\xi_i,\mu_r)  {} &=  \frac{ \int_{0}^{\mu_r} \left [ (\tilde{\xi_i}(\mu'))^3 - \xi_i^3 \right ] d\mu' }{\int_{0}^{1} \left [ \tilde{\xi_i}(\mu') \right ]^3 d\mu' } \;\;,
\end{aligned} \\[4ex]
& \begin{aligned}
\label{eqn:Ki2k}
K_{i,2k}(\xi_i,\mu_r) {} &=  - \left ( \frac{3}{2k+3} \right ) \\
& \frac{ \int_{0}^{\mu_r} \left [ \left ( (\tilde{\xi_i}(\mu'))^{\, 2k+3} - \xi_i^{\, 2k+3} \right ) P_{2k}(\mu') \right ] d\mu' }{\int_{0}^{1} \left [ \tilde{\xi_i}(\mu') \right ]^3 d\mu' }, \\ 
&  \textrm{for} \; k \ge 1 \;\;,
\end{aligned} \\[4ex]
& \begin{aligned}
\label{eqn:Ji2k}
J_{i,2k}  {} &=  \\
& - \left ( \frac{3}{2k+3} \right ) \frac{ \int_{0}^{1} \left [ (\tilde{\xi_i}(\mu'))^{\, 2k+3}  P_{2k}(\mu') \right ] d\mu' }{\int_{0}^{1} \left [ \tilde{\xi_i}(\mu') \right ]^3 d\mu' } \;\;,
\end{aligned}
\end{align}
\end{linenomath*}
and
\begin{linenomath*}
\begin{align}
& \begin{aligned}
\label{eqn:Mi}
M_i {} &= \frac{4 \pi \delta \rho_i a_i^3}{3}  \int_{0}^{1} \left [ \tilde{\xi_i}(\mu') \right ]^3 d\mu' \;\;.
\end{aligned}
\end{align}
\end{linenomath*}
A subscript $i$ indicates the quantity was calculated for the $i$th spheroid. Note that $\mu_{r,i}=\mu_{r,i}(r)$ is different for each spheroid, $i$, and each point $r$. For the shape iteration, the above series of equations is solved using the Newton-Raphson method for each $\mu$ point on each equipotential surface. A step size of $\delta \xi$ is used to calculate the gradient of Equation \ref{eqn:all_start} at each step in the Newton-Raphson iteration. The mass distribution of the structure is not changed during the shape iteration, with the whole structure only updated at the end of the iteration. In order to make the iterative equation numerically tractable we truncate the series at a maximum spherical harmonic degree of $2k$~$=$~$2k_{\rm max}$.

The notation we have used varies slightly from that used by \cite{Hubbard2013}. In our formulation $\xi_i$~$=$~$r/a_i$, but \cite{Hubbard2013} normalizes the $\xi$ to the outermost equatorial radii, $a_0$.
Also the $J_{i,2k}$ defined here for each spheroid are not the contribution of each spheroid to the gravitational moments, unlike in \cite{Hubbard2013}. The total gravitational moments of the structure are given by 
\begin{linenomath*}
\begin{align}
\begin{aligned}
J_{n}  = \sum_{i=0}^{N_{\rm lay}-1} \frac{M_i}{M} \left ( \frac{a_i}{a_0} \right )^{n} J_{i,n} \;\;.
\end{aligned}
\end{align}
\end{linenomath*}
These changes in notation have been made to allow the layers to remain independent of each other for ease of calculation.

The equations used in HERCULES are much more complicated than in the CMS model \citep{Hubbard2013} as extra terms are needed to allow the expression for potential to converge for very oblate bodies.
In the CMS model, all the integrals are over a fixed interval $0 \rightarrow 1$ allowing the use of Gaussian quadrature points for $\mu$ and highly precise integration.
In the iterative equation used in the HERCULES code, there is not a fixed domain of integration and so Gaussian quadrature points cannot be used. Instead, we use equally spaced $\mu$ but this decreases the precision that is computationally feasible. 

\begin{table}[t!]
{\renewcommand{\arraystretch}{1.3}
\centering
\caption{The standard HERCULES parameters used in this work. Briefly: $n_{\rm int}$ is the maximum number of shape iterations; $\chi_{\rm toll}$ is the tolerance for the shape iteration; $n_{\rm int}^{\xi}$ is the maximum number of conservation iterations; $\chi_{\rm toll}^{\xi}$ is the tolerance for the conservation iteration; $\delta \xi$ is the step size used to calculate gradients in the shape iteration; $N_{\rm lay}$ is the number of concentric spheroids; $N_{\rm lay}^{\rm mat}$ is the number of layers used to describe each material; $M_{\rm j}$ is the mass of each material; $2k_{\rm max}$ is the highest spherical harmonic degree term used in the iterative equation; $N_{\mu}$ is the number of points used to describe each equipotential; and $p_{\rm min}$ is the bounding pressure of the body. The division of mass and layers between material layers is shown for a body with a single layer of silicate.}
\begin{tabular}{l | l | l}
\label{sup:tab:HERCULES_params}
Parameter & Value & Units \\ \hline
$n_{\rm int}$ & 200 & \\ 
$\chi_{\rm toll}$ & $10^{-8}$ & \\
$n_{\rm int}^{\xi}$ & 200 & \\
$\chi_{\rm toll}^{\xi}$ & $10^{-10}$ & \\
$\delta \xi $ & $10^{-2}$ & \\
$N_{\rm lay}$ & 100 & \\
$N_{\rm lay}^{\rm mat}$ & \{ 80, 20 \} & \\
$M_{\rm j}$ & \{ $4.1729748$, $1.81499$ \} & $10^{24}$~kg \\
$k_{\rm max}$ & 6 & \\
$N_{\mu}$ & 1000 & \\
$p_{\rm min}$ & $10^6$ & Pa
\end{tabular}
}
\end{table}

%%%%%%%%%%%%%%%%%%%%%%%%%%%%%%%%%
\subsubsection{Calculation of pressure}
\label{sup:sec:press_calc}

During the conservation iteration, the pressure in each layer of the structure is calculated and used to update the density of each layer.
The pressure in the structure is calculated assuming hydrostatic equilibrium. 
Since each of the layers is constant density, the pressure at the top of a layer can be calculated as follows \citep{Hubbard2013}
\begin{linenomath*}
\begin{align}
\label{eqn:press}
p_i=p_{i-1}+\rho_{i-1}\left[ U(a_i, 0) - U(a_{i-1}, 0) \right ] \;\;.
\end{align}
\end{linenomath*}
In HERCULES the pressure at the top of the uppermost layer is specified, $p_0$~$=$~$p_{\rm min}$, as a bounding pressure and the pressure at the top of all subsequent layers can be determined sequentially.
The pressure at the center of the body is given by
\begin{linenomath*}
\begin{align}
p_{\rm core}=p_{N_{\rm lay}-1}+\rho_{N_{\rm lay}-1}\left[ U(0, 0) - U(a_{N_{\rm lay}-1}, 0) \right ] \;\;,
\end{align}
\end{linenomath*}
which is calculated for reference and used for some convergence criteria (see \S\ref{sup:sec:convergence}).

The full density, $\rho_i$, of each layer is determined from pressure by using the EOS corresponding to the relevant material layer. The pressure used to pass to the EOS is the average of the pressures at the top and bottom of the layer. 
This approximates the pressure in the middle of the layer, assuming that the variation in potential is roughly linear over the region between layers.
From $\rho_i$, the density of each spheroid, $\delta \rho_i$, is calculated from the outside, inwards.

%%%%%%%%%%%%%%%%%%%%%%%%%%%%%%%%%
\subsubsection{Mass conservation}
\label{sup:sec:Mconc}

We conserve the mass of each of the material layers by scaling the radius of each layer. 
The radii of layers in a material are scaled by a constant factor to conserve the mass of that material.
The mass of a material after scaling is
\begin{linenomath*}
\begin{align}
\begin{aligned}
M_{j} {} & =\sum_{i<i_{out}} \delta\rho_i(\lambda_j V_{out}^j -\lambda_{j+1} V_{in}^j) \\
& +\sum_{i_{out}\le i<i_{in}} \delta \rho_i(\lambda_j V_{i} -\lambda_{j+1} V_{in}^j) \;\;,
\end{aligned}
\end{align}
\end{linenomath*}
where $M_j$ is the mass of layer $j$, $i_{out}$ is the index of the outermost layer in the material layer $j$, and $i_{in}$ is the index of the shell on the inside boundary of the material layer (note that this layer is not of material $j$). $V_{out}^j$ and $V_{in}^j$ are the volume of the outside and inside layers, and $\lambda_j$ is the volume scaling factor for material $j$. 
Rewriting this expression
\begin{linenomath*}
\begin{align}
\begin{aligned}
M_{j} {}&= \lambda_j \left ( \sum_{i<i_{out}} \delta\rho_i  V_{out}^j +\sum_{i_{out}\le i<i_{in}} \rho_i V_{i}\right ) \\ & -\lambda_{j+1} \left ( \sum_{i<i_{out}} \delta\rho_i V_{in}^j +\sum_{i_{out}\le i<i_{in}} \rho_i V_{in}^j \right ) \;\;,
\end{aligned}
\end{align}
\end{linenomath*}
which can be rearranged to give
\begin{linenomath*}
\begin{align}
\begin{aligned}
 \lambda_j = \frac{ M_{j}+ \lambda_{j+1} \left ( \sum_{i<i_{in}} \rho_i V_{in}^j  \right )}{ \left ( \sum_{i<i_{out}} \rho_i  V_{out}^j +\sum_{i_{out}\le i<i_{in}} \rho_i V_{i}\right ) } \;\;.
\end{aligned}
\end{align}
\end{linenomath*}
We can solve for each $\lambda_{j}$ sequentially from inside out in the body. For the innermost material
\begin{linenomath*}
\begin{align}
\begin{aligned}
 \lambda_{N_{mat}}= \frac{ M_{N_{mat}}}{ \left ( \sum_{i<i_{out}} \rho_i  V_{out}^{N_{mat}} +\sum_{i_{out}\le i<i_{in}} \rho_i V_{i}\right ) } \;\;,
\end{aligned}
\end{align}
\end{linenomath*}
which requires no prior knowledge of the other $\lambda_j$.

From Equation \ref{eqn:Mi} the volume of a spheroid scales as to the third power of $a_i$. 
The scaled $a_i$, $a'_i$, are therefore given by 
\begin{linenomath*}
\begin{align}
\begin{aligned}
a_i'=\lambda^{\frac{1}{3}_j} a_i \;\;.
\end{aligned}
\end{align}
\end{linenomath*}
Each of the layers in the body are then assigned the new scaled equatorial radii. Since the radii of all the points on an equipotential surface, $\tilde{\xi}_i(\mu)$, in HERCULES are normalized by $a_i$, scaling $a_i$ automatically scales the entire volume of the spheroid, with the shape remaining constant.
This routine conserves mass but must be iterated over to give a self consistent pressure and density structure.

In some cases, due to the different scalings of the material layers, the above routine causes equipotential surfaces to change relative positions i.e. $a_{i-1}$~$<$~$a_{i}$. 
Such an arrangement is not acceptable in HERCULES.
The radii of any crossing layers are altered so that
\begin{linenomath*}
\begin{align}
a_i=a_i'+1.005 \times (a_i'-a_{i-1}') \;\;,
\end{align}
\end{linenomath*}
where $a_i'$ are the equatorial radii calculated from scaling the layers in the above routine.
A single crossing layer can cause several layers to have their radii altered which is done sequentially from the outside in. HERCULES also ensures the mass conservation scaling does not lead to large gaps between layers. 
If
\begin{linenomath*}
\begin{align}
a_{i_{\rm in}-1}-a_{i_{\rm in}}>1.05 \times (a_{i_{\rm in}-2}-a_{i_{\rm in}-1}) \;\;,
\end{align}
\end{linenomath*}
then the the radii of layers in the upper material are altered. 
The radii of all the layers in the upper material are moved inwards such that
\begin{linenomath*}
\begin{align}
a_{i_{\rm in}-1}-a_{i_{\rm in}}=a_{i_{\rm in}-2}-a_{i_{\rm in}-1} \;\;.
\end{align}
\end{linenomath*}
In both the above cases where the radii of the structure must be altered from the calculated positions from mass conservation, mass is not conserved on that iterative step. 
Over several iterations the structure will come to an arrangement where the radii scaling does not cause such corrections and mass conservation will be achieved.

%%%%%%%%%%%%%%%%%%%%%%%%%%%%%%%%%
\subsubsection{AM conservation}
\label{sup:sec:Lconc}

The version of HERCULES used in this work assumes that a body is corotating, i.e., the material in the body has a single angular velocity, $\omega_{\rm rot}$. The angular velocity required to conserve the AM of the body is given by
\begin{linenomath*}
\begin{align}
\begin{aligned}
\label{eqn:AMconc}
\omega_{\rm rot} = \frac{L_{\rm tot}}{I} ,
\end{aligned}
\end{align}
\end{linenomath*}
where $I$ is the moment of inertia of the whole body and $L_{\rm tot}$ is the desired AM of the body. 
The total moment of inertia is given by the sum of the moments of inertia of each of the concentric spheroids.
The moment of inertia about the $z$ axis of the $i$th constant density spheroid is defined as
\begin{linenomath*}
\begin{align}
\begin{aligned}
I_i=\int_{\mathcal{V}} \frac{2}{3} \delta \rho_i r'^2 (1-P_2(\mu)) d\mathcal{V'} \;\;,
\end{aligned}
\end{align}
\end{linenomath*}
where $\mathcal{V}$ is the volume of the spheroid.
In our notation this becomes
\begin{linenomath*}
\begin{align}
\begin{aligned}
I_i=\frac{4 \pi}{3} \delta \rho_i  \int_{-1}^{1} \int_0^{\tilde{r_i}(\mu')} (r')^4 (1-P_2(\mu')) dr' d\mu' \;\;.
\end{aligned}
\end{align}
\end{linenomath*}
We can use the symmetry of the spheroids and integrate over $r$ to give
\begin{linenomath*}
\begin{align}
\begin{aligned}
\label{eqn:I_spheroid}
I_i=\frac{8 \pi}{15} \delta \rho_i  \int_{0}^{1}  (\tilde{r_i}(\mu'))^5 (1-P_2(\mu')) d\mu' \;\;.
\end{aligned}
\end{align}
\end{linenomath*}
The total moment of inertia is then
\begin{linenomath*}
\begin{align}
\begin{aligned}
\label{eqn:I_total}
I=\sum_{i=0}^{N_{\rm lay}} I_i.
\end{aligned}
\end{align}
\end{linenomath*}
Once the moment of inertia have been calculated the new $\omega_{\rm rot}$ that conserves AM can be found from equation \ref{eqn:AMconc}.

%%%%%%%%%%%%%%%%%%%%%%%%%%%%%%%%%
\subsubsection{Convergence criterion}
\label{sup:sec:convergence}

Here we outline the convergence criteria for HERCULES for each of the iterative loops.
The convergence criteria for the main iteration to find the equilibrium structure is based on the equatorial potential of the spheroids. 
The iteration is stopped when
\begin{linenomath*}
\begin{align}
\left | \frac{ \sum_{i=0}^{N_{\rm lay}} \left [ U^n(a_i^n, 0) - U^{n-1}(a_i^{n-1}, 0)\right ] }{N_{\rm lay} U^n(a_0^n, 0)} \right | < \chi_{\rm toll} \;\;,
\end{align}
\end{linenomath*}
where $n$ indicates the current iteration.
This is similar to requiring that the average fractional change in equatorial potential for the equipotential layers is less than a given tolerance. 

In the shape iteration, the Newton-Raphson method is repeated for each point until
\begin{linenomath*}
\begin{align}
\left | \left ( \xi_i^{n} -\xi_i^{n-1} \right ) \left ( \frac{a_i}{a_0} \right ) \right | < \chi_{\rm toll}^{\xi} \;\;,
\end{align}
\end{linenomath*}
where $n$ indicates the current iteration. In other words, the iteration for each point is continued until the change in the absolute radius of that point is small.

The convergence iteration is repeated until the fractional change in the core pressure is less than a certain value, $\chi_{\rm toll}$, i.e.
\begin{linenomath*}
\begin{align}
\left | p_{\rm core}^n - p_{\rm core}^{n-1} \right | < \chi_{\rm toll} \;\;.
\end{align}
\end{linenomath*}
Note that $\chi_{\rm toll}$ is the same tolerance used for the main iteration.

For each iterative loop has a maximum number of iterations ($n_{\rm int}^{\xi}$ for the shape iteration and $n_{\rm int}$ otherwise) that are allowed before the iteration exits and continues.
The parameters used for this study are given in Table~\ref{sup:tab:HERCULES_params}.

%xxxxxxxxxxxxxxxxxxxxxxxxxxxxxxxxxxxxxxxxxxxxxxxxxxxxxxxxxxxxxxxxxxxxxxxxxxxxxxxxxxxxxxxxxxxxxxxxxxxxxxxxxxxxxxxxxxxxxxxxxxxxxxxxxxxxxxxx
%xxxxxxxxxxxxxxxxxxxxxxxxxxxxxxxxxxxxxxxxxxxxxxxxxxxxxxxxxxxxxxxxxxxxxxxxxxxxxxxxxxxxxxxxxxxxxxxxxxxxxxxxxxxxxxxxxxxxxxxxxxxxxxxxxxxxxxxx
%xxxxxxxxxxxxxxxxxxxxxxxxxxxxxxxxxxxxxxxxxxxxxxxxxxxxxxxxxxxxxxxxxxxxxxxxxxxxxxxxxxxxxxxxxxxxxxxxxxxxxxxxxxxxxxxxxxxxxxxxxxxxxxxxxxxxxxxx
\section{Comparison of HERCULES to other structure models}
\label{sup:sec:HERCULES_compare}

\begin{figure}
\centering
\includegraphics[scale=0.8333333]{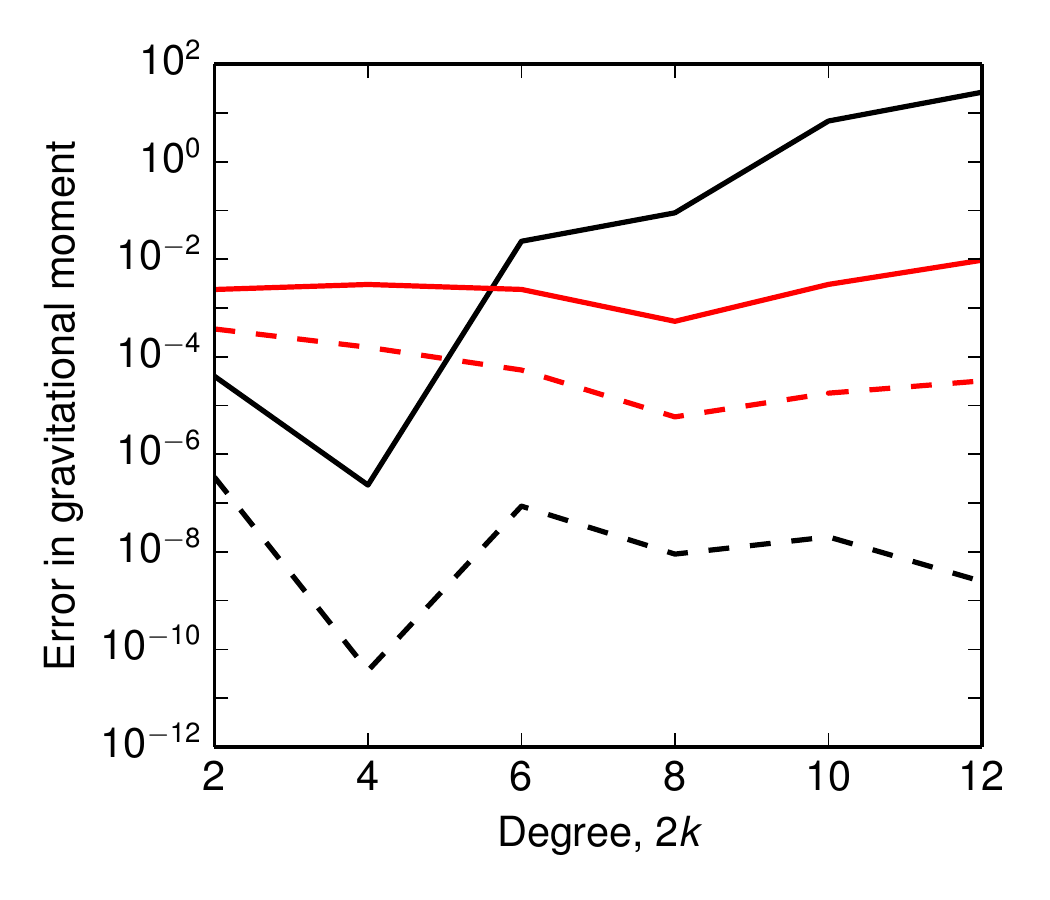}
\caption[]{Reproducibility of Maclaurin spheroid gravitational moments using HERCULES. Plotted here are the errors in the gravitational moments, as a function of degree, for equilibrium structures calculated using HERCULES for a constant density body relative to the analytical solution for a Maclaurin spheroid. Shown are examples for a spheroid with minor flattening (black, aspect ratio~$=$~0.98, $\ell$~$=$~$0.2$) and extreme rotational flattening (red, aspect ratio~$=$~0.47, $\ell$~$=$~$1.9$). The equipotential surfaces were calculated using HERCULES with $N_{\mu}$~$=$~$1000$ and $k_{\rm max}$~$=$~$6$. Both absolute error (dashed lines) and fractional error (solid lines) are shown. The large fractional error in the high degree gravitational moments for the minor flattening case is due to the magnitude of the moments being very small. The majority of the error for the extreme flattening case is due to neglecting higher degree terms. }
\label{sup:fig:HERCULES_Maclauren}
\end{figure}

We tested the structures found by HERCULES against previous calculations for rotating bodies. First, we compared the structure of constant density spheroids calculated by HERCULES to the analytical results for Maclaurin spheroids. The surface of a Maclaurin spheroid is given by
\begin{linenomath*}
\begin{align}
\tilde{r}^2 (\mu)=\frac{a^2}{1+\ell^2 \mu^2} \;\;,
\end{align}
\end{linenomath*}
where the ellipticity is
\begin{linenomath*}
\begin{align}
\ell^2=\frac{a^2}{b^2} -1 \;\;.
\end{align}
\end{linenomath*}
\citet{Hubbard2012} showed that the gravitational moments for a Maclaurin spheroid are
\begin{linenomath*}
\begin{align}
J_{2 n}=\frac{3 (-1)^{1+n}}{(2n+1)(2n+3)} \left ( \frac{\ell^2}{1 + \ell^2} \right )^n \;\;.
\end{align}
\end{linenomath*}
We calculated the surfaces of two spheroids, with varying rotation rates and hence different aspect ratios and ellipticities. We used the same parameters as used in the rest of this work (Table~\ref{sup:tab:HERCULES_params}); $N_{\mu}$~$=$~$1000$, and including terms up to spherical harmonic degree 12 ($k_{\rm max}$~$=$~$6$).
For these parameters the HERCULES code is able to calculate the gravitational moments of Maclaurin spheroids with extreme rotational flattening to better than 1\% (Figure~\ref{sup:fig:HERCULES_Maclauren}).
The maximum fractional error in finding the surface, i.e., $\tilde{r}(\mu)$, of the two ellipsoids shown in Figure~\ref{sup:fig:HERCULES_Maclauren} is on the order $10^{-6}$ and $10^{-3}$ for the slow and fast rotation cases respectively.
The fractional errors in the calculated potential at the surface of the ellipsoid are on the order $10^{-8}$ and $10^{-3}$.
The majority of the discrepancy for the high flattening case comes from not including higher order terms in the expansion, as can be shown by running simulations with higher $k_{\rm max}$.

We also tested the structure calculated by HERCULES for a planet with a density structure described by a polytrope of index one that has moderate rotational flattening. The barotrope for such a structure is
\begin{linenomath*}
\begin{align}
p=K\rho^2 \;\;,
\end{align}
\end{linenomath*}
where $K$ is a constant scaling factor set to conserve mass for a body with a given equatorial radius. Rotating planets obeying this barotrope have been extensively studied \citep[e.g.][]{Zharkov1978,Hubbard1975} and, in particular, \citet{Hubbard2013} calculated the gravitational moments for the degree-1 polytrope case using the CMS model, allowing us to make direct comparisons. The gravitational moments of a planet with the mass, equatorial radius and rotational period of Jupiter calculated using a 3rd-order expansion in $q$ \citep{Zharkov1978,Hubbard1975} and the CMS model with varying numbers of concentric spheroids \citep{Hubbard2012} are shown in Table~\ref{sup:tab:HERCULES_compare1}. $q$ is a dimensionless parameter that describes the relative magnitude of rotational and gravitational forces in a body and is given by
\begin{linenomath*}
\begin{align}
q=\frac{\omega_{\rm rot} a_0^3}{G M} .
\end{align}
\end{linenomath*}
$q$ is commonly used in small parameter expansions to approximate the structure of rotating bodies. For comparison, Table~\ref{sup:tab:HERCULES_compare2} shows the gravitational moments for the same body calculated using HERCULES for varying number and arrangement of spheroids. The bounding pressure was set to zero, but otherwise we used the standard set of parameters as elsewhere in this study. To give an indication of the relative differences between the calculated gravitational moments we have included in both tables the difference between each value and those found using the CMS model with $N_{\rm lay}$~$=$~$512$. 

\begin{table*}
\centering
{\renewcommand{\arraystretch}{1.3}
\centering
\caption{A comparison of literature values of the gravitational moments calculated for a Jupiter-like body using a 3rd-order expansion in $q$ \citep{Zharkov1978,Hubbard1975} and the CMS model with varying numbers of concentric spheroids \citep{Hubbard2013}. $q$~$=$~$0.089195487$ for all bodies. We present both the absolute value of each moment and the fractional difference of that calculation to the value found using the CMS model with $N_{\rm lay}$~$=$~$512$. }
\begin{tabular}{l c c c c c }
\label{sup:tab:HERCULES_compare1}
 & \multicolumn{1}{c}{CMS ($N_{lay}$~$=$~$512$)} & \multicolumn{2}{c}{3rd-order theory} & \multicolumn{2}{c}{CMS ($N_{lay}$~$=$~$256$)}  \\
Quantity  & Reference value & Value & \% difference & Value & \% difference   \\
\hline
$\hphantom{-}J_2 \times 10^2$ &	1.3989253 & 1.3994099 & 0.0346 & 1.3991574 & 0.0166   \\	
$-J_4 \times 10^4$ & 5.3187997 & 5.3871087 &	1.2843 &	5.3203374 &	0.0289   \\
$\hphantom{-}J_6 \times 10^5$ &	3.0122356 & 3.9972442 &	32.7003 &	3.0133819 &	0.0381   \\
$-J_8 \times 10^5$ & 2.1324628 & - & - & 2.1334136	& 0.0446  \\
$\hphantom{-}J_{10} \times 10^7$ &	1.7409925 & - & - & 1.7418428 &	0.0488  \\
$-J_{12} \times 10^8$ &	1.5685327 & - & - & 1.5693324 &	0.0510  \\
\end{tabular}
}
\end{table*}

\begin{sidewaystable*}
\centering
{\renewcommand{\arraystretch}{1.3}
\centering
\caption{A comparison of the gravitational moments calculated for a Jupiter-like body by HERCULES using different numbers and arrangements of spheroids. All other parameters for HERCULES were the same as the standard parameters used elsewhere in this paper with the exception of the bounding pressure which was set to zero.  $q$~$=$~$0.089195487$ for all bodies. We present both the absolute value of each moment and the fractional difference of that calculation to the value found using the CMS model with $N_{\rm lay}$~$=$~$512$ (see Table~\ref{sup:tab:HERCULES_compare1}). }
\begin{tabular}{l c c c@{\hspace{20pt}} c c c c c c}
\label{sup:tab:HERCULES_compare2}
& \multicolumn{2}{c}{Equal layer spacing} & & \multicolumn{6}{c}{Logarithmic layer spacing} \\
 & \multicolumn{2}{c}{$N_{lay}$~$=$~$100$} & & \multicolumn{2}{c}{$N_{lay}$~$=$~$4100$} & \multicolumn{2}{c}{$N_{lay}$~$=$~$256$} & \multicolumn{2}{c}{$N_{lay}$~$=$~$512$} \\
 \cline{2-3} \cline{5-10}
Quantity  & Value & \% difference & & Value & \% difference & Value & \% difference & Value & \% difference \\
\hline
$\hphantom{-}J_2 \times 10^2$ & 1.4333451 &	2.4604 & &	1.4018542 &	0.2094 &	1.3995710 &	0.0462 & 1.3990313 & 0.0076 \\	
$-J_4 \times 10^4$ & 5.5831640 &	4.9704 & &	5.3394223 &	0.3877 &	5.3221134 &	0.0623 & 5.3192414 & 0.0083 \\
$\hphantom{-}J_6 \times 10^5$ & 3.2388140 &	7.5219 &&	3.0289739 &	0.5557 &	3.0143152 &	0.0690 & 3.0124791 & 0.0081 \\
$-J_8 \times 10^5$ & 2.3475085 &	10.0844 &&	2.1478449 &	0.7213 &	2.1340524 &	0.0745 & 2.1326885 & 0.0106 \\
$\hphantom{-}J_{10} \times 10^7$ & 1.9497025 &	11.9880 &&	1.7553814 &	0.8265 &	1.7414453 &	0.0260 & 1.7406042 & 0.0223 \\
$-J_{12} \times 10^8$ & 1.7441342 &	11.1953 &&	1.5807215 &	0.7771 &	1.5662394 &	0.1462 & 1.5669045 & 0.1038 \\

\end{tabular}
}
\end{sidewaystable*}

The agreement between HERCULES and previous results for polytropic structures depends strongly on the number and arrangement of concentric spheroids used. For $100$ layers evenly spaced in radius, the gravitational moments are larger than those found using the CMS method by several percent. The gravitational moments are most sensitive to the outer layers of the structure and for $100$ evenly spaced layers the scale height at the top of the polytropic structure is poorly sampled. A large ($>\sim$1000) number of evenly spaced layers would be needed to resolve the scale height in the outer regions of the structure. As an alternative, we used logarithmically spaced layers, so that the layers are concentrated at larger radii. For logarithmically spaced layers the disagreement between HERCULES and the previously calculated results fall to a few tenths of a percent even with just $100$ layers. The structure is also resolved to much lower pressure. The agreement continues to improve with the addition of more logarithmically spaced layers and becomes similar to, or better than, the agreement between the CMS model using $256$ and $512$ layers. From this we conclude that the HERCULES code performs well in comparison to previous models of rotating planets.

In this work we used a number of material layers with the evenly spaced concentric layers within each material. We concentrated layers in the outer material layers to give better resolution at the edge of the structure. For structures that are vapor in their outer regions, such as the majority of the structures considered in this work, the scale height in the outer regions is large and the resolution we have used is sufficient to resolve the outer structure. In cases that are mostly condensed at the surface the scale height becomes comparable to the radial resolution. In such cases the lack of resolution in the outer layers can cause small (a few percent) errors in the properties of the outer structure. This can be seen by comparing structures found using different radial resolutions (see \S\ref{sup:sec:HSSL_calculation}). As our work focuses on mostly vaporized bodies this does not significantly affect our results but is important to bare in mind for future studies of condensed planets using HERCULES.

%xxxxxxxxxxxxxxxxxxxxxxxxxxxxxxxxxxxxxxxxxxxxxxxxxxxxxxxxxxxxxxxxxxxxxxxxxxxxxxxxxxxxxxxxxxxxxxxxxxxxxxxxxxxxxxxxxxxxxxxxxxxxxxxxxxxxxxxx
%xxxxxxxxxxxxxxxxxxxxxxxxxxxxxxxxxxxxxxxxxxxxxxxxxxxxxxxxxxxxxxxxxxxxxxxxxxxxxxxxxxxxxxxxxxxxxxxxxxxxxxxxxxxxxxxxxxxxxxxxxxxxxxxxxxxxxxxx
%xxxxxxxxxxxxxxxxxxxxxxxxxxxxxxxxxxxxxxxxxxxxxxxxxxxxxxxxxxxxxxxxxxxxxxxxxxxxxxxxxxxxxxxxxxxxxxxxxxxxxxxxxxxxxxxxxxxxxxxxxxxxxxxxxxxxxxxx
\section{HERCULES parameter sensitivity tests}
\label{sup:sec:HSSL_calculation}

The HERCULES parameters used in this study, unless otherwise noted, are given in Table~\ref{sup:tab:HERCULES_params}. These parameters were chosen as a compromise between accuracy and computational efficiency to allow us to explore a wide parameter space.
Here we examine the effect of a few of these parameters on our results.

\begin{figure}
\centering
\includegraphics[scale=0.8333333]{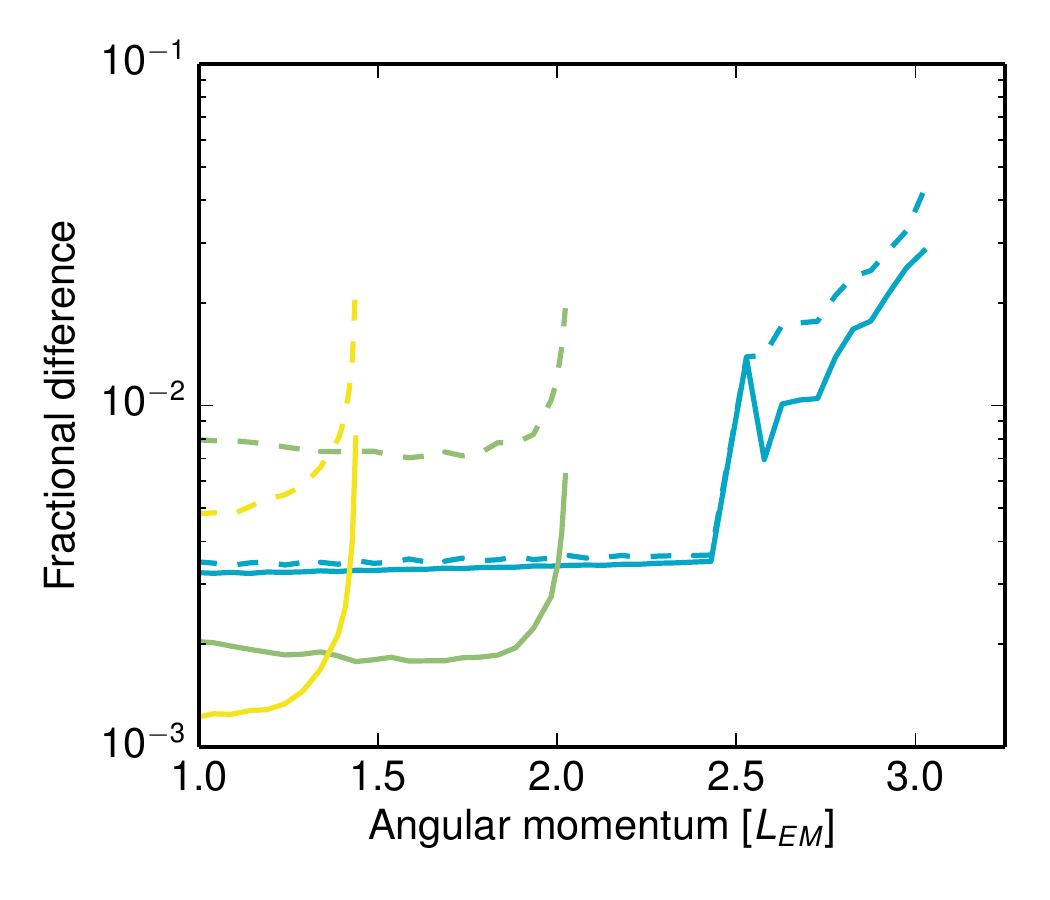}
\caption[]{The radii of bodies calculated using HERCULES has a small dependence on the number of concentric layers used. The difference is largest for the most extended planets and those with small scale heights in the outer structure. Plotted are the fractional difference in equatorial radii of HERCULES bodies with 50 (dashed lines) and 100 (solid lines) concentric layers relative to a body with 200 concentric layers. All bodies were Earth mass with isentropic silicate thermal profiles with specific entropies of 4.5, 5.5 and 6.75~kJ~K$^{-1}$~kg$^{-1}$ (see Figure~\ref{fig:HSSL_R-L} for color bar).}
\label{sup:fig:R_resolution}
\end{figure}

\begin{figure}
\centering
\includegraphics[scale=0.8333333]{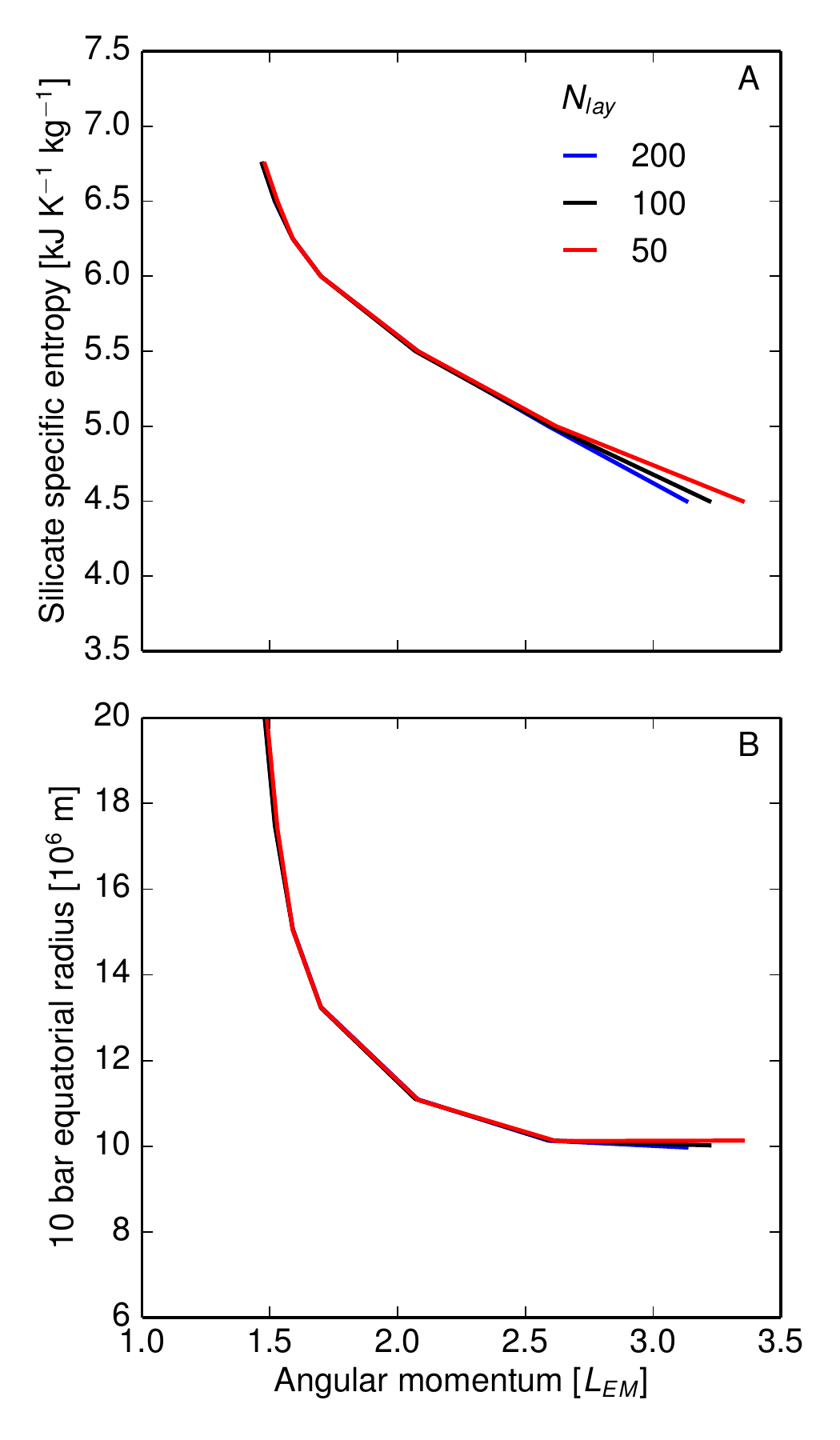}
\caption[]{The change in the CoRoL found using HERCULES with a varying number of concentric layers is not significant. Shown is the CoRoL found using HERCULES with a different number of concentric layers (colored lines) for Earth-mass bodies with isentropic silicate thermal profiles of varying specific entropy.}
\label{sup:fig:HERCULES_resolution}
\end{figure}

We tested the sensitivity of the calculated equatorial radii of individual bodies to the number of concentric layers used, $N_{\rm lay}$.
We ran sets of corotating, Earth-mass bodies with constant entropy mantles using $N_{\rm lay}$~$=$~$50$ and $N_{\rm lay}$~$=$~$100$ and compared them to our results using $N_{\rm lay}$~$=$~$200$ (see Table~\ref{sup:tab:HSSL}).
The equatorial radius of bodies varied by up to a few percent between the different resolutions (Figure~\ref{sup:fig:R_resolution}).
The difference is largest for the most oblate bodies, close to the CoRoL.
This increase is due to the extended structure increasing the sensitivity of the model to the radial resolution. 
The difference is particularly large for the $S_{\rm lower}$~$=$~$4.5$~kJ~K$^{-1}$~kg$^{-1}$ isentropic silicate thermal profile, as the very small scale height of the condensate dominated atmosphere is not resolved well with equally spaced layers.
The number of layers that are used to model this region of the structure can therefore substantially change the equatorial radius, and so the structure is sensitive to the radial resolution. The number of concentric layers used also has little effect on the CoRoL boundary (Figure~\ref{sup:fig:HERCULES_resolution}). Again, the only exception is for the $S_{\rm lower}$~$=$~$4.5$~kJ~K$^{-1}$~kg$^{-1}$ isentropic silicate thermal profile as the CoRoL is sensitive to the equatorial radius of a body. In all cases, the small dependence of the calculated structures and the CoRoL on the number of concentric layers is not significant for this work.

\begin{figure}
\centering
\includegraphics[scale=0.83333333]{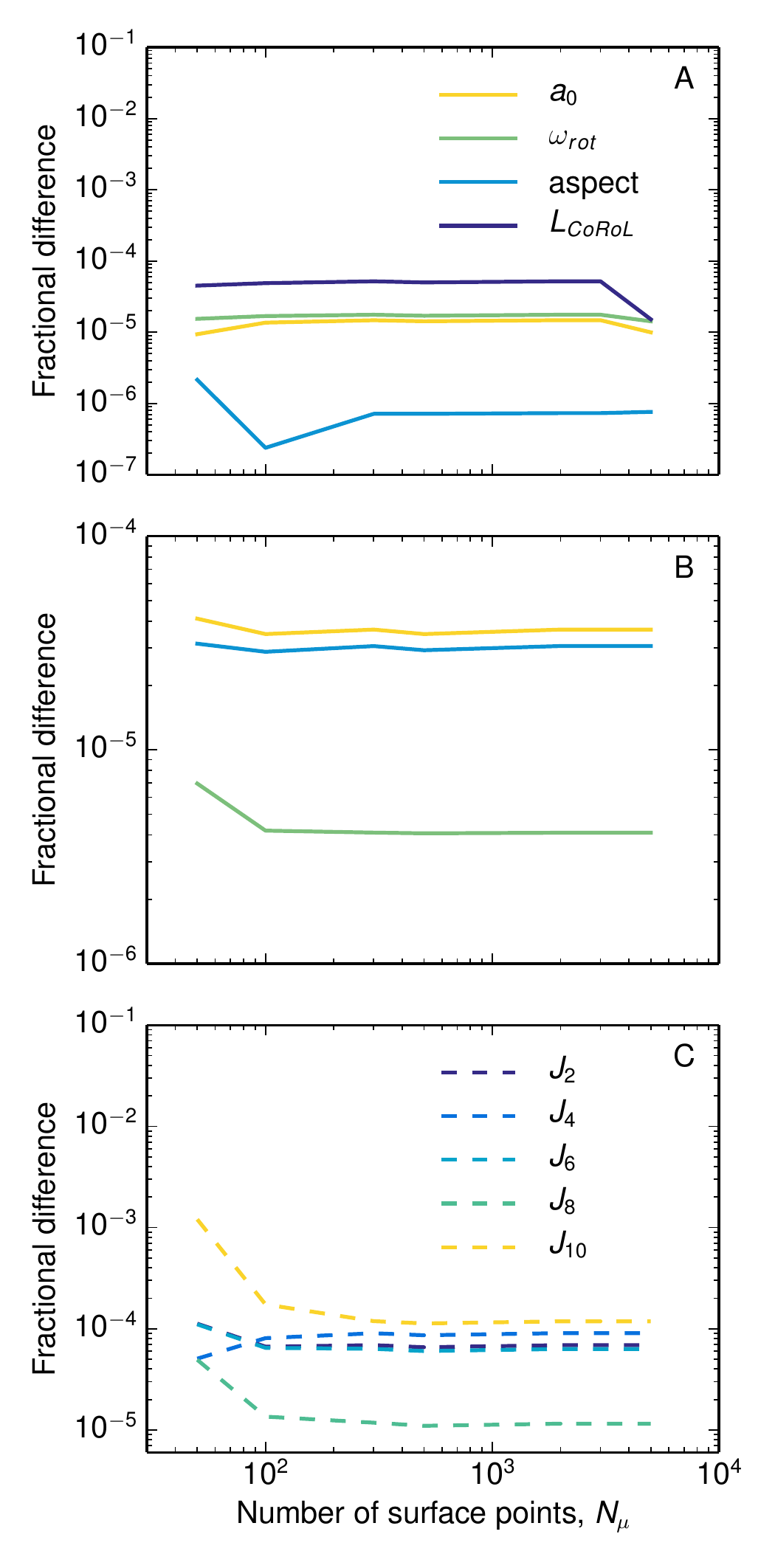}
\caption[]{The properties of structures and the CoRoL calculated using HERCULES with different numbers of surface points ($N_{\mu}$) are similar to those calculated using $N_{\mu}$~$=$~$1000$, the value used in this study. Shown are the fractional change in the properties at the CoRoL (A) and for a body just below the CoRoL with $L_{\rm tot}$~$=$~$1.45 L_{\rm EM}$ (B,C) for bodies with constant entropy silicate thermal profiles (I) and a silicate specific entropy of $S_{\rm lower}$~$=$~$6.5$~kJ~K$^{-1}$~kg$^{-1}$ in comparison to those calculated with $N_{\mu}$~$=$~$1000$. Lines are for different properties as given in the legends. The legend is the same in A and B.}
\label{sup:fig:HERCULES_Nmu}
\end{figure}

In HERCULES, a number of points, $N_{\mu}$, is used to describe each equipotential surface. For this study we have chosen to use $N_{\mu}$~$=$~$1000$. To test whether the value of $N_{\mu}$ used affects the calculated structure, we have used HERCULES to calculate the properties of bodies with constant entropy mantle thermal profile (class I) with a specific entropy of $S_{\rm lower}$~$=$~$6.5$~kJ~K$^{-1}$~kg$^{-1}$ using a range of $N_{\mu}$. The location of the CoRoL and the extrapolated properties at the CoRoL vary little compared to the $N_{\mu}$~$=$~$1000$ case for sufficiently large values of $N_{\mu}$ (Figure~\ref{sup:fig:HERCULES_Nmu}A). The properties of bodies below the CoRoL also does not vary significantly with different numbers of surface points (Figure~\ref{sup:fig:HERCULES_Nmu}B, C). We are therefore confident that the $N_{\mu}$ used in this study is sufficient for our purposes.

\begin{figure}
\centering
\includegraphics[scale=0.8333333]{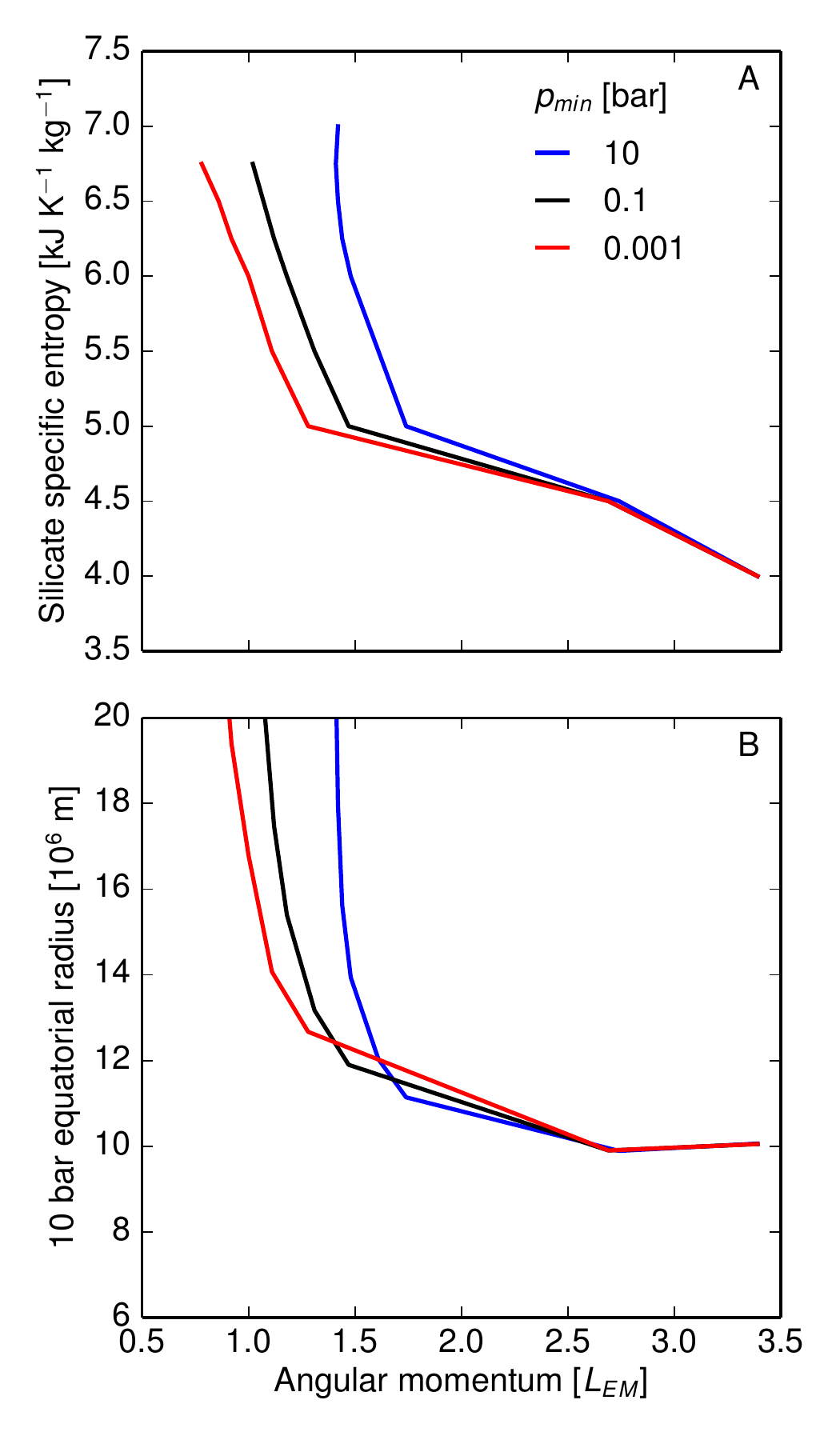}
\caption[]{The CoRoL is sensitive to the outermost thermal structure of a body. Shown is the CoRoL calculated using the same thermal structure but with varying bounding pressures used in the HERCULES code. The bodies are Earth mass with vapor atmosphere silicate thermal profiles (class II).}
\label{sup:fig:HSSL_pressure}
\end{figure}

\begin{figure}
\centering
\includegraphics[scale=0.8333333]{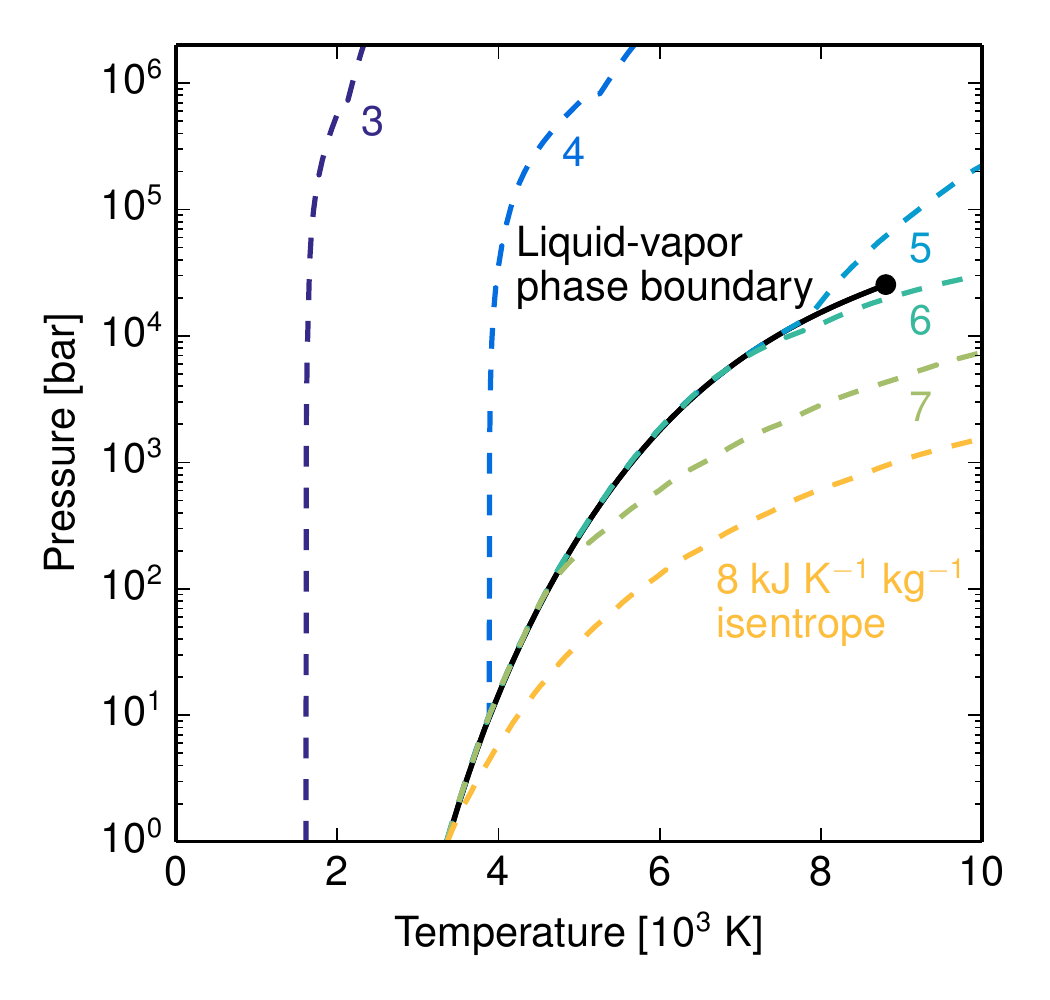}
\caption[]{Isentropes for the M-ANEOS derived forsterite EOS used in this work in pressure-temperature space. Each colored line is an isentrope for the specific entropy given by the number of the same color in kJ~K$^{-1}$~kg$^{-1}$. The black line is the liquid-vapor phase boundary. The black dot is the critical point.
}
\label{sup:fig:profiles_PT}
\end{figure}

\begin{figure}
\centering
\includegraphics[scale=0.8333333]{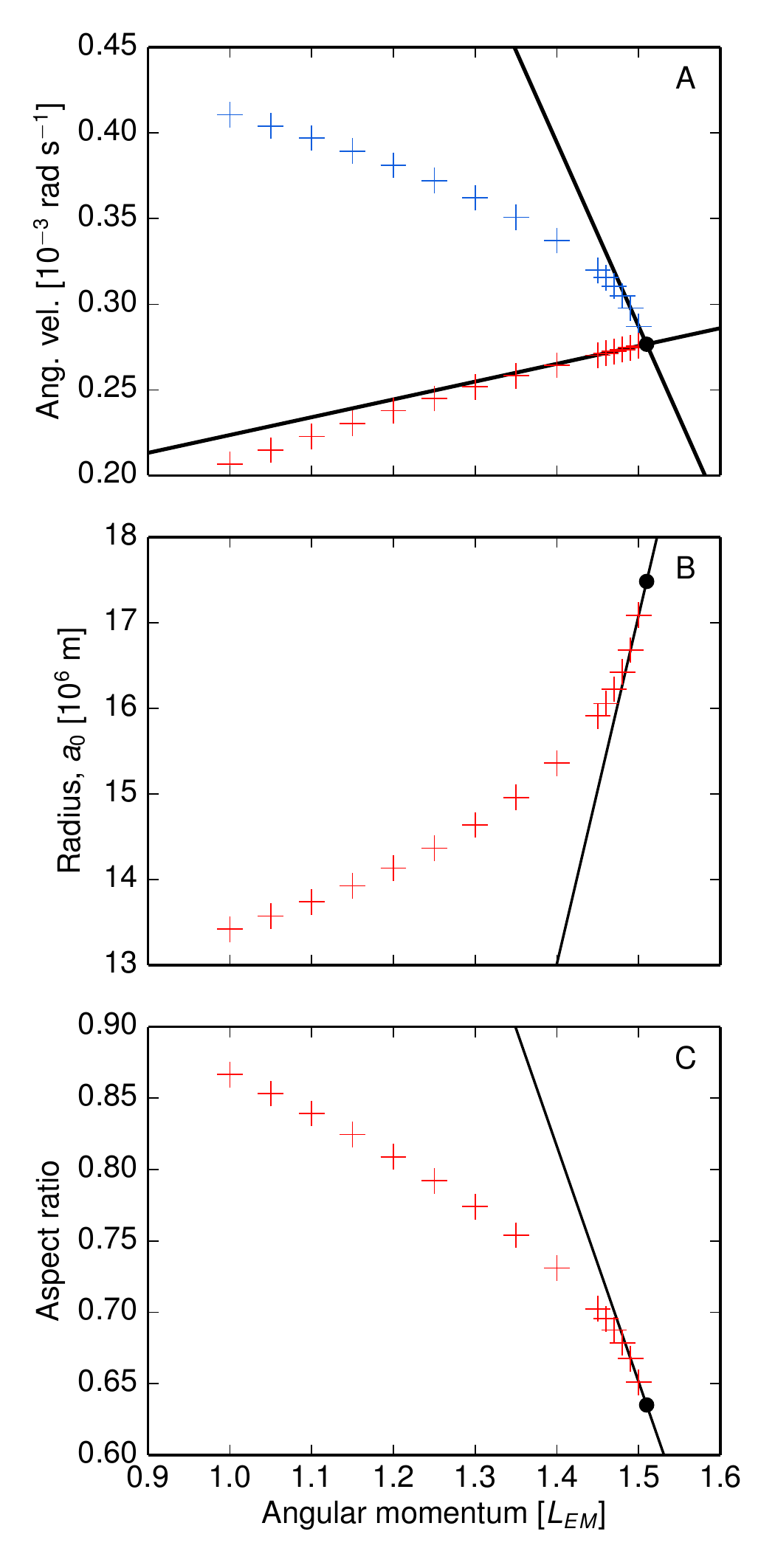}
\caption[]{An example of finding the CoRoL by extrapolation in AM (A). The Keplerian (blue) and corotating (red) angular velocities for the stable structures below the CoRoL are linearly extrapolated (black lines) to find the CoRoL where they intersect (black dot). The other properties of the body, such as the equatorial radius (B) and aspect ratio (C), at the CoRoL are then found by extrapolation.}
\label{sup:fig:CoRoLfind}
\end{figure}

We have also considered the effect of the exterior bounding pressure used in our model on the CoRoL.
We find that the CoRoL varies significantly based on the bounding pressure, particularly for high-entropy planets where the scale height of the outer portion of the planet is large (Figure~\ref{sup:fig:HSSL_pressure}).
This demonstrates once again the sensitivity of the CoRoL to the thermal structure of the planet.
The thermal profiles used in this study did not consider the intricacies of the structure at low pressure where the effects of radiative transfer could be significant \citep{Lock2016LPSC}.
Given the sensitivity of planetary structure to the thermal structure at low pressure, there must be further work done on understanding the thermal structure of partially vaporized bodies.
In this work, we chose the bounding pressure of $p_{\rm min}$~$=$~$10$~bar to allow easy comparison with SPH simulations where the numerical smoothing tends only to extend to moderate pressures. 

%xxxxxxxxxxxxxxxxxxxxxxxxxxxxxxxxxxxxxxxxxxxxxxxxxxxxxxxxxxxxxxxxxxxxxxxxxxxxxxxxxxxxxxxxxxxxxxxxxxxxxxxxxxxxxxxxxxxxxxxxxxxxxxxxxxxxxxxx
%xxxxxxxxxxxxxxxxxxxxxxxxxxxxxxxxxxxxxxxxxxxxxxxxxxxxxxxxxxxxxxxxxxxxxxxxxxxxxxxxxxxxxxxxxxxxxxxxxxxxxxxxxxxxxxxxxxxxxxxxxxxxxxxxxxxxxxxx

\section{A Modified Specific Energy for Giant Impacts, $Q_{\rm S}$}
\label{sup:sec:Qs}

A modified specific impact energy has been developed to estimate impact-induced melting \citep{Stewart2015} and atmospheric loss \citep{Stewart2014} for impacts between similar-sized bodies. The formulation is a variation on the specific energy developed in \citet{Leinhardt2012} that includes more geometric effects to estimate the energy deposited within the final body.

The specific impact energy, $Q_{\rm S}$, is defined by 
\begin{linenomath*}
\begin{equation}
Q_{\rm S} = Q'_{\rm R} \left ( 1 + \frac{M_{\rm p}}{M_{\rm t}} \right ) (1-b) \;\;,
\label{eqn:QS}
\end{equation}
\end{linenomath*}
where $Q'_{\rm R}$ is a center of mass specific impact energy modified to include only the interacting mass of the projectile (see
\cite{Leinhardt2012}). $M_{\rm p}$ and $M_{\rm t}$ are the mass of the
projectile and target respectively, and $b$ is the impact parameter.
$Q'_{\rm R}$ is given by
\begin{linenomath*}
\begin{equation}
  Q'_{\rm R} = \frac{\mu_{\alpha}}{\mu} Q_{\rm R} \;\;.
  \label{eqn:QR_prime}
\end{equation}
\end{linenomath*}
Note that there is a typographical error in the definition of $Q'_{\rm R}$ in equation 13 of \cite{Leinhardt2012}.
$Q_{\rm R}$ is the unmodified center of mass specific impact energy,
\begin{linenomath*}
\begin{equation}
  Q_{\rm R} = \frac{\mu V_{\rm i}^2}{2 M_{\rm tot}} \;\;.
  \label{eqn:QR}
\end{equation}
\end{linenomath*}
The reduced mass is defined as 
\begin{linenomath*}
\begin{equation}
  \mu = \frac{ M_{\rm p} M_{\rm t}}{ M_{\rm tot}} \;\;,
  \label{eqn:mu}
\end{equation}
\end{linenomath*}
and, to consider only the interacting fraction of the projectile, a modified reduced mass is used, given by
\begin{linenomath*}
\begin{equation}
  \mu_{\alpha} = \frac{ \alpha M_{\rm p} M_{\rm t}}{\alpha M_{\rm p} + M_{\rm t}} \;\;.
  \label{eqn:mu_alpha}
\end{equation}
\end{linenomath*}
$M_{\rm tot}$~$=$~$M_{\rm p}+M_{\rm t}$, $V_{\rm i}$ is the impact
velocity and $\alpha$ is the mass fraction of the projectile that is involved in the collision.  
$\alpha$ is defined as
\begin{linenomath*}
\begin{equation}
 \alpha = \frac{m_{\rm interact}}{M_{\rm p}} = \frac{3 R_{\rm p} l^2 -l^3}{4R_{\rm p}^3}  \;\;,
 \end{equation}
 \end{linenomath*}
where $m_{\rm interact}$ is the interacting projectile mass, $R_{\rm t}$ and $R_{\rm p}$ are the radii of the target and projectile and $B$~$=$~$(R_{\rm t}+R_{\rm p}) b$.   
$l$ is the projected length of the projectile overlapping the target, 
\begin{linenomath*}
\begin{equation}
l = \begin{dcases*}
      R_{\rm t}+R_{\rm p}-B   & when $B+R_{\rm p} > R_{\rm t}$\\
	2R_{\rm p}   & when $B+R_{\rm p} \leq R_{\rm t}$	
        \end{dcases*} \;\;.
\end{equation}
\end{linenomath*}
If $B+R_{\rm p}$~$\leq$~$R_{\rm t}$ then the whole projectile is interacting with the target and $\alpha$~$=$~$1$.

Each of the terms in the definition of $Q_{\rm S}$ (Equation \ref{eqn:QS})
takes into account a factor that affects how efficiently energy is
coupled into the shock pressure field in the impacting bodies. Thus the parameter is proportional to the entropy increase in the impacting bodies. The first factor corrects for the fact
that a larger volume is shocked to peak pressure for more equal sized
impacts.  The second factor accounts for the fact that grazing impacts
less efficiently couple impact energy into the target body, leading to
a smaller volume reaching the highest shock pressures. 

%xxxxxxxxxxxxxxxxxxxxxxxxxxxxxxxxxxxxxxxxxxxxxxxxxxxxxxxxxxxxxxxxxxxxxxxxxxxxxxxxxxxxxxxxxxxxxxxxxxxxxxxxxxxxxxxxxxxxxxxxxxxxxxxxxxxxxxxx
%xxxxxxxxxxxxxxxxxxxxxxxxxxxxxxxxxxxxxxxxxxxxxxxxxxxxxxxxxxxxxxxxxxxxxxxxxxxxxxxxxxxxxxxxxxxxxxxxxxxxxxxxxxxxxxxxxxxxxxxxxxxxxxxxxxxxxxxx
%xxxxxxxxxxxxxxxxxxxxxxxxxxxxxxxxxxxxxxxxxxxxxxxxxxxxxxxxxxxxxxxxxxxxxxxxxxxxxxxxxxxxxxxxxxxxxxxxxxxxxxxxxxxxxxxxxxxxxxxxxxxxxxxxxxxxxxxx
%Figures not associated with a section

%xxxxxxxxxxxxxxxxxxxxxxxxxxxxxxxxxxxxxxxxxxxxxxxxxxxxxxxxxxxxxxxxxxxxxxxxxxxxxxxxxxxxxxxxxxxxxxxxxxxxxxxxxxxxxxxxxxxxxxxxxxxxxxxxxxxxxxxx
%xxxxxxxxxxxxxxxxxxxxxxxxxxxxxxxxxxxxxxxxxxxxxxxxxxxxxxxxxxxxxxxxxxxxxxxxxxxxxxxxxxxxxxxxxxxxxxxxxxxxxxxxxxxxxxxxxxxxxxxxxxxxxxxxxxxxxxxx
%xxxxxxxxxxxxxxxxxxxxxxxxxxxxxxxxxxxxxxxxxxxxxxxxxxxxxxxxxxxxxxxxxxxxxxxxxxxxxxxxxxxxxxxxxxxxxxxxxxxxxxxxxxxxxxxxxxxxxxxxxxxxxxxxxxxxxxxx

\begin{table}
\caption[SPH simulation results]{Summary of SPH impact simulations and properties of their post-impact states. For each impact, the table includes: an index number; impact classification number; target mass $M_{\rm t}$; number of SPH particles in target, $N_{\rm t}$; target equatorial radius, $R_{\rm t}$; target angular momentum, $L_{\rm t}$; target angular velocity $\omega_{\rm t}$; target spin period, $T_{\rm t}$; projectile mass, $M_{\rm p}$; number of SPH particles in projectile, $N_{\rm p}$; projectile radius, $R_{\rm p}$; projectile angular momentum, $L_{\rm p}$; projectile angular velocity, $\omega_{\rm p}$; projectile spin period, $T_{\rm p}$; impact velocity, $V_{\rm i}$; impact parameter, $b$; modified specific energy, $Q_{\rm S}$; final simulation time; bound mass of post-impact structure, $M_{\rm bnd}$; bound mass angular momentum, $L_{\rm bnd}$; angular velocity of the dense ($\rho$~$>$~$1000$~kg~m$^{-3}$) region of post-impact structure, $\omega_{\rho}$; spin period of dense region, $T_{\rho}$; mass outside Roche limit in unmodified post-impact structure, $M_{\rm Roche}$; mass outside Roche limit after the conventional analysis following the method of \citet{Cuk2012}, $M_{\rm Roche}^{\rm conv}$; average specific entropy of silicate in the post-impact structure, $S_{\rm avg}$; average specific entropy of lowest pressure 50 wt\% of silicate in post-impact structure, $S_{50\%}$; average specific entropy of lowest pressure 25 wt\% of silicate in post-impact structure, $S_{25\%}$; and post-impact structure dynamical class. The impact classification number indicates the overall geometry of the event, which we have collected into 4 groups: (0) small impactors onto pre-rotating targets \citep{Cuk2012}, (1) canonical Moon-forming impacts \citep{Canup2001,Canup2004}, (2) similar mass grazing impacts \citep{Canup2012}, and (3) other grazing impacts that span the parameter space of giant impacts from \citet{Quintana2016}. The impacts that formed the example structures shown in Figures~\ref{fig:MAD} and \ref{fig:PI_pressure} are indicated by stars.}
\label{sup:tab:impacts}
\end{table}

%xxxxxxxxxxxxxxxxxxxxxxxxxxxxxxxxxxxxxxxxxxxxxxxxxxxxxxxxxxxxxxxxxxxxxxxxxxxxxxxxxxxxxxxxxxxxxxxxxxxxxxxxxxxxxxxxxxxxxxxxxxxxxxxxxxxxxxxx
%xxxxxxxxxxxxxxxxxxxxxxxxxxxxxxxxxxxxxxxxxxxxxxxxxxxxxxxxxxxxxxxxxxxxxxxxxxxxxxxxxxxxxxxxxxxxxxxxxxxxxxxxxxxxxxxxxxxxxxxxxxxxxxxxxxxxxxxx
%xxxxxxxxxxxxxxxxxxxxxxxxxxxxxxxxxxxxxxxxxxxxxxxxxxxxxxxxxxxxxxxxxxxxxxxxxxxxxxxxxxxxxxxxxxxxxxxxxxxxxxxxxxxxxxxxxxxxxxxxxxxxxxxxxxxxxxxx 
\begin{table}
\caption[CoRoL boundaries]{Summary of the corotation limit, CoRoL, for bodies of varying thermal profiles, masses, resolutions and bounding pressures. For each body, the table includes: planet mass, $M$; mass fraction of lower silicate, $f_{\rm lower}$; bounding pressure of structure, $p_{\rm min}$; number of points used to describe each surface, $_{\mu}$; number of core layers, $N_{\rm lay}^{\rm core}$; number of lower silicate layers, $N_{\rm lay}^{\rm lower}$; number of upper silicate layers, $N_{\rm lay}^{\rm upper}$; core entropy, $S_{\rm core}$; thermal profile class; lower silicate specific entropy, $S_{\rm lower}$; upper silicate specific entropy, $S_{\rm upper}$; the AM step used to find the CoRoL; AM of the CoRoL boundary, $L_{\rm CoRoL}$; equatorial radius of body at CoRoL, $a_{\rm CoRoL}$; aspect ratio of body at CoRoL; angular velocity at CoRoL, $\omega_{\rm CoRoL}$; and spin period at CoRoL, $T_{\rm CoRoL}$. The properties of the body at CoRoL reported are found by extrapolation. The different thermal profiles are discussed in \S\ref{sec:profiles}.}  
\label{sup:tab:HSSL}
\end{table}

%xxxxxxxxxxxxxxxxxxxxxxxxxxxxxxxxxxxxxxxxxxxxxxxxxxxxxxxxxxxxxxxxxxxxxxxxxxxxxxxxxxxxxxxxxxxxxxxxxxxxxxxxxxxxxxxxxxxxxxxxxxxxxxxxxxxxxxxx
%xxxxxxxxxxxxxxxxxxxxxxxxxxxxxxxxxxxxxxxxxxxxxxxxxxxxxxxxxxxxxxxxxxxxxxxxxxxxxxxxxxxxxxxxxxxxxxxxxxxxxxxxxxxxxxxxxxxxxxxxxxxxxxxxxxxxxxxx
%xxxxxxxxxxxxxxxxxxxxxxxxxxxxxxxxxxxxxxxxxxxxxxxxxxxxxxxxxxxxxxxxxxxxxxxxxxxxxxxxxxxxxxxxxxxxxxxxxxxxxxxxxxxxxxxxxxxxxxxxxxxxxxxxxxxxxxxx
\section*{Supplementary References}
\bibliographystyle{elsarticle-harv} 
%IF NEEDED USE A SEPERATE BIB FILE
\bibliography{References}

\end{document}